\documentclass[a4paper,UKenglish,cleveref, autoref, thm-restate]{lipics-v2021}

\hideLIPIcs  

\title{Improved Circular Dictionary Matching} 

\author{Nicola {Cotumaccio}}{University of Helsinki, Finland}{nicola.cotumaccio@helsinki.fi}{https://orcid.org/0000-0002-1402-5298}{}

\authorrunning{N. Cotumaccio} 

\Copyright{Nicola Cotumaccio} 



\category{} 

\relatedversion{} 

\funding{Funded by the Helsinki Institute for Information
Technology (HIIT).}

\acknowledgements{I would like to thank Kunihiko Sadakane for introducing me to the topic.}

\nolinenumbers 

\usepackage{array}
\usepackage{algorithmic}
\usepackage{mathrsfs}
 \usepackage{multirow}
 \usepackage{blindtext}
\usepackage{hyperref}
\usepackage{graphicx}
\usepackage{todonotes}
\usepackage{tabularx}

\usetikzlibrary{arrows,automata}
\usepackage{tikz}
\usepackage{tikz-cd}

\usepackage{wrapfig}

\newcommand{\lcp}{\mathop{\mathsf{lcp}}}
\newcommand{\LCP}{\mathsf{LCP}}
\newcommand{\Str}{\mathsf{Str}}
\newcommand{\Suff}{\mathsf{Suff}}
\newcommand{\SA}{\mathsf{SA}}

\newcommand{\Cdm}{\mathsf{Cdm}}
\newcommand{\pred}{\mathsf{pred}}
\newcommand{\bws}{\mathsf{bws}}
\newcommand{\BWT}{\mathsf{BWT}}
\newcommand{\back}{\mathsf{back}}
\newcommand{\access}{\mathsf{access}}
\newcommand{\rank}{\mathsf{rank}}
\newcommand{\select}{\mathsf{select}}
\newcommand{\rmq}{\mathsf{rmq}}
\newcommand{\prev}{\mathsf{prev}}
\newcommand{\follow}{\mathsf{follow}}
\newcommand{\parent}{\mathsf{parent}}
\newcommand{\lca}{\mathsf{lca}}
\newcommand{\rightmost}{\mathsf{rightmost}}
\newcommand{\leftmost}{\mathsf{leftmost}}
\newcommand{\toaux}{\mathsf{toaux}}
\newcommand{\fromaux}{\mathsf{fromaux}}
\newcommand{\nma}{\mathsf{nma}}
\newcommand{\open}{\mathsf{open}}
\newcommand{\frominter}{\mathsf{frominter}}
\newcommand{\tointer}{\mathsf{tointer}}
\newcommand{\Len}{\mathsf{Len}}

\EventEditors{Paola Bonizzoni and Veli M\"{a}kinen}
\EventNoEds{2}
\EventLongTitle{36th Annual Symposium on Combinatorial Pattern Matching (CPM 2025)}
\EventShortTitle{CPM 2025}
\EventAcronym{CPM}
\EventYear{2025}
\EventDate{June 17--19, 2025}
\EventLocation{Milan, Italy}
\EventLogo{}
\SeriesVolume{331}
\ArticleNo{19}

\begin{document}

\maketitle

\begin{abstract}
The circular dictionary matching problem is an extension of the classical dictionary matching problem where every string in the dictionary is interpreted as a circular string: after reading the last character of a string, we can move back to its first character. The circular dictionary matching problem is motivated by applications in bioinformatics and computational geometry.

In 2011, Hon et al. [ISAAC 2011] showed how to efficiently solve circular dictionary matching queries within compressed space by building on Mantaci et al.'s eBWT and Sadakane's compressed suffix tree. The proposed solution is based on the assumption that the strings in the dictionary are all distinct and non-periodic, no string is a circular rotation of some other string, and the strings in the dictionary have similar lengths.

In this paper, we consider \emph{arbitrary} dictionaries, and we show how to solve circular dictionary matching queries in $ O((m + occ) \log n) $ time within compressed space using $ n \log \sigma (1 + o(1)) + O(n) + O(d \log n) $ bits, where $ n $ is the total length of the dictionary, $ m $ is the length of the pattern, $ occ $ is the number of occurrences, $ d $ is the number of strings in the dictionary and $ \sigma $ is the size of the alphabet. Our solution is based on an extension of the suffix array to arbitrary dictionaries and a sampling mechanism for the LCP array of a dictionary inspired by recent results in graph indexing and compression.
\end{abstract}


\section{Introduction}\label{sec:introduction}

The Burrows-Wheeler transform (BWT) \cite{burrows1994} and the FM-index \cite{ferragina2005jacm} support pattern matching on the compressed representation of a \emph{single} string. If we want to encode a \emph{collection} $ \mathcal{T} $ of strings, we may append a distinct end-of-string $ \$_i $ to each string and store the Burrows-Wheeler Transform of the concatenation of the strings. This approach is only a naive extension of the BWT and can significantly increase the size of the alphabet. In 2007, Mantaci et al. introduced the eBWT \cite{mantaci2007}, a more sophisticated and elegant extension of the BWT to a collection of strings. When sorting all suffixes of all strings in the collection, Mantaci et al. define the mutual order between two suffixes $ T_i $ and $ T_j $ to be the mutual order of $ T_i^\omega $ and $ T_j^\omega $ in the lexicographic order, where $ T_i^\omega $ and $ T_j^\omega $ are the infinite strings obtained by concatenating $ T_i $ with itself and $ T_j $ with itself infinitely many times (see \cite{cenzato2023dcc} for a recent paper on different variants of the eBWT). From the definition of the eBWT we obtain that, if we extend the backward search mechanism of the FM-index to multisets of strings, we are not matching a pattern $ P $ against all suffixes $ T_i $'s, but against all strings $ T_i^\omega $'s. Equivalently, we are interpreting each string in the collection as a \emph{circular} string in which, after reaching the last character of the string, we can go back to its first character.

In 2011, Hon et al. applied the framework of the eBWT to solve \emph{circular dictionary matching} queries \cite{hon2011isaac}, even if they explicitly spotted the relationship between their techniques (which extends Sadakane's compressed suffix tree \cite{sadakane2007} to a collection of strings) and the eBWT only in a subsequent paper \cite{hon2012cpm}. The circular dictionary matching problem is an extension of the \emph{dictionary matching problem}, which admits a classical solution based on Aho-Corasick automata \cite{ahocorasick1975}, as well as more recent solutions within compressed space \cite{belazzogui2010cpm, hon2013tcs}. Other variations of the same problem include dictionary matching with gaps \cite{amir2015, hon2018dictionary, amir2019} or mismatches \cite{gawrychowski2019streaming}, dictionary matching in the streaming model \cite{clifford2015}, dynamic dictionary matching \cite{golan2019} and internal dictionary matching \cite{charalampopoulos2021}. The circular dictionary matching problem is motivated by some applications, see \cite{iliopoulos2008}. In bioinformatics, the genome of herpes simplex virus (HSV-1) and many other viruses exists as circular strings \cite{strang2005circularization}, and microbial samples collected from the environment directly without isolating and culturing the samples have circular chromosomes \cite{eisen2007environmental, simon2011metagenomic}.  In computational geometry, a polygon can be stored by listing its vertices in clockwise order. The circular dictionary matching problem has also been studied from other perspectives (average-case behavior \cite{iliopoulos2017}, approximate variants \cite{charalampopoulos2022}).

\subsection{Our Contribution}

Consider a dictionary $ \mathcal{T} = (T_1, T_2, \dots, T_d) $ of total length $ n $ on an alphabet of size $ \sigma $. In \cite{hon2011isaac}, Hon et al. show that, by storing a data structure of $ n \log \sigma (1 + o(1)) + O(n) + O(d \log n) $ bits, it is possible to solve circular dictionary matching queries in $ O((m + occ) \log^{1 + \epsilon} n) $ time, where $ m $ is the length of the pattern. This result holds under three assumptions: (i) the $ T_h $'s are distinct and not periodic, (ii) no $ T_h $ is a circular rotation of some other $ T_{h'} $, and (iii) the $ T_h $'s have bounded aspect ratio, namely, $ (\max_{h = 1}^d T_h) / (\min_{h = 1}^d T_h) = O(1) $. Assumption (ii) was made explicit only in the subsequent paper \cite{hon2012cpm}, where Hon et al. also mention that, in an extension of the paper, they will show how to remove assumptions (i-ii). Assumption (iii) is required to store a space-efficient data structure supporting longest common prefix (LCP) queries. In \cite{hon2013cpm}, Hon et al. sketched a new data structure for LCP queries that uses $ O(n + d \log n) $ bits without requiring assumption (iii), but all details are again deferred to an extended version.

The main result of our paper is Theorem \ref{theor:main}. In particular, we give two main contributions.
\begin{itemize}
    \item We obtain results that are valid for an arbitrary dictionary $ \mathcal{T} $, without imposing any restriction. To this end, in Section \ref{sec:suffixarray} we introduce a more general compressed suffix array for an arbitrary collection of strings. In particular, we support LCP functionality within only $ O(n) + o(n \log \sigma) $ bits by using a sampling mechanism inspired by recent results \cite{cotumaccio2023spire, conte2022dcc} on Wheeler automata.
    \item We provide a self-contained proof of our result and a full example of the data structure for solving circular dictionary matching queries. The original paper \cite{hon2011isaac} contains the main ideas to extend Sadakane's compressed suffix tree to a multiset of strings, but it is very dense and proofs are often only sketched or missing, which may explain why additional properties to potentially remove assumptions (i - iii) were only observed in subsequent papers. We will provide an intuitive explanation of several steps by using graphs (and in particular cycles, see Figure \ref{fig:graphexample}), consistently with a research line that has shown how the suffix array, the Burrows-Wheeler transform and the FM-index admit a natural interpretation in the graph setting \cite{gagie2017, alanko2020soda, cotumacciosoda2021, cotumaccio2023jacm, cotumaccio2022dcc, alanko2024cpm, alanko2022cpm, cotumaccio2024pnse, cotumaccio2024stacs}.
\end{itemize}
We will also show that the time bound of Theorem \ref{theor:main} can be improved when the number of occurrences is large.

The paper is organized as follows. In Section \ref{sec:preliminaries} we introduce the circular dictionary matching problem. In Section \ref{sec:compressedsuffix} we define our compressed suffix tree. In Section \ref{sec:solvingcirculad} we present an algorithm for circular dictionary matching. Due to space constraints, most proofs and some auxiliary results can be found in the appendix.

\section{Preliminaries}\label{sec:preliminaries}

\begin{figure}
\captionsetup[subfigure]{justification=centering}
\centering
    \begin{subfigure}[b]{1 \textwidth}
    \centering
        \scalebox{0.9}{
\begin{tabularx}{1 \textwidth}{|c|>{\centering\arraybackslash}X>{\centering\arraybackslash}X>{\centering\arraybackslash}X>{\centering\arraybackslash}X>{\centering\arraybackslash}X>{\centering\arraybackslash}X|>{\centering\arraybackslash}X>{\centering\arraybackslash}X>{\centering\arraybackslash}X>{\centering\arraybackslash}X>{\centering\arraybackslash}X|>{\centering\arraybackslash}X>{\centering\arraybackslash}X>{\centering\arraybackslash}X|}
\hline
$ \mathcal{T} $ & $ a $ & $ b $ & $ c $ & $ a $ & $ b $ & $ c $ & $ b $ & $ c $ & $ a $ & $ b $ & $ c $ & $ c $ & $ a $ & $ b $  \\
\hline
$ k $ & $ 1 $ & $ 2 $ & $ 3 $ & $ 4 $ & $ 5 $ & $ 6 $ & $ 7 $ & $ 8 $ & $ 9 $ & $ 10 $ & $ 11 $ & $ 12 $ & $ 13 $ & $ 14 $ \\
\hline
$ B_1 [k] $ & $ 1 $ & $ 0 $ & $ 0 $ & $ 0 $ & $ 0 $ & $ 0 $ & $ 1 $ & $ 0 $ & $ 0 $ & $ 0 $ & $ 0 $ & $ 1 $ & $ 0 $ & $ 0 $ \\
\hline
\end{tabularx}
}
\caption{}
\label{fig:examplebasic}
\end{subfigure}

\bigskip

\begin{subfigure}[b]{1\textwidth}
    \centering
	\scalebox{0.9}{
\begin{tabularx}{1 \textwidth}{|>{\centering\arraybackslash}X>{\centering\arraybackslash}X>{\centering\arraybackslash}X>{\centering\arraybackslash}X>{\centering\arraybackslash}X>{\centering\arraybackslash}X>{\centering\arraybackslash}X>{\centering\arraybackslash}X>{\centering\arraybackslash}X>{\centering\arraybackslash}X|>{\centering\arraybackslash}X|>{\centering\arraybackslash}X|c|c|c|c|}
\hline
\multicolumn{10}{|c|}{$ \mathcal{S}_j $} & $ D_j $ & $ j $ & $ \LCP[j] $ & $ \BWT[j] $ & $ \BWT^*[j] $ & $ B_2[j] $ \\
\hline
$ \color{orange}{a} $ & $ \color{orange}{b} $ & $ \color{orange}{c} $ & $ \color{orange}{a} $ & $ \color{orange}{b} $ & $ \color{orange}{c} $ & $ a $ & $ b $ & $ c $ & $ \dots $ & $ 1 $ & \multirow{3}{*}{$ 1 $} & \multirow{3}{*}{$ $} & \multirow{3}{*}{$ c $} & \multirow{3}{*}{$ a $} & \multirow{3}{*}{$ 1 $} \\
$ \color{orange}{a} $ & $ \color{orange}{b} $ & $ \color{orange}{c} $ & $ \color{orange}{a} $ & $ \color{orange}{b} $ & $ \color{orange}{c} $ & $ a $ & $ b $ & $ c $ & $ \dots $ & $ 4 $ & & & & & \\
$ \color{orange}{a} $ & $ \color{orange}{b} $ & $ \color{orange}{c} $ & $ a $ & $ b $ & $ c $ & $ a $ & $ b $ & $ c $ & $ \dots $ & $ 13 $ & & & & & \\
\hline
$ \color{orange}{a} $ & $ \color{orange}{b} $ & $ \color{orange}{c} $ & $ \color{orange}{b} $ & $ \color{orange}{c} $ & $ a $ & $ b $ & $ c $ & $ b $ & $ \dots $ & $ 9 $ & \multirow{1}{*}{$ 2 $} & \multirow{1}{*}{$ 3 $} & \multirow{1}{*}{$ c $} & \multirow{1}{*}{$ a $} & \multirow{1}{*}{$ 0 $} \\
\hline
$ \color{orange}{b} $ & $ \color{orange}{c} $ & $ \color{orange}{a} $ & $ \color{orange}{b} $ & $ \color{orange}{c} $ & $ \color{orange}{a} $ & $ b $ & $ c $ & $ a $ & $ \dots $ & $ 2 $ & \multirow{3}{*}{$ 3 $} & \multirow{3}{*}{$ 0 $} & \multirow{3}{*}{$ a $} & \multirow{3}{*}{$ b $} & \multirow{3}{*}{$ 1 $} \\
$ \color{orange}{b} $ & $ \color{orange}{c} $ & $ \color{orange}{a} $ & $ \color{orange}{b} $ & $ \color{orange}{c} $ & $ \color{orange}{a} $ & $ b $ & $ c $ & $ a $ & $ \dots $ & $ 5$ & & & & & \\
$ \color{orange}{b} $ & $ \color{orange}{c} $ & $ \color{orange}{a} $ & $ b $ & $ c $ & $ a $ & $ b $ & $ c $ & $ a $ & $ \dots $ & $ 14 $ & & & & & \\
\hline
$ \color{orange}{b} $ & $ \color{orange}{c} $ & $ \color{orange}{a} $ & $ \color{orange}{b} $ & $ \color{orange}{c} $ & $ b $ & $ c $ & $ a $ & $ b $ & $ \dots $ & $ 7 $ & \multirow{1}{*}{$ 4 $} & \multirow{1}{*}{$ 5 $} & \multirow{1}{*}{$ c $} & \multirow{1}{*}{$ b $} & \multirow{1}{*}{$ 0 $} \\
\hline
$ \color{orange}{b} $ & $ \color{orange}{c} $ & $ \color{orange}{b} $ & $ \color{orange}{c} $ & $ \color{orange}{a} $ & $ b $ & $ c $ & $ b $ & $ c $ & $ \dots $ & $ 10 $ & \multirow{1}{*}{$ 5 $} & \multirow{1}{*}{$ 2 $} & \multirow{1}{*}{$ a $} & \multirow{1}{*}{$ b $} & \multirow{1}{*}{$ 0 $} \\
\hline
$ \color{orange}{c} $ & $ \color{orange}{a} $ & $ \color{orange}{b} $ & $ \color{orange}{c} $ & $ \color{orange}{a} $ & $ \color{orange}{b} $ & $ c $ & $ a $ & $ b $ & $ \dots $ & $ 3 $ & \multirow{3}{*}{$ 6 $} & \multirow{3}{*}{$ 0 $} & \multirow{3}{*}{$ b $} & \multirow{3}{*}{$ c $} & \multirow{3}{*}{$ 1 $} \\
$ \color{orange}{c} $ & $ \color{orange}{a} $ & $ \color{orange}{b} $ & $ \color{orange}{c} $ & $ \color{orange}{a} $ & $ \color{orange}{b} $ & $ c $ & $ a $ & $ b $ & $ \dots $ & $ 6 $ & & & & & \\
$ \color{orange}{c} $ & $ \color{orange}{a} $ & $ \color{orange}{b} $ & $ c $ & $ a $ & $ b $ & $ c $ & $ a $ & $ b $ & $ \dots $ & $ 12 $ & & & & & \\
\hline
$ \color{orange}{c} $ & $ \color{orange}{a} $ & $ \color{orange}{b} $ & $ \color{orange}{c} $ & $ \color{orange}{b} $ & $ c $ & $ a $ & $ b $ & $ c $ & $ \dots $ & $ 8 $ & \multirow{1}{*}{$ 7 $} & \multirow{1}{*}{$ 4 $} & \multirow{1}{*}{$ b $} & \multirow{1}{*}{$ c $} & \multirow{1}{*}{$ 0 $} \\
\hline
$ \color{orange}{c} $ & $ \color{orange}{b} $ & $ \color{orange}{c} $ & $ \color{orange}{a} $ & $ \color{orange}{b} $ & $ c $ & $ b $ & $ c $ & $ a $ & $ \dots $ & $ 11 $ & \multirow{1}{*}{$ 8 $} & \multirow{1}{*}{$ 1 $} & \multirow{1}{*}{$ b $} & \multirow{1}{*}{$ c $} & \multirow{1}{*}{$ 0 $} \\
\hline
\end{tabularx}
}
\caption{}
\label{fig:maintable}
\end{subfigure}
\caption{Consider the dictionary $ \mathcal{T} = (abcabc, bcabc, cab) $, our running example. In (b), the strings $ \mathcal{S}_j $'s are sorted lexicographically, and every block identifies strings $ T_k $'s that correspond to the same $ \mathcal{S}_j $. Each string $ T_k $ is the orange prefix of $ \mathcal{S}_j $. For example, we have $ 4 \in D_1 $, $ \mathcal{T}_4 = abcabc $ and $ \mathcal{S}_1 = (abc)^\omega $.}
\label{fig:referencetobothmain}
\end{figure}

Let $ \Sigma $ be a finite alphabet of size $ \sigma = |\Sigma| $. We denote by $ \Sigma^* $ the set of all finite strings on $ \Sigma $ (including the empty string $ \epsilon $) and by $ \Sigma^\omega $ the set of all countably infinite strings on $ \Sigma $. For example, if $ \Sigma = \{a, b \} $, then $ abbba $ is a string in $ \Sigma^* $ and $ ababab\dots $ is a string in $ \Sigma^\omega $. If $ T \in \Sigma^* $, we denote by $ |T| $ the length of $ T $. If $ 1 \le i \le |T| $, we denote by $ T[i] $ the $ i $-th character of $ T $, and if $ 1 \le i \le j \le |T| $, we define $ T[i, j] = T[i] T[i + 1] \dots T[j - 1] T[j] $. If $ i > j $, let $ T[i, j] = \epsilon $. Analogously, if $ T \in \Sigma^\omega $, then $ T[i] $ is the $ i $-th character of $ T $, we have $ T[i, j] = T[i] T[i + 1] \dots T[j - 1] T[j] $ for every $ j \ge i \ge 1 $, and if $ i > j $ we have $ T[i, j] = \epsilon $.

If $ T \in \Sigma^* $, then for every integer $ z \ge 1 $ the string $ T^z \in \Sigma^* $ is defined by $ T^z = T T \dots T $, where $ T $ is repeated $ z $ times, and the string $ T^\omega \in \Sigma^\omega $ is defined by $ T^\omega = T T T \dots $, where $ T $ is repeated infinitely many times. Given $ T_1, T_2 \in \Sigma^* \cup \Sigma^\omega $, let $ \lcp(T_1, T_2) $ be the length of the longest common prefix between $ T_1 $ and $ T_2 $.

We say that a string $ T \in \Sigma^* $ is \emph{primitive} is for every $ S \in \Sigma^* $ and for every integer $ z \ge 1 $, if $ T = S^z  $, then $ z = 1 $ and $ T = S $. For every $ T \in \Sigma^* $, there exists exactly one primitive string $ R \in \Sigma^* $ and exactly one integer $ z \ge 1 $ such that $ T = R^z $ (see \cite[Prop. 1.3.1]{lothaire1997}; we say that $ R $ is the \emph{root} of $ T $ and we write $ R = \rho(T) $.

Let $ T \in \Sigma^* $ a string of length $ n $. We define the following queries on $ T $: (i) $ \access (T, i) $: given $ 1 \le i \le n $, return $ T[i] $; (ii) $ \rank_c (T, i) $: given $ c \in \Sigma $ and $ 1 \le i \le n $, return $ |\{1 \le h \le i \;|\; T[h] = c \}| $; (iii) $ \select_c (T, i) $: given $ c \in \Sigma $ and $ 1 \le i \le  \rank_c (T, n) $, return the unique $ 1 \le j \le n $ such that $ T[j] = c $ and $ |\{1 \le h \le j \;|\; T[h] = c \}| = i $. To handle limit cases easily, it is expedient to define $ \select_c (T, \rank_c (T, n) + 1) = n + 1 $. A bit array of length $ n $ can be stored using a data structure (called \emph{bitvector}) of $ n + o(n) $ bits that supports $ \access $, $ \rank $ and $ \select $ queries in $ O(1) $ time \cite{navarro2016book}.

We consider a fixed total order $ \preceq $ on $ \Sigma $ and we extend it to $ \Sigma^* \cup \Sigma^\omega $ \emph{lexicographically}. For every $ T_1, T_2 \in \Sigma^* \cup \Sigma^\omega $, we write $ T_1 \prec T_2 $ if $ (T_1 \preceq T_2) \land (T_1 \not = T_2) $. In our examples (see e.g. Figure \ref{fig:referencetobothmain}), $ \Sigma $ is a subset of the English alphabet and $ \preceq $ is the usual order on letters. In our data structures, $ \Sigma = \{0, 1, \dots, |\Sigma| - 1 \} $ and $ \preceq $ is the usual order such that $ 0 \prec 1 \prec \dots \prec |\Sigma| - 1 $.

\subsection{Circular Dictionary Matching}\label{sec:circularpatternmatching}

Let $ T, P \in \Sigma^* $ and $ P $. For every $ 1 \le i \le |P| - |T| + 1 $, we say that $ T $ \emph{occurs} in $ P $ at position $ i $ if  $ P[i, i + |T| - 1] = T $.

Let $ T \in \Sigma^* $ and let $ 1 \le i \le |T| $. The \emph{circular suffix} $ T_i $ of $ T $ is the string $ T[i, |T|]T[1, i - 1] $. For example, if $ T = cab $, then $ T_1 = cab $, $ T_2 = abc $ and $ T_3 = bca $.

Let $ d \ge 1 $, and let $ T_1, T_2, \dots, T_d \in \Sigma^* $ be nonempty strings. We say that $ \mathcal{T} = (T_1, T_2, \dots, T_d) $ is a \emph{dictionary}. We assume that the alphabet $ \Sigma $ is \emph{effective}, that is, every character in $ \Sigma $ occurs in some $ T_j $ (our results could be extended to larger alphabets by using standard tools such as dictionaries \cite{navarro2016book, cotumaccio2023jacm}). Define the \emph{total length} $ n $ of $ \mathcal{T} $ to be $ n = \sum_{k = 1}^d |T_k| $. In the example of Figure \ref{fig:examplebasic}, we have $ d = 3 $ and $ n = 14 $. For every $ 1 \le k \le n $, the circular suffix $ \mathcal{T}_k $ of $ \mathcal{T} $ is the string $ T_f(g) $, where $ f $ is the largest integer in $ \{1, 2, \dots, d \} $ such that $ \sum_{h = 1}^{f - 1} |T_h| < k $ and $ 1 \le g \le |T_f| $ is such that $ k = (\sum_{h = 1}^{f - 1} |T_h|) + g $. In other words, the circular suffix $ \mathcal{T}_k $ is the circular suffix starting at position $ k $ in the concatenation $ T_1 T_2 \dots T_d $. In Figure \ref{fig:referencetobothmain}, if $ k = 13 $, we have $ f  = 3 $, $ g = 2 $ and $ \mathcal{T}_{13} = abc $. Given a pattern $ P \in \Sigma^* $, where $ |P| = m $, define:
\begin{equation*}
    \Cdm (\mathcal{T}, P) = \{(i, k) \in \{1, 2, \dots, m \} \times \{1, 2, \dots, n \} \;|\; \text{ $ \mathcal{T}_k $ occurs in $ P $ at position $ i $} \}
\end{equation*}
and let $ occ = |\Cdm (\mathcal{T}, P)| $. For example, if $ P = abcbca $, in the example of Figure \ref{fig:referencetobothmain} we have $ \Cdm (\mathcal{T}, P) = \{(1, 9), (1, 13), (2, 10), (4, 14) \} $ and $ occ = 4 $.

The main result of this paper is the following.

\begin{theorem}\label{theor:main}
    Let $ \mathcal{T} = (T_1, T_2, \dots, T_d) $ be a dictionary of total length $ n $. Then, $ \mathcal{T} $ can be encoded using a data structure of $ n \log \sigma (1 + o(1)) + O(n) + O(d \log n) $ bits such that, given a string $ P $ of length $ m $, we can compute $ \Cdm(\mathcal{T}, P) $  in $ O((m + occ) \log n) $ time.
\end{theorem}

We can also improve the time bound in Theorem \ref{theor:main} to $ O(m \log n + \min \{occ \log n, n \log n + occ \}) $, without needing to know the value $ occ $ in advance. This is useful if $ occ $ is large.

Following Hon et al. \cite{hon2011isaac}, we will build a data structure that extends Sadakane's suffix tree to $ \mathcal{T} $. We will achieve the space bound in Theorem \ref{theor:main} as follows. (i) We will store the FM-index of $ \mathcal{T} $ using $ n \log \sigma (1 + o(1)) + O(n) $ bits, see Theorem \ref{theor:FMindex}. (ii) We will introduce a notion of suffix array (Section \ref{sec:suffixarray}) and a notion of longest comment prefix (LCP) array (Section \ref{sec:lcp}). Storing the suffix array and LCP array explicitly would require $ O(n \log n) $ bits so, in both cases, we will only store a sample (Theorem \ref{theor:sampledsuffixarray} and Theorem \ref{theor:lcpsampling}), leading to a compressed representation of both arrays. (iii) We will store some auxiliary data structures ($ O(n) $ bits in total): 8 bitvectors (called $ B_1 $-$ B_8 $), a data structure supporting range minimum queries on the LCP array (see Section \ref{sec:lcp}), a data structure storing the topology of the suffix tree (see Section \ref{sec:topologysuffixtree}), a data structure storing the topology of the auxiliary tree $ \Suff^* (\mathcal{T}) $ (see Section \ref{sec:solvingcirculad}), and a data structure supporting range minimum queries on the auxiliary array $ \Len $ (see Section \ref{sec:solvingcirculad}).

\section{The Compressed Suffix Tree of $ \mathcal{T} $}\label{sec:compressedsuffix}

In this section, we extend Sadakane's compressed suffix tree to the dictionary $ \mathcal{T} $. To this end, we will introduce a BWT-based encoding of $ \mathcal{T} $ (Section \ref{sec:BWTFM}), a compressed suffix array (Section \ref{sec:suffixarray}), an LCP array (Section \ref{sec:lcp}), and the topology of the suffix tree (Section \ref{sec:topologysuffixtree}).

\subsection{The Burrows-Wheeler Transform and the FM-index of $ \mathcal{T} $}\label{sec:BWTFM}

Let us consider our dictionary $ \mathcal{T} $ (recall that $ n = \sum_{k = 1}^d |T_k| $). We can naturally define a bijective function $ \phi $ from the set $ \{1, 2, \dots, n \} $ to the set $ \{(f, g) \;|\; 1 \le f \le d, 1 \le g \le |T_f| \} $ such that, if $ \phi(k) = (f, g) $, then the $ k $-th character of the concatenation $ T_1 T_2 \dots T_d $ is the $ g $-th character of $ T_f $. For example, in Figure \ref{fig:referencetobothmain} we have $ \phi(13) = (3, 2) $. Let us define a bitvector $ B_1 $ to compute $ \phi $ and $ \phi^{-1} $ in $ O(1) $ time. $ B_1 $ is the bitvector of length $ n $ that contains exactly $ d $ ones such that, for every $ 1 \le h \le d $, the number of consecutive zeros after the $ h $-th one is $ |T_h| - 1 $ (see Figure \ref{fig:examplebasic}). Assume that $ \phi (k) = (f, g) $. Given $ k $, we have $ f = \rank_1 (B_1, k) $ and $ g = k - \select_1 (B_1, f) + 1 $. Conversely, given $ f $ and $ g $, we have $ k = \select_1 (B_1, f) + g - 1 $. Note that in $ O(1) $ time we can also compute $ |T_h| $ for every $ 1 \le h \le d $, because $ |T_h| = \select_1 (B_1, h + 1) - \select_1 (B_1, h) $.

To exploit the circular nature of the dictionary, we will not sort the finite strings $ \mathcal{T}_k $'s, but we will sort the infinite strings $ \mathcal{T}_k^\omega $'s, and the suffix tree of $ \mathcal{T} $ will be the trie of the $ \mathcal{T}_k $'s. Notice that we may have $ \mathcal{T}_k = \mathcal{T}_{k'} $ for distinct $ 1 \le k, k' \le n $ (in Figure \ref{fig:referencetobothmain}, we have $ \mathcal{T}_1 = \mathcal{T}_4 = \mathcal{T}_{13} = (abc)^\omega $). The next lemma shows that this happens exactly when $ \mathcal{T}_k $ and $ \mathcal{T}_{k'} $ have the same root.

\begin{lemma}\label{lem:roots}
    Let $ \mathcal{T} $ be a dictionary, and let $ 1 \le k, k' \le n $. Then, $ \mathcal{T}_k^\omega = \mathcal{T}_{k'}^\omega $ if and only if $ \rho(\mathcal{T}_k) = \rho(\mathcal{T}_{k'}) $.
\end{lemma}

The next step is to sort the ``suffixes'' $ \mathcal{T}_k $'s. Since in general the $ \mathcal{T}_k $'s are not pairwise distinct, we will use an ordered partition (see Figure \ref{fig:maintable}).

\begin{definition}
    Let $ \mathcal{T} $ be a dictionary.
    \begin{itemize}
        \item Let $ \mathcal{D} = (D_1, D_2, \dots, D_{n'}) $ be the ordered partition of $ \{1, 2, \dots, n \} $ such that, for every $ k, k' \in \{1, 2, \dots, n \} $, (i) if $ k $ and $ k' $ are in the same $ D_j $, then $ \mathcal{T}_k^\omega = \mathcal{T}_{k}^\omega $ and (ii) if $ k \in D_j $, $ k' \in D_{j'} $ and $ j < j' $, then $ \mathcal{T}_k^\omega \prec \mathcal{T}_{k'}^\omega $.
        \item For every $ 1 \le j \le n' $, let $ \mathcal{S}_j = \mathcal{T}_k^\omega $, where $ 1 \le k \le n $ is such that $ k \in D_j $.
    \end{itemize}
\end{definition}

Note that $ \mathcal{S}_j $ is well defined because it does not depend on the choice of $ k \in D_j $ by the definition of $ \mathcal{D} $. Note also that $ \mathcal{S}_1 \prec \mathcal{S}_2 \prec \dots \prec \mathcal{S}_{n'} $.

Let $ 1 \le k \le n $, and assume that $ \phi(k) = (f, g) $. Define $ \pred (k) $ as follows: (i) if $ g \ge 2 $, then $ \pred (k) = k - 1 $, and (ii) if $ g = 1 $, then $ \pred (k) = \phi^{-1}(f, |T_f|) $. In other words, $ \pred(k) $ equals $ k - 1 $, modulo staying within the same string of $ \mathcal{T} $. In Figure \ref{fig:examplebasic}, we have $ \pred(9) = 8 $, $ \pred(8) = 7 $ but $ \pred(7) = 11 $. For every $ 1 \le k \le n $, we can compute $ \pred(k) $ in $ O(1) $ time by using the bitvector $ B_1 $: if $ B_1[k] = 0 $, then $ \pred(k) = k - 1 $, and if $ B_1[k] = 1 $, then $ \pred(k) = \select_1 (\rank_1(B_1, k) + 1) - 1 $.

If $ U \subseteq \{1, 2, \dots, n \} $, define $ \pred (U) = \bigcup_{k \in U} \pred(k) $. Since the function $ \pred $ is a premutation of the set $ \{1, 2, \dots, n \} $, we always have $ |\pred(U)| = |U| $ for every $ U \subseteq \{1, 2, \dots, n \} $. Let us prove that $ \pred $ yields a permutation of the $ D_j $'s.

\begin{lemma}\label{lem:goodbehaviordj}
    Let $ \mathcal{T} $ be a dictionary, and let $ 1 \le j \le n' $. Then, there exists $ 1 \le j' \le n $ such that $ \pred(D_j) = D_{j'} $. Moreover, if $ c = \mathcal{S}_{j'} [1] $, we have $ \mathcal{S}_{j'} = c \mathcal{S}_j $.
\end{lemma}

For every $ 1 \le j \le n' $, let $ \psi(j) = j' $, where $ \pred(D_j) = D_{j'} $ as in Lemma \ref{lem:goodbehaviordj}. Notice that $ \psi $ is permutation of $ \{1, 2, \dots, n' \} $ because it is subjective: indeed, for every $ 1 \le j' \le n' $, if we pick any $ k' \in D_{j'} $, and consider $ 1 \le k \le n'$ such that $ k' = \pred (k) $ and $ 1 \le j \le n' $ such that $ k \in D_j $, then  $ \pred(D_j) = D_{j'} $. Moreover, for every $ 1 \le j \le n' $ define $ \mu (j) = \mathcal{S}_{\psi(j)}[1] $. By Lemma \ref{lem:goodbehaviordj}, we know that $ \mathcal{S}_{\psi(j)} = \mu (j) \mathcal{S}_j $ for every $ 1 \le j \le n' $.

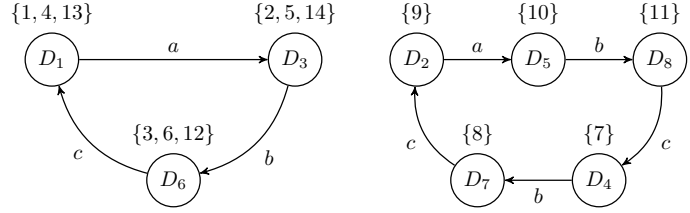
\begin{figure}
	\centering
	\scalebox{0.8}{
\begin{tikzpicture}[->,>=stealth', semithick, auto, scale=1]
\node[state, label=above:{$ \{1, 4, 13 \} $}] (1)    at (0,0)	{$ D_1 $};
\node[state, label=above:{$ \{2, 5, 14 \} $}] (3)    at (4,0)	{$  D_3 $};
\node[state, label=above:{$ \{3, 6, 12 \} $}] (6)    at (2,-2)	{$  D_6 $};

\node[state, label=above:{$ \{9 \} $}] (2)    at (6,0)	{$  D_2 $};
\node[state, label=above:{$ \{10 \} $}] (5)    at (8,0)	{$  D_5 $};
\node[state, label=above:{$ \{11 \} $}] (8)    at (10,0)	{$  D_8 $};
\node[state, label=above:{$ \{7 \} $}] (4)    at (9, -2)	{$  D_4 $};
\node[state, label=above:{$ \{8 \} $}] (7)    at (7, -2)	{$  D_7 $};

\draw (1) edge [] node [] {$ a $} (3);
\draw (3) edge [bend left] node [] {$ b $} (6);
\draw (6) edge [bend left ] node [] {$ c $} (1);
\draw (2) edge [] node [] {$ a $} (5);
\draw (5) edge [] node [] {$ b $} (8);
\draw (8) edge [bend left] node [] {$ c $} (4);
\draw (4) edge [] node [] {$ b $} (7);
\draw (7) edge [bend left] node [] {$ c $} (2);

\end{tikzpicture}
}

\caption{The graph $ G_\mathcal{T} $ for the dictionary $ \mathcal{T} = (abcabc, bcabc, cab) $ of Figure \ref{fig:referencetobothmain}.}
\label{fig:graphexample}
\end{figure}

We can visualize $ \psi $ by drawing a (directed edge-labeled) graph $ G_\mathcal{T} = (V, E) $, where $ V = \{D_j \;|\; 1 \le j \le n' \} $ and $ E = \{(D_{\psi (j)}, D_j, \mu (j) \;|\; 1 \le j \le n' \} $, see Figure \ref{fig:graphexample}. Since $ \psi $ is a permutation of $ \{1, 2, \dots, n' \} $, then every node of $ G $ has exactly one incoming edge and exactly one ongoing edge, so $ G $ is the disjoint union of some cycles. Moreover, for every $ 1 \le j \le n' $ the infinite string that we can read starting from node $ D_j $ and following the edges of the corresponding cycle is $ \mathcal{S}_j $, because we know that $ \mathcal{S}_{\psi(j)} = \mu (j) \mathcal{S}_j $ for every $ 1 \le j' \le n' $.  For example, in Figure \ref{fig:graphexample} the infinite string that we can read starting from $ D_3 $ is $ \mathcal{S}_3 = (bca)^\omega  $.

We will not explicitly build the graph $ G_\mathcal{T} $ to solve circular dictionary matching queries, but $ G_\mathcal{T} $ naturally captures the cyclic nature of $ \mathcal{T} $ and will help us detect some properties of the $ D_j $'s. For example, since we know that the infinite string starting from node $ D_j $ is $ \mathcal{S}_j $, and $ \mathcal{S}_1 \prec \mathcal{S}_2 \prec \dots \prec \mathcal{S}_{n'} $, we can easily infer the following properties (see Figure \ref{fig:graphexample}): if we consider two edges $ (D_{j'_1}, D_{j_1}, c) $ and $ (D_{j'_2}, D_{j_2}, d) $, then (i) if $ j'_1 < j'_2 $, then $ c \preceq d $, and (ii) if $ c = d $ and $ j'_1 < j'_2 $, then $ j_1 < j_2 $. We can formalize these properties in our graph-free setting as follows.

\begin{lemma}\label{lem:axioms1and2}
    Let $ \mathcal{T} $ be a dictionary. let $ 1 \le j_1, j'_1, j_2, j'_2 \le n' $ and $ c, d \in \Sigma $ such that $ \mathcal{S}_{j'_1} = c \mathcal{S}_{j_1} $ and $ \mathcal{S}_{j'_2} = d \mathcal{S}_{j_2} $.
    \begin{enumerate}
        \item (Property 1) If $ j'_1 < j'_2 $, then $ c \preceq d $.
        \item (Property 2) If $ c = d $ and $ j'_1 < j'_2 $, then $ j_1 < j_2 $.
    \end{enumerate}
\end{lemma}

We now want to extend the \emph{backward search} \cite{ferragina2005jacm} to our dictionary $ \mathcal{T}$. We will again use $ G_\mathcal{T} $ to guide our intuition. Given $ U \subseteq \{1, 2, \dots, n' \} $ and given $ c \in \Sigma $, let $ \back(U, c) $ be the set of all $ 1 \le j' \le n' $ such that there exists an edge $ (D_{j'}, D_j, c) $ for some $ j \in U $. For example, in Figure \ref{fig:graphexample} we have $ \back(\{6, 7, 8 \}, b) = \{3, 4, 5 \} $. Notice that $ \{6, 7, 8 \} $ is convex, and $ \{3, 4, 5 \} $ is also convex. This is true in general: $ \back(U, c) $ is always convex if $ U $ is convex. Indeed, if $ j'_1 < j' < j'_2 $ and $ j'_1, j'_2 \in \back(U, c) $, then from the properties of $ G_\mathcal{T} $ mentioned before Lemma \ref{lem:axioms1and2} we first conclude that the edge leaving node $ D_{j'} $ is labeled with $ c $, and then we infer that the node $ D_j $ reached by this edge must be such that $ j \in U $, which implies $ j' \in \back(U, c) $. We can now formally define $ \back(U, c) $ in our graph-free setting and prove that $ \back(U, c) $ is convex if $ U $ is convex.

\begin{definition}
    Let $ \mathcal{T} $ be a dictionary, let $ U \subseteq \{1, 2, \dots, n' \} $ and let $ c \in \Sigma $. Define $ \back(U, c) = \{1 \le j' \le n \;|\; \text{there exists $ j \in U $ such that $ \mathcal{S}_{j'} = c \mathcal{S}_j $} \} $.
\end{definition}

Note that, if $ U_1 \subseteq U_2 \subseteq \{1, 2, \dots, n' \} $ and $ c \in \Sigma $, then $ \back(U_1, c) \subseteq \back(U_2, c) $.

\begin{lemma}\label{lem:convex}
    Let $ \mathcal{T} $ be a dictionary, let $ U \subseteq \{1, 2, \dots, n' \} $ and let $ c \in \Sigma $. If $ U $ is convex, then $ \back(U, c) $ is convex.
\end{lemma}

Let us define the Burrows-Wheeler transform (BWT) of $ \mathcal{T} $.

\begin{definition}
    Let $ \mathcal{T} $ be a dictionary. Define the string $ \BWT \in \Sigma^* $ of length $ n' $ such that $ \BWT[j] = \mu (j) $ for every $ 1 \le j \le n' $.
\end{definition}

In other words, $ \BWT[j] $ is the label of the edge \emph{reaching} node $ D_j $ for every $ 1 \le j \le n' $ (see Figure \ref{fig:maintable} and Figure \ref{fig:graphexample}). The data structure that we will define in Theorem \ref{theor:FMindex} is based on two sequences, $ \BWT^* $ and $ B_2 $, that are related to $ \BWT $ (see Figure \ref{fig:maintable}). The sequence $ \BWT^* $ is obtained from $ \BWT $ by sorting its elements. We know that for every pair of edges $ (D_{j'_1}, D_{j_1}, c) $ and $ (D_{j'_2}, D_{j_2}, d) $, if $ j'_1 < j'_2 $, then $ c \preceq d $, so $ \BWT^*[j] $ is the label of the edge \emph{leaving} node $ D_j $ for every $ 1 \le j \le n' $, which implies $ \BWT^*[j] = \mathcal{S}_j[1] $ for every $ 1 \le j \le n' $. The bitvector $ B_2 $ has length $ n' $ and is obtained by marking with 1 the beginning of each equal-letter run in $ \BWT^* $. Formally, for every $ 1 \le j \le n' $, we have $ B_2[j] = 1 $ if and only if $ j = 1 $ or $ (j > 2) \land (\BWT^*[j - 1] \not = \BWT^*[j]) $. Since the alphabet is effective, the set $ \{ \select_1 (B_2, c), \select_1 (B_2, c) + 1, \select_1 (B_2, c) + 2, \dots, \select_1 (B_2, c + 1) - 2, \select_1 (B_2, c + 1) - 1 \} $ consists of all $ 1 \le j' \le n' $ such that the edge leaving node $ D_{j'} $ is labeled with $ c $.

We can now extend the properties of the Burrows-Wheeler Transform and the FM-index to $ \mathcal{T} $. In particular, we will show that $ \BWT $ if an encoding of $ G_\mathcal{T} $. This result shows that $ \BWT $ is related to the eBWT of Mantaci et al. \cite{mantaci2007}, but there are two main differences: (1) we do not need assumptions (i-ii) in Section \ref{sec:introduction} to define $ \BWT $ (in particular, the $ T_h $'s need not be primitive), and (2) $ \BWT $ is an encoding of $ G_\mathcal{T} $ but not of $ \mathcal{T} $ (namely, $ \BWT $ encodes the cyclic nature of the $ T_h $'s and not a specific circular suffix of each $ T_h $). To extend the FM-index to $ \mathcal{T} $, we will once again base our intuition on $ G_\mathcal{T} $.

Recall that two (directed edge-labeled) graphs $ G_1 = (V_1, E_1) $ and $ G_2 = (V_2, E_2) $ are \emph{isomorphic} if there exists a bijection $ f $ from $ V_1 $ to $ V_2 $ such that for every $ u, v \in V_1 $ and for every $ c \in \Sigma $ we have $ (u, v, c) \in E_1 $ if and only if $ (f(u), f(v), c) \in E_2 $. In other words, two graphs are isomorphic if they are the same graph up to renaming the nodes.

\begin{theorem}\label{theor:FMindex}
    Let $ \mathcal{T} $ be a dictionary.
    \begin{enumerate}
        \item $ \BWT $ is an encoding of $ G_\mathcal{T} $, that is, given $ \BWT $, we can retrieve $ G_\mathcal{T} $ up to isomorphism.
        \item There exists a data structure encoding $ G_\mathcal{T} $ of $ n' \log \sigma (1 + o(1)) + O(n') $ bits that supports the following queries in $ O(\log \log \sigma) $ time: (i) $ \access $, $\rank $, and $ \select $ queries on $ \BWT $, (ii) $ \bws(\ell, r, c) $: given $ 1 \le \ell \le r \le n' $ and $ c \in \Sigma $, decide if $ \back(\{\ell, \ell + 1, \dots, r \}, c) $ is empty and, if it is nonempty, return $ \ell' $ and $ r' $ such that $ \back(\{\ell, \ell + 1, \dots, r \}, c) = \{\ell', \ell' + 1, \dots, r' \} $, (iii) $ \prev(j) $: given $ 1 \le j \le n' $, return $ 1 \le j' \le n' $ such that $ \pred (D_j) = D_{j'} $, and (iv) $ \follow(j') $: given $ 1 \le j' \le n' $, return $ 1 \le j \le n' $ such that $ \pred (D_j) = D_{j'} $.
    \end{enumerate}
\end{theorem}

Note that, even if $ \BWT $ is an encoding of $ G_\mathcal{T} $, it is not an encoding of $ \mathcal{T} $ (in particular, from $ G_\mathcal{T} $ we cannot recover $ \mathcal{T} $). This is true even when all $ D_j $'s are singletons because $ G_\mathcal{T} $ only stores circular strings, so we cannot retrieve the bitvector $ B_1 $ the marks the beginning of each string (we cannot even retrieve the specific enumeration $ T_1 $, $ T_2 $, $ \dots $, $ T_d $ of the strings in $ \mathcal{T} $).

\subsection{The Compressed Suffix Array of $ \mathcal{T} $}\label{sec:suffixarray}

We know that $ \mathcal{D} = (D_1, D_2, \dots, D_{n'}) $ is an ordered partition of $ \{1, 2, \dots, n \} $, and we know that $ \mathcal{S}_1 \prec \mathcal{S}_2 \prec \dots \prec \mathcal{S}_{n'} $. The suffix array of $ \mathcal{T} $ should be defined in such a way that we can answer the following query: given  $ 1 \le j \le n' $, return the set $ D_j $.

In the following, we say that a position $ 1 \le k \le n $ \emph{refers} to the string $ T_h $ if $ h = \rank_1(B_1, k) $. Every position refers to exactly one string. Notice that a set $ D_j $ may contain elements that refer to the same string $ T_h $. In Figure \ref{fig:referencetobothmain}, we have $ 2, 5 \in D_3 $, and both $ 2 $ and $ 5 $ refer to $ T_1 $. This happens because $ T_1 $ is not a primitive string. Let us show that, if we know \emph{any} element of $ D_j $ referring to $ T_h $, we can retrieve \emph{all} elements of $ D_j $ referring to $ T_h $.

\begin{figure}
    \centering
        \scalebox{0.9}{
\begin{tabularx}{0.8 \textwidth}{|c|>{\centering\arraybackslash}X>{\centering\arraybackslash}X>{\centering\arraybackslash}X>{\centering\arraybackslash}X>{\centering\arraybackslash}X>{\centering\arraybackslash}X>{\centering\arraybackslash}X>{\centering\arraybackslash}X>{\centering\arraybackslash}X>{\centering\arraybackslash}X>{\centering\arraybackslash}X|}
\hline
$ t $ & $ 1 $ & $ 2 $ & $ 3 $ & $ 4 $ & $ 5 $ & $ 6 $ & $ 7 $ & $ 8 $ & $ 9 $ & $ 10 $ & $ 11 $ \\
\hline
$ B_3[t] $ & $ 1 $ & $ 0 $ & $ 0 $ & $ 1 $ & $ 0 $ & $ 0 $ & $ 0 $ & $ 0 $ & $ 1 $ & $ 0 $ & $ 0 $ \\
\hline
$ \SA [t] $ & $ 1 $ & $ 13 $ & $ 9 $ & $ 2 $ & $ 14 $ & $ 7 $ & $ 10 $ & $ 3 $ & $ 12 $ & $ 8 $ & $ 11 $ \\
\hline
$ B_4 [t] $ & $ 1 $ & $ 0 $ & $ 1 $ & $ 1 $ & $ 0 $ & $ 1 $ & $ 1 $ & $ 1 $ & $ 0 $ & $ 1 $ & $ 1 $ \\
\hline
$ \SA^* [t] $ & $ 1 $ & $ 9 $ & $ 14 $ & $ 7 $ & $ 3 $ & $ 12 $ & $ 11 $ & & & & \\
\hline
$ B_5[t] $ & $ 1 $ & $ 0 $ & $ 1 $ & $ 0 $ & $ 1 $ & $ 1 $ & $ 0 $ & $ 1 $ & $ 1 $ & $ 0 $ & $ 1 $ \\
\hline
$ \Len [t] $ & $ 6 $ & $ 3 $ & $ 5 $ & $ 6 $ & $ 3 $ & $ 5 $ & $ 5 $ & $ 6 $ & $ 3 $ & $ 5 $ & $ 5 $ \\
\hline
\end{tabularx}
}

\caption{The compressed suffix array of the dictionary $ \mathcal{T} = (abcabc, bcabc, cab) $ in Figure \ref{fig:referencetobothmain}. We use the sampling factor $ s = 2 $.}
\label{fig:suffixarray}
\end{figure}

For every $ 1 \le h \le d $, let $ \rho_h $ be the root of $ T_h $. Then, $ |T_h| $ is divisible by $ |\rho_h| $, and for every $ 1 \le k \le n $ the root $ \rho (\mathcal{T}_k) $ of $ \mathcal{T}_k $ is $ \mathcal{T}_k[1, \rho_h] $, where $ k $ refers to string $ T_h $. For every $ 1 \le h \le  d $, let $ k_h = \select_1 (B_1, h) $ be the index in $ \{1, 2, \dots, n \} $ corresponding to the first character of $ T_h $.

Fix $ 1 \le j \le n' $ and $ 1 \le h \le d $, and let $ k \in D_j $ \emph{any} position referring to $ T_h $. For every $ k' $ referring to $ T_h $, we have $ k' \in D_j $ if and only if $ \rho (\mathcal{T}_k) = \rho (\mathcal{T}_{k'}) $ (by Lemma \ref{lem:roots}), if and only if $ \mathcal{T}_k[1, \rho_h] = \mathcal{T}_{k'}[1, \rho_h] $, if and only if $ |k' - k| $ is a multiple of $ |\rho_h| $. Then, if $ G $ is the set of \emph{all} elements of $ D_j $ referring to $ T_h $, we have $ G = \{k_h + (k - k_h + w |\rho_h| \mod |T_h|) \;|\; 0 \le w \le (|T_h| / |\rho_h|) - 1 \} $. For example, if Figure \ref{fig:maintable}, if we consider $ j = 3 $ and $ h = 1 $ and we pick $ k = 5 $, then $ 5 \in D_3 $, $ 5 $ refers to $ T_1 $, $ \rho_1 = abc $, $ |\rho_1| = 3 $, $ |T_1| = 6 $, $ k_1 = 1 $ and $ G = \{2, 5 \} $.

To compute $ G $ we need to compute $ |T_h| $ and $ |\rho_h| $. We have already seen that we can compute $ |T_h| $ in $ O(1) $ time using $ B_1 $ (see Section \ref{sec:BWTFM}), and we now store a bitvector $ B_3 $ to compute $ |\rho_h| $ in $ O(1) $ time (see Figure \ref{fig:suffixarray}). The bitvector $ B_3 $ contains exactly $ d $ ones, we have $ B_3[1] = 1 $, and for every $ 1 \le h \le d $ the number of consecutive zeros after the $ h $-th one is $ |\rho_h| - 1 $. Then, $ |\rho_h| = \select_1 (B_3, h + 1) - \select_1 (B_3, h) $ for every $ 1 \le h \le d$.

Let us define a suffix array for $ \mathcal{T} $. Let $ \sim $ be the equivalence relation on $ \{1, 2, \dots, n \} $ such that for every $ 1 \le k, k^* \le n $ we have $ k \sim k^* $ if and only if $ k $ and $ k^* $ belong to the same $ D_j $ and refer to the same string $ T_h $. Fix $ 1 \le j \le n' $. Notice that $ D_j $ is the union of some $ \sim $-equivalence classes. Let $ D'_j $ be the subset of $ D_j $ obtained by picking the smallest element of each $ \sim $-equivalence class contained in $ D_j $. In Figure \ref{fig:maintable}, for $ j = 3 $, the partition of $ D_3 $ induced by $ \sim $ is $ \{\{2, 5 \}, \{14 \}\} $, so $ D'_3 = \{2, 14 \} $. Then $ \SA $ is the array obtained by concatenating the elements in $ D'_1 $, the elements in $ D'_2 $, $ \dots $, the elements in $ D'_{n'} $ (see Figure \ref{fig:suffixarray}), where the elements in $ D'_{j} $ are sorted from smallest to largest. Equivalently, the elements in each $ D'_{j} $ are sorted according to the indexes of the strings to which they refer (by definition, two distinct elements of $ D'_{j} $ cannot refer to the same string). We also define a bitvector $ B_4 $ to mark the beginning of each $ D'_{j} $ (see again Figure \ref{fig:suffixarray}). More precisely, $ B_4 $ contains exactly $ n' $ ones and for every $ 1 \le j \le n' $ the number of consecutive zeros after the $ j $-th one is $ |D'_{j}| - 1 $.

Fix $ 1 \le k \le n $, where $ k \in D_j $ refers to string $ T_h $.
Note that there exists $ t $ such that $ \SA[t] = k $ if and only if $ k \in D'_j $, if and only if $ k_h \le k \le k_h + |\rho_h| - 1 $. Moreover, for every $ k \in D_{j'} $ we have that $ [k]_\sim = \{k + w |\rho_h| \;|\; 0 \le w \le (|T_h| / |\rho_h|) - 1 \} $ is the set of all elements of $ D_j $ referring to $ T_h $. In particular, $ \SA $ is an array of length $ n^* = \sum_{j = 1}^{n'} |D'_{j}| = \sum_{h = 1}^d |\rho_h| $ (in Figure \ref{fig:suffixarray}, we have $ n^* = 11 $). 

The suffix array $ \SA $ has the desired property: given $ 1 \le j \le n' $, we can compute the set $ D_j $ in $ O(|D_j|) $ time as follows. We first retrieve $ D'_j $ by observing that its elements are stored in $ \SA[t] $, $ \SA[t + 1] $, $ \SA[t + 2] $, $ \dots $, $ \SA[t'] $, where $ t = \select_1 (B_4, j) $ and $ t' = \select_1 (B_4, j + 1) - 1 $. Then, for every $ k \in D'_j $, we know that $ k $ refers to string $ T_h $, where $ h = \rank_1 (B_1, k) $, and we can compute the set $ [k]_\sim $ of all elements of $ D_j $ referring to $ T_h $ as shown above. Then, we have $ D_j = \bigcup_{k \in D'_j} [k]_\sim $ and the union is disjoint.

\begin{figure}
\captionsetup[subfigure]{justification=centering}
	\centering 
\begin{subfigure}[b]{0.49 \textwidth}
        \centering
        \scalebox{0.9}{
\begin{tabularx}{0.8 \textwidth}{|c|>{\centering\arraybackslash}X>{\centering\arraybackslash}X>{\centering\arraybackslash}X>{\centering\arraybackslash}X>{\centering\arraybackslash}X>{\centering\arraybackslash}X>{\centering\arraybackslash}X|}
\hline
$ j $ & $ 2 $ & $ 3 $ & $ 4 $ & $ 5 $ & $ 6 $ & $ 7 $ & $ 8 $ \\
\hline
$ \LCP[j] $ & $ 3 $ & $ 0 $ & $ 5 $ & $ 2 $ & $ 0 $ & $ 4 $ & $ 1 $ \\
\hline
$ B_6[j] $ & $ 1 $ & $ 0 $ & $ 0 $ & $ 0 $ & $ 0 $ & $ 0 $ & $ 1 $ \\
\hline
$ \LCP^*[j] $ & $ 3 $ & $ 1 $ & & & & & \\
\hline
\end{tabularx}
}
\caption{}
\label{subfig:tablelcp}
\end{subfigure}
\begin{subfigure}[b]{0.49 \textwidth} 
        \centering
        \scalebox{0.8}{
\begin{tikzpicture}[->,>=stealth', semithick, auto, scale=1]
\node[state] (4)    at (0, 0)	{$  D_4 $};
\node[state] (7)    at (2, 0)	{$  D_7 $};
\node[state, fill=yellow] (2)    at (2, 2)	{$  D_2 $};
\node[state] (5)    at (4,2)	{$  D_5 $};
\node[state, fill=yellow] (8)    at (6,2)	{$  D_8 $};
\node[state] (3)    at (6,0)	{$  D_3 $};
\node[state] (6)    at (4,0)	{$  D_6 $};
\draw (4) edge [] node [] {} (7);
\draw (7) edge [] node [] {} (2);
\draw (2) edge [] node [] {} (5);
\draw (5) edge [] node [] {} (8);
\draw (8) edge [] node [] {} (3);
\end{tikzpicture}
}
\caption{}
\label{subfig:samplelcp}
\end{subfigure}

\caption{The LCP array of the dictionary $ \mathcal{T} = (abcabc, bcabc, cab) $ in Figure \ref{fig:referencetobothmain}. We use the sampling factor $ s = 2 $. The graph in (b) is the graph $ Q $ used to determine which values will be sampled. Yellow nodes are the sampled nodes.}
\end{figure}
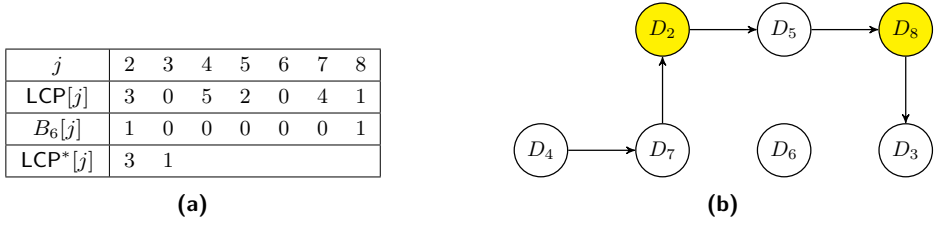

Storing $ \SA $ explicitly can require up to $ n \log n $ bits, so to achieve the space bound in Theorem \ref{theor:main} we need a sampling mechanism similar to the compressed suffix array of a string: we will sample some entries and we will reach a sampled entry by using the backward search to navigate the string (see \cite{navarro2016book}). More precisely, in our setting we will use the query $ \prev(j) $ of Theorem \ref{theor:FMindex} to navigate $ G_\mathcal{T} $ in a backward fashion. Note that in the conventional case of a string, if we start from the end of the string and we apply the backward search, we can reach each position of the string, but in our setting the graph $ G_\mathcal{T} $ is the disjoint union of some cycles (see Figure \ref{fig:graphexample}), and the backward search does not allow navigating from a cycle to some other cycle. This means that we will need to sample at least one value per cycle. In the worst case, we have $ d $ cycles (one for each $ T_h $), so in the worst case we need to sample at least $ d $ values.

After choosing an appropriate sampling factor $ s $, we store the sampled values in an array $ \SA^* $ (respecting the order in $ \SA $), and we use a bitvector $ B_5 $ to mark the entries of $ \SA[k] $ that have been sampled (see Figure \ref{fig:suffixarray}). We finally obtain the following theorem.

\begin{theorem}\label{theor:sampledsuffixarray}
    Let $ \mathcal{T} $ be a dictionary. By storing $ o(n \log \sigma) + O(n) + O(d \log n) $ bits, we can compute each entry $ \SA[t] $ in $ O(\log n) $ time.
\end{theorem}

We conclude that, by storing the data structure of Theorem \ref{theor:sampledsuffixarray}, for every $ 1 \le j \le n' $ we can compute $ D_j $ in $ O(|D_j| \log n) $ time.

\subsection{The LCP Array of $ \mathcal{T} $}\label{sec:lcp}

Let us define the longest common prefix (LCP) array of $ \mathcal{T} $. We know that $ \mathcal{S}_1 \prec \mathcal{S}_2 \prec \dots \prec \mathcal{S}_{n'} $, so it is natural to give the following definition (see Figure \ref{fig:maintable} and Figure \ref{subfig:tablelcp}).

\begin{definition}
    Let $ \mathcal{T} $ be a dictionary. Let $ \LCP = \LCP[2, n'] $ be the array such that $ \LCP[j] = \lcp(\mathcal{S}_{j - 1}, \mathcal{S}_j) $ for every $ 2 \le j \le n' $.
\end{definition}

Note that each $ \LCP[j] $ is finite because the $ \mathcal{S}_j $'s are pairwise distinct. Let us prove a stronger result: $ \LCP[j] \le n' - 2 $ for every $ 2 \le j \le n' $ (Lemma \ref{lem:upperbound}). This implies that we can store an entry $ \LCP[j] $ using at most $ \log n $ bits.

\begin{lemma}\label{lem:upperbound}
    Let $ \mathcal{T} $ be a dictionary. Then, $ \LCP[j] \le n' - 2 $ for every $ 2 \le j \le n' $.
\end{lemma}

Storing the LCP array explicitly can require $ n \log n $ bits, so to achieve the space bound of Theorem \ref{theor:main} we need a sampling mechanism similar to the one for suffix arrays in Section \ref{sec:suffixarray}.

Recall that, given an array $ A[1, n] $, we can define \emph{range minimum queries} as follows: for every $ 1 \le i \le j \le n $, $ \rmq_A (i, j) $ returns an index $ i \le k \le j $ such that $ A[k] = \min_{i \le h \le j} A[h] $. There exists a data structure of $ 2n + o(n) $ bits that can solve a query $ \rmq_A (i, j) $ in $ O(1) $ time \emph{without accessing $ A $} \cite{fischer2011scicomp}; moreover the data structure always returns the \emph{smallest} $ i \le k \le j $ such that $ A[k] = \min_{i \le h \le j} A[h] $.

Let us store a data structure for solving range minimum queries $ \rmq_{\LCP} $ on $ \LCP $ in $ O(1) $ time. We will use this data structure to retrieve each entry of the LCP array from our sampled LCP array. The main idea is to use the sampling mechanism in \cite{cotumaccio2023spire}: an auxiliary graph $ Q $ allows determining which values will be sampled based on a sampling parameter $ s $ (see Figure \ref{subfig:samplelcp}). Then, we define an array $ \LCP^* $ that stores the sampled values (respecting the order in $ \LCP $, see Figure \ref{subfig:tablelcp}), and a bitvector $ B_6 $ to mark the sampled entries of $ \LCP $ (see Figure \ref{subfig:tablelcp}). By choosing an appropriate $ s $, we obtain the following result.

\begin{theorem}\label{theor:lcpsampling}
    Let $ \mathcal{T} $ be a dictionary. By storing $ o(n \log \sigma) + O(n) $ bits, we can compute each entry $ \LCP[j] $ in $ O(\log n) $ time.
\end{theorem}

\subsection{The Topology of the Suffix Tree of $ \mathcal{T} $}\label{sec:topologysuffixtree}

\begin{figure}
\captionsetup[subfigure]{justification=centering}
	\centering
    \begin{subfigure}[b]{0.6 \textwidth}
        \centering
        \scalebox{0.75}{
\begin{tikzpicture}[->,>=stealth', semithick, auto, scale=1]
\node[] (18)    at (0,0)	{$ \color{orange}{[1, 8]} $};
\node[] (12)    at (-4,-2)	{$ [1, 2] $};
\node[] (35)    at (0,-2)	{$ [3, 5] $};
\node[] (68)    at (4,-2)	{$ [6, 8] $};
\node[] (11)    at (-5,-4)	{$ [1, 1] $};
\node[] (22)    at (-3,-4)	{$ \color{orange}{[2, 2]} $};
\node[] (34)    at (-1,-4)	{$ [3, 4] $};
\node[] (55)    at (1,-4)	{$ [5, 5] $};
\node[] (67)    at (3,-4)	{$ [6, 7] $};
\node[] (88)    at (5,-4)	{$ [8, 8] $};
\node[] (33)    at (-2,-6)	{$ \color{orange}{[3, 3]} $};
\node[] (44)    at (0,-6)	{$ \color{orange}{[4, 4]} $};
\node[] (66)    at (2,-6)	{$ [6, 6] $};
\node[] (77)    at (4,-6)	{$ \color{orange}{[7, 7]} $};
\draw (18) edge [above left] node [] {$ abc $} (12);
\draw (18) edge [] node [] {$ bc $} (35);
\draw (18) edge [] node [] {$ c $} (68);
\draw (12) edge [left] node [] {$ abcabc\dots $} (11);
\draw (12) edge [] node [] {$ bcabcb\dots $} (22);
\draw (35) edge [left] node [] {$ abc $} (34);
\draw (35) edge [] node [] {$ bcabcbc\dots $} (55);
\draw (68) edge [left] node [] {$ abc $} (67);
\draw (68) edge [] node [] {$ bcabcbca\dots $} (88);
\draw (34) edge [left] node [] {$ abca\dots $} (33);
\draw (34) edge [] node [] {$ bcab\dots $} (44);
\draw (67) edge [left] node [] {$ abcab\dots $} (66);
\draw (67) edge [] node [] {$ bcabc\dots $} (77);
\end{tikzpicture}
}
\caption{}
\label{subfig:suffixtreeexample}
\end{subfigure}
    \begin{subfigure}[b]{0.39 \textwidth}
        \centering
        \scalebox{0.75}{
\begin{tikzpicture}[->,>=stealth', semithick, auto, scale=1]
\node[] (18)    at (0,0)	{$ [1, 8] $};
\node[] (22)    at (-3,-2)	{$ [2, 2] $};
\node[] (33)    at (-1,-2)	{$ [3, 3] $};
\node[] (44)    at (1, -2)	{$ [4, 4] $};
\node[] (77)    at (3, -2)	{$ [7, 7] $};
\draw (18) edge [] node [] {} (22);
\draw (18) edge [] node [] {} (33);
\draw (18) edge [] node [] {} (44);
\draw (18) edge [] node [] {} (77);
\end{tikzpicture}
}
\caption{}
\label{subfig:treeuseful}
\end{subfigure}

\bigskip

\begin{subfigure}[b]{1 \textwidth}
        \centering
        \scalebox{0.9}{
\begin{tabularx}{\textwidth}{|c|>{\centering\arraybackslash}X>{\centering\arraybackslash}X>{\centering\arraybackslash}X>{\centering\arraybackslash}X>{\centering\arraybackslash}X>{\centering\arraybackslash}X>{\centering\arraybackslash}X>{\centering\arraybackslash}X>{\centering\arraybackslash}X>{\centering\arraybackslash}X>{\centering\arraybackslash}X>{\centering\arraybackslash}X>{\centering\arraybackslash}X>{\centering\arraybackslash}X>{\centering\arraybackslash}X>{\centering\arraybackslash}X>{\centering\arraybackslash}X>{\centering\arraybackslash}X>{\centering\arraybackslash}X>{\centering\arraybackslash}X>{\centering\arraybackslash}X>{\centering\arraybackslash}X>{\centering\arraybackslash}X>{\centering\arraybackslash}X>{\centering\arraybackslash}X>{\centering\arraybackslash}X>{\centering\arraybackslash}X>{\centering\arraybackslash}X|}
\hline
$ t $ & $ 1 $ & $ 2 $ & $ 3 $ & $ 4 $ & $ 5 $ & $ 6 $ & $ 7 $ & $ 8 $ & $ 9 $ & $ 0 $ & $ 1 $ & $ 2 $ & $ 3 $ & $ 4 $ & $ 5 $ & $ 6 $ & $ 7 $ & $ 8 $ & $ 9 $ & $ 0 $ & $ 1 $ & $ 2 $ & $ 3 $ & $ 4 $ & $ 5 $ & $ 6 $ & $ 7 $ & $ 8 $ \\
\hline
$ Z_{\Suff (\mathcal{T})}[t] $ & $ 1 $ & $ 1 $ & $ 1 $ & $ 0 $ & $ 1 $ & $ 0 $ & $ 0 $ & $ 1 $ & $ 1 $ & $ 1 $ & $ 0 $ & $ 1 $ & $ 0 $ & $ 0 $ & $ 1 $ & $ 0 $ & $ 0 $ & $ 1 $ & $ 1 $ & $ 1 $ & $ 0 $ & $ 1 $ & $ 0 $ & $ 0 $ & $ 1 $ & $ 0 $ & $ 0 $ & $ 0 $ \\
\hline
$ B_7[t] $ & $ 0 $ & $ 0 $ & $ 1 $ & $ 0 $ & $ 1 $ & $ 0 $ & $ 0 $ & $ 0 $ & $ 0 $ & $ 1 $ & $ 0 $ & $ 1 $ & $ 0 $ & $ 0 $ & $ 1 $ & $ 0 $ & $ 0 $ & $ 0 $ & $ 0 $ & $ 1 $ & $ 0 $ & $ 1 $ & $ 0 $ & $ 0 $ & $ 1 $ & $ 0 $ & $ 0 $ & $ 0 $ \\
\hline
$ B_8[t] $ & $ 1 $ & $ 0 $ & $ 0 $ & $ 0 $ & $ 1 $ & $ 1 $ & $ 0 $ & $ 0 $ & $ 0 $ & $ 1 $ & $ 1 $ & $ 1 $ & $ 1 $ & $ 0 $ & $ 0 $ & $ 0 $ & $ 0 $ & $ 0 $ & $ 0 $ & $ 0 $ & $ 0 $ & $ 1 $ & $ 1 $ & $ 0 $ & $ 0 $ & $ 0 $ & $ 0 $ & $ 1 $ \\
\hline
$ Z_{\Suff^* (\mathcal{T})}[t] $ & $ 1 $ & $ 1 $ & $ 0 $ & $ 1 $ & $ 0 $ & $ 1 $ & $ 0 $ & $ 1 $ & $ 0 $ & $ 0 $ & & & & & & & & & & & & & & & & & & \\
\hline
\end{tabularx}
}
\caption{}
\label{subfig:tabletrees}
\end{subfigure}
\caption{(a) The suffix tree $ \Suff(\mathcal{T}) $ of the dictionary $ \mathcal{T} = (abcabc, bcabc, cab) $ in Figure \ref{fig:referencetobothmain}. Marked nodes are orange. (b) The tree $ \Suff^* (\mathcal{T}) $. (c) The bit arrays to navigate $ \Suff(\mathcal{T}) $ and $ \Suff^* (\mathcal{T}) $. The first row compactly includes all indexes $ t $ from $ 1 $ to $ 28 $.}
\end{figure}

We can now introduce the suffix tree $ \Suff (\mathcal{T}) $ of our dictionary $ \mathcal{T} $ (the full details are discussed in the appendix). Refer to Figure \ref{subfig:suffixtreeexample} for an example. The (infinite) set of all finite strings that we can read starting from the root is $ \mathcal{P} = \{\mathcal{S}_j[1, t] \;|\; 1 \le j \le n', t \ge 0 \} $. Every node is an interval that describes the strings that can be read starting from the root and reading a nonempty prefix of the last edge. For example, the set of all strings $ P $ for which $ P $ is a prefix of $ \mathcal{S}_j $ if and only if $ 3 \le j \le 4 $ is $ \{bca, bcab, bcabc \} $ and, in fact, the set of all strings that reach node $ [3, 4] $ is $ \{bca, bcab, bcabc \} $. The suffix tree contains exactly $ n' $ leaves (namely, $ [j, j] $ for every $ 1 \le j \le n' $), and the set $ \mathcal{N} $ of all nodes has size at most $ 2n' - 1 $. The set of all strings reaching a node $ [\ell, r] $ is finite if and only if $ [\ell, r] $ is an internal node, and in this case the longest string reaching $ [\ell, r] $ has length $ \lambda_{\ell, r} = \lcp (\mathcal{S}_\ell, \mathcal{S}_r) $, which can be computed by using the LCP array (for example, $ \lambda_{3, 4} = \lcp (\mathcal{S}_3, \mathcal{S}_4) = 5 $). The suffix tree $ \mathcal{T} $ is an ordinal tree, and any ordinal tree $ T $ with $ t $ nodes can be encoded using its \emph{balanced paranthesis} representation $ Z_T[1, 2t] $, a bit array that we can build incrementally as follows. Visit all nodes of the tree in depth-first search order (respecting the order of the children of each node) starting from the root. Every time we encounter a new node $ u $ we append an $ 1 $, and as soon as we leave the subtree rooted at $ u $ we add a $ 0 $. Hence $ Z_T[1, 2t] $ is a bit array containing exactly $ t $ values equal to $ 1 $ and $ t $ values equal to $ 0 $ (see Figure \ref{subfig:tabletrees} for the balanced parenthesis representation $ Z_{\Suff (\mathcal{T})} $ of $ \Suff (\mathcal{T}) $). Navarro and Sadakane showed how to support navigational queries on the balanced representation of an ordinal tree in compact space and constant time (see Theorem \ref{theor:balancedparenthesisnavarrosadakane} in the appendix), and we store such a data structure for $ \Suff (\mathcal{T}) $ (which requires $ O(n) $ bits). To correctly switch to the balanced parenthesis representation of $ \Suff (\mathcal{T}) $, we also store a bitvector $ B_7 $ of length $ 2|\mathcal{N}| $ such that for every $ 1 \le t \le 2|\mathcal{N}| $ we have $ B_7 [t] = 1 $ if and only if (i) $ Z_{\Suff (\mathcal{T})}[t] = 1 $ and (ii) the index $ t $ corresponds to a leaf of $ \Suff (\mathcal{T}) $ (see Figure \ref{subfig:tabletrees}).

\section{Solving Circular Dictionary Matching Queries}\label{sec:solvingcirculad}

In the following, we consider a pattern $ P = P[1, m] $ of length $ m $, and we want to compute $ \Cdm (\mathcal{T}, P) $.

Fix $ 1 \le i \le m $. We need to compute all $ 1 \le k \le n $ such that $ \mathcal{T}_k $ occurs in $ P $ at position $ i $. In general, if $ k \in D_j $, then $ \mathcal{S}_j = \mathcal{T}_k^\omega $, hence $ \mathcal{T}_k = \mathcal{S}_j[1, |\mathcal{T}_k|]  \in \mathcal{P} $, which means that $ \mathcal{T}_k $ identifies a node of $ \Suff (\mathcal{T}) $. This means that, if $ k $ occurs in $ P $ at position $ i $, then there exists $ i' \ge i $ such that $ P[i, i'] \in \mathcal{P} $, and then every prefix of $ P[i, i'] $ is also in $ \mathcal{P} $ (because every prefix of a string in $ \mathcal{P} $ is also in $ \mathcal{P} $). Consequently, by considering the \emph{largest} $ i^* $ such that $ P[i, i^*] \in \mathcal{P} $, we know that we only need to consider $ P[i, i^*] $ to compute all $ 1 \le k \le n $ such that $ \mathcal{T}_k $ occurs in $ P $ at position $ i $. We can then give the following definition.

\begin{definition}
    Let $ \mathcal{T} $ be a dictionary, and let $ P \in \Sigma^* $. For every $ 1 \le i \le m $, let $ t_i $ be the largest integer such that $ i - 1 \le t_i \le m $ and $ P[i, t_i] \in \mathcal{P} $.
\end{definition}

Note that $ t_i $ is well-defined for every $ 1 \le i \le m $ because $ P[i, i - 1] = \epsilon \in \mathcal{P} $. 

Fix $ 1 \le i \le m $, $ 1 \le j \le n' $ and $ k \in D_j $. If $ j \in  [\ell_{P[i, t_i]}, r_{P[i, t_i]}] $, then $ P[i, t_i] $ is a prefix of $ \mathcal{S}_j = \mathcal{T}_k^\omega $ and so $ \mathcal{T}_k $ occurs in $ P $ at position $ i $ if and only $ |\mathcal{T}_k| \le |P[i, t_i]| $. If $ j \not \in  [\ell_{P[i, t_i]}, r_{P[i, t_i]}] $, we know that $ P[i, t_i] $ is not a prefix of $ \mathcal{T}_k^\omega $, but it might still be true that a prefix of $ P[i, t_i] $ is equal to $ \mathcal{T}_k $. Let $ P[i, i^*] $ be the \emph{longest} prefix of $ P[i, t_i] $ that is a prefix of $ \mathcal{T}_k^\omega $ (or equivalently, let $ i^* $ be the largest integer such that $ j \in  [\ell_{P[i, i^*]}, r_{P[i, i^*]}] $). Then, $ \mathcal{T}_k $ occurs in $ P $ at position $ i $ if and only $ |\mathcal{T}_k| \le |P[i, i^*]| $. To compute $ |P[i, i^*]| $, first notice that every prefix of $ P[i, t_i] $ reaches an ancestor of $ [\ell_{P[i, t_i]}, r_{P[i, t_i]}] $, and every string reaching a strict ancestor of $ [\ell_{P[i, t_i]}, r_{P[i, t_i]}] $ is a strict prefix of $ P[i, t_i] $. As a consequence, we can first compute the nearest ancestor $ [\ell', r'] $ of $ [\ell_{P[i, t_i]}, r_{P[i, t_i]}] $ for which $ j \in [\ell, r] $, and then $ |P[i, i^*]| $ is the length $ \lambda_{\ell', r'} $ of the largest string reaching node $ [\ell', r'] $.

The following lemma captures our intuition.

\begin{lemma}\label{lem:climbingsuffixtree}
    Let $ \mathcal{T} $ be a dictionary, let $ P \in \Sigma^* $, let $ 1 \le i \le m $ and let $ 1 \le j \le n' $.
    \begin{enumerate}
        \item Assume that $ j \in  [\ell_{P[i, t_i]}, r_{P[i, t_i]}] $. Then, for every $ k \in D_j $ we have that $ \mathcal{T}_k $ occurs in $ P $ at position $ i $ if and only if $ |\mathcal{T}_k| \le t_i - i + 1 $.
        \item Assume that $ j \not \in  [\ell_{P[i, t_i]}, r_{P[i, t_i]}] $. Let $ [\ell, r] $ be the nearest ancestor of $ [\ell_{P[i, t_i]}, r_{P[i, t_i]}] $ in $ \Suff (\mathcal{T}) $ for which $ j \not \in [\ell, r] $. Then, $ [\ell, r] $ is not the root of $ \Suff (\mathcal{T}) $. Moreover, let $ [\ell', r'] $ be the parent of $ [\ell, r] $ in $ \Suff (\mathcal{T}) $. Then, $ j \in [\ell', \ell - 1] \cup [r + 1, r'] $, and for every $ k \in D_j $ we have that $ \mathcal{T}_k $ occurs in $ P $ at position $ i $ if and only if $ |\mathcal{T}_k| \le \lambda_{\ell', r'} $.
    \end{enumerate}
\end{lemma}

Fix $ 1 \le i \le m $. To implement Lemma \ref{lem:climbingsuffixtree}, we should navigate $ \Suff (\mathcal{T}) $ starting from node $ [\ell_{P[i, t_i]}, r_{P[i, t_i]}] $, so we need to know $ \ell_{P[i, t_i]} $ and $ r_{P[i, t_i]} $. Moreover, if $ j \in  [\ell_{P[i, t_i]}, r_{P[i, t_i]}] $ and $ k \in D_j $, we know that $ \mathcal{T}_k $ occurs in $ P $ at position $ i $ if and only if $ |\mathcal{T}_k| \le t_i - i + 1 $, so we also need to compute $ t_i $. Lemma \ref{lem:computing3mvalues} in the appendix presents an $ O(m \log n) $ algorithm to compute all the $ t_i $'s, all the $ \ell_{P[i, t_i]} $'s and all the $ r_{P[i, t_i]} $'s. The algorithm shares many similarities with Ohlebusch et al.'s algorithm for computing matching statistics using the FM-index and the LCP array of a string \cite{ohlebusch2010}.

For every $  1 \le i \le m $, let $ \Cdm_i (\mathcal{T}, P) $ be the set of all $ 1 \le k \le n $ such that $ \mathcal{T}_k $ occurs in $ P $ at position $ i $, and let $ occ_i = |\Cdm_i (\mathcal{T}, P)| $. Computing $ \Cdm (\mathcal{T}, P) $ is equivalent to computing $ \Cdm_i (\mathcal{T}, P) $ for every $ 1 \le i \le m $, and $ occ = \sum_{i = 1}^m occ_i $. In view of Theorem \ref{theor:main}, it will be sufficient to show that we can compute $ \Cdm_i (\mathcal{T}, P) $ in $ O((1 + occ_i) \log n) $ time, because then we can compute each $ \Cdm (\mathcal{T}, P) $ in $ \sum_{i = 1}^m O((1 + occ_i) \log n) = O((m + occ) \log n) $ time.

Fix $ 1 \le i \le m $. Lemma \ref{lem:climbingsuffixtree} suggests that, to compute  $ \Cdm_i (\mathcal{T}, P) $, we can proceed as follows. We start from node $ [\ell_{P[i, t_i]}, r_{P[i, t_i]}] $ is $ \Suff (\mathcal{T}) $, we consider every $ j \in [\ell_{P[i, t_i]}, r_{P[i, t_i]}] $ and every $ k \in D_j $, and we decide whether $ \mathcal{T}_k $ occurs in $ P $ at position $ i $ (we will see later how to do this efficiently). Next, we navigate from node $ [\ell_{P[i, t_i]}, r_{P[i, t_i]}] $ up to the root. Every time we move from a node $ [\ell, r] $ to its parent $ [\ell', r'] $, we consider every $ j \in [\ell', \ell - 1] \cup [r + 1, r'] $ and every $ k \in D_j $, and we decide whether $ \mathcal{T}_k $ occurs in $ P $ at position $ i $ (again, we will see later how to do this efficiently). To navigate $ \Suff (\mathcal{T}) $ from $ [\ell_{P[i, t_i]}, r_{P[i, t_i]}] $ to the root, in the worst case we visit many nodes --- say $ \Omega (\log n) $ nodes. If each of these steps leads to discovering at least one element in $ \Cdm_i (\mathcal{T}, P) $ we can hope to achieve the bound $ O((1 + occ_i) \log n) $, but in the worst case many steps are not successful, $ occ_i $ is small and we cannot achieve the bound $ O((1 + occ_i) \log n) $.

Notice that $ i $ only determines the initial node $ [\ell_{P[i, t_i]}, r_{P[i, t_i]}] $, but once we are navigating $ \Suff (\mathcal{T}) $ up to the root, we do not need $ i $ to assess whether we have found an element of $ \Cdm_i (\mathcal{T}, P) $, because $ \mathcal{T}_k $ occurs in $ P $ at position $ i $ if and only if $ |\mathcal{T}_k| \le \lambda_{\ell', r'} $, and $ \lambda_{\ell', r'} $ does not depend on $ i $. This means that we can determine which nodes will allow discovering an element of $ \Cdm_i (\mathcal{T}, P) $ \emph{before} knowing the pattern $ P $ (that is, at indexing time). We can then give the following definition which, crucially, does not depend on $ i $ or $ P $ (see Figure \ref{subfig:suffixtreeexample}).

\begin{definition}
    Let $ \mathcal{T} $ be a dictionary, and let $ [\ell, r] \in \mathcal{N} $. We say that $ [\ell, r] $ is \emph{marked} if one of the following is true: (i) $ [\ell, r] = [1, n'] $ (i.e., $ [\ell, r] $ is the root of $ \Suff (\mathcal{T}) $), or (ii) $ [\ell, r] \not = [1, n'] $ (i.e., $ [\ell, r] $ is not the root of $ \Suff (\mathcal{T}) $) and, letting $ [\ell', r'] $ be the parent of $ [\ell, r] $ in $ \Suff (\mathcal{T}) $, there exists $ j \in [\ell', \ell - 1] \cup [r + 1, r'] $ and there exists $ k \in D_j $ such that $ |\mathcal{T}_k| \le \lambda_{\ell', r'} $.
\end{definition}

When we navigate $ \Suff (\mathcal{T}) $ from $ [\ell_{P[i, t_i]}, r_{P[i, t_i]}] $ to the root, we should skip all non-marked nodes. Notice the set of all marked nodes induces a new tree structure. More precisely, consider the ordinal tree $ \Suff^*(\mathcal{T}) $ with root $ [1, n'] $ defined as follows. The set $ \mathcal{N}^* $ of nodes is the set of all marked nodes in $ \mathcal{N} $; in particular, the root $ [1, n'] $ of $ \Suff (\mathcal{T}) $ belongs to $ \mathcal{N}^* $. We can now build $ \Suff^*(\mathcal{T}) $ incrementally. We traverse all nodes of $ \Suff(\mathcal{T})$ in depth-first search order, respecting the order of the children of each node (this is the same order used for the balanced parenthesis representation of $ \Suff(\mathcal{T})$). The first marked node that we encounter is $ [1, n'] $, which will be root of $ \Suff^*(\mathcal{T}) $. Then, every time we encounter a marked node $ [\ell, r] $, let $ [\ell', r'] $ be the nearest strict ancestor of $ [\ell, r] $ in $ \Suff (\mathcal{T}) $ that is marked, namely, $ [\ell', r'] $ is the first marked node that we encounter after $ [\ell, r] $ in the (backward) path from $ [\ell, r] $ to $ [1, n'] $ in $ \Suff (\mathcal{T}) $. Then, $ [\ell', r'] $ will be the parent of $ [\ell, r] $ in $ \Suff^*(\mathcal{T}) $, and if $ [\ell', r'] $ has already been assigned $ q \ge 0  $ children, then $ [\ell, r] $ will be its $ (q + 1) $-th smallest child. See Figure \ref{subfig:treeuseful} for an example, and see Figure \ref{subfig:tabletrees} for the balanced parenthesis representation of $ \Suff^*(\mathcal{T}) $. In addition to the data structure of Theorem \ref{theor:balancedparenthesisnavarrosadakane} for the tree $ \Suff(\mathcal{T}) $, we also store the data structure of Theorem \ref{theor:balancedparenthesisnavarrosadakane} for the tree $ \Suff^*(\mathcal{T}) $, which also requires $ O(n) $ bits.

We also remember which nodes of $ \Suff (\mathcal{T}) $ are marked by  using a bitvector $ B_8 $ of length $ 2 |\mathcal{N}| $ such that for every $ 1 \le t \le 2 |\mathcal{N}| $ we have $ B_8[t] = 1 $ if and only if $ Z_{\Suff (\mathcal{T})}[t] $ is one of the two values corresponding to a marked node of $ \Suff (\mathcal{T}) $. More precisely, we build $ B_8 $ as follows. We visit all nodes of $ \Suff (\mathcal{T)} $ in depth-first search order, respecting the order of the children of each node. Every time we encounter a new node $ u $ we append an $ 1 $ if $ u $ is marked and a $ 0 $ if $ u $ is not marked, and as soon as we leave the subtree rooted at $ u $ we add a $ 1 $ if $ u $ is marked and a $ 0 $ if $ u $ is not marked (see Figure \ref{subfig:tabletrees}). By using $ B_8 $, in $ O(1) $ time (i) we can move from $ \Suff (\mathcal{T}) $ to $ \Suff^* (\mathcal{T}) $ and from $ \Suff^* (\mathcal{T}) $ to $ \Suff (\mathcal{T}) $ and (ii) we can determine the nearest marked ancestor of a node (see Lemma \ref{lem:fromtreetoauxiliary} in the appendix).

Define the array $ \Len $ of length $ n^* = \sum_{h = 1}^d |\rho_h| $ (the length of $ \SA $) such that $ \Len[t] = |\mathcal{T}_{\SA[t]}| $ for every $ t $ (see Figure \ref{fig:suffixarray}). We will not store $ \Len $, but only a data structure supporting range minimum queries on $ \Len $ in $ O(1) $ time, which requires $ O(n) $ bits (see Section \ref{sec:lcp}).

We now have all the ingredients to prove Theorem \ref{theor:main}, our main claim. As we have seen, it will suffice to compute each $ \Cdm_i (\mathcal{T}, P) $ in $ O((1 + occ_i) \log n) $. We start from node $ [\ell_{P[i, t_i]}, r_{P[i, t_i]}] $, and for every $ k \in \bigcup_{\ell_{P[i, t_i]} \le j \le r_{P[i, t_i]}} D_j $ we determine whether $ \mathcal{T}_k $ occurs in $ P $ at position $ i $. By Lemma \ref{lem:climbingsuffixtree}, we need to check whether $ |\mathcal{T}_k| \le t_i - i + 1 $, so we repeatedly solve range minimum queries on $ \Len $ starting from the interval $ [\ell_{P[i, t_i]}, r_{P[i, t_i]}] $ to find all $ k $'s for which $ \mathcal{T}_k $ occurs in $ P $ at position $ i $ in time proportional to the number of such occurrences. Next, we compute the lowest marked ancestor of $ [\ell_{P[i, t_i]}, r_{P[i, t_i]}] $, and we use $ \Suff^* (\mathcal{T}) $ to determine all marked ancestors of $ [\ell_{P[i, t_i]}, r_{P[i, t_i]}] $. Let $ [\ell, r] $ be one of these ancestors, and let $ [\ell', r'] $ be its parent in $ \Suff (\mathcal{T}) $. For every $ k \in (\bigcup_{\ell' \le j \le \ell - 1} D_j) \cup (\bigcup_{r + 1 \le j \le r'} D_j) $, we determine whether $ \mathcal{T}_k $ occurs in $ P $ at position $ i $. By Lemma \ref{lem:climbingsuffixtree}, we need to check whether $ |\mathcal{T}_k| \le \lambda_{\ell', r'} $, so we repeatedly solve range minimum queries on $ \Len $ starting from the intervals $ [\ell', \ell - 1] $ and $ [r + 1, r'] $ to find all $ k $'s for which $ \mathcal{T}_k $ occurs in $ P $ at position $ i $ in time proportional to the number of such occurrences.

The details of the proof of Theorem \ref{theor:main} are in the appendix (see also Remark \ref{rem:occlarge} for large values of $ occ$). Note that we cannot directly infer that our data structure is an encoding of $ \mathcal{T} $ from Theorem \ref{theor:FMindex} because the graph $ G_\mathcal{T} $ is not sufficient to retrieve $ \mathcal{T} $.

\section{Conclusions and Future Work}

We have shown how to improve and extend previous results on circular dictionary matching. In the literature, much effort has been devoted to designing construction algorithms for the data structure of Hon et al. \cite{hon2011isaac} and, implicitly, the eBWT of Mantaci et al. \cite{mantaci2007}. All available approaches first build the eBWT and the suffix array of circular strings, and then use the suffix array of circular strings to build the remaining auxiliary data structures. In \cite{hon2012cpm}, Hon et al. showed that the eBWT and the suffix array can be built in $ O(n \log n) $ time using $ O(n \log \sigma + d \log n) $ bits of working space. Bannai et al. \cite{bannai2021cpm} improved the time bound to $ O(n) $ by showing that the bijective BWT can be built in linear time (Bonomo et al. \cite[Section 6]{bonomo2014} showed how to reduce in linear time the problem of computing the eBWT to the problem of computing the bijective BWT). A more direct algorithm was proposed by Boucher et al. \cite{boucher2021spire}. However, all these algorithms are still based on assumptions (i-ii) in Section \ref{sec:introduction} and cannot immediately applying to our setting in which we consider an arbitrary dictionary. After building the eBWT and the suffix array, it is possible to build the remaining auxiliary data structure in $ O(n \log n) $ time using $ O(n \log \sigma + d \log n) $ bits of working space \cite{hon2013cpm}. Some proofs in \cite{hon2013cpm} are only sketched, but it is not too difficult to show that, if we do not insist on achieving $ O(n \log \sigma + d \log n) $ bits of working space, it is possible to build the remaining auxiliary data structure from the eBWT and the suffix array in linear time.

In a companion paper, we will show that the data structure of Theorem \ref{theor:main} can be built in $ O(n) $ time. The main technical issue is understanding how to remove assumptions (i-ii) in Section \ref{sec:introduction} when building the suffix array. The algorithm by Boucher et al.'s \cite{boucher2021spire} is a recursive algorithm based on the SAIS algorithm \cite{nong2011} for building the suffix array. SAIS divides all suffixes into two categories (S-type and L-type). We will show that, to remove assumptions (i-ii), we will use three (and not two) categories, building on a recent recursive algorithm \cite{cotumaccio2023isaac} for constructing the ``suffix array'' of a deterministic automaton.

The natural question is whether the data structure of Theorem 1 (or a similar data structure with the same functionality) can be built in $ O(n) $ time within $ O(n \log \sigma + d \log n) $ bits of working space. We conjecture that this is possible but, thinking of the case $ d = 1 $, this problem should be at least as difficult as building the compressed suffix tree of a string in $ O(n) $ time and $ O(n \log \sigma) $ working bits, which requires advanced techniques \cite{munro2017soda, belazzougui2020}.

\newpage

\bibliography{lipics-v2021-sample-article}

\appendix

\section{Proofs from Section \ref{sec:compressedsuffix}}

\subsection{Proofs from Section \ref{sec:BWTFM}}

\noindent{\textbf{Statement of Lemma \ref{lem:roots}}.}
 Let $ \mathcal{T} $ be a dictionary, and let $ 1 \le k, k' \le n $. Then, $ \mathcal{T}_k^\omega = \mathcal{T}_{k'}^\omega $ if and only if $ \rho(\mathcal{T}_k) = \rho(\mathcal{T}_{k'}) $.

\begin{proof}
    The proof is a rewriting of \cite[Prop. 5]{mantaci2007}. Let $ R = \rho(\mathcal{T}_k) $ and $ R' = \rho(\mathcal{T}_{k'}) $, and let $ z, z' \ge 1 $ integers such that $ \mathcal{T}_k = R^z $ and $ \mathcal{T}_{k'} = R^{z'} $. Notice that $ \mathcal{T}_k^\omega = (R^z)^\omega = R^\omega $; analogously, $ \mathcal{T}_{k'}^\omega = (R')^\omega $.

    We have to show that $ R^\omega = (R')^\omega $ if and only if $ R = R' $. The implication $ (\Leftarrow) $ is immediate, so let us prove $ (\Rightarrow) $. If $ R^\omega = R^\omega $, then by Fine and Wilf's theorem (see \cite[Prop. 1.3.5]{lothaire1997}) there exist $ S \in \Sigma^* $ and integers $ b, b' \ge 1 $ such that $ R = S^{b} $ and $ R' = S^{b'} $. Since $ R $ and $ R $ are roots and so primitive strings, we have $ b = b' = 1 $ and we obtain $ R = S = R' $.
\end{proof}

\begin{lemma}\label{lem:consistencypredecessors}
    Let $ \mathcal{T} $ be a dictionary, and let $ 1 \le k, k_1 \le n $. Then, $ k $ and $ k_1 $ belong to the same $ D_j $ if and only if $ \pred(k) $ and $ \pred(k_1) $ belong to the same $ D_{j'} $.
\end{lemma}

\begin{proof}
    $ (\Rightarrow) $ Since $ k $ and $ k_1 $ belong to the same $ D_j $, we know that $ \mathcal{T}_k^\omega = \mathcal{T}_{k_1}^\omega $. By Lemma \ref{lem:roots}, we know that $ \mathcal{T}_k $ and $ \mathcal{T}_{k_1} $ have the same root, so if we define $ R = \rho(\mathcal{T}_k) = \rho(\mathcal{T}_{k_1}) $, there exist integers $ z, z_1 \ge 1 $ such that $ \mathcal{T}_k = R^z $ and $ \mathcal{T}_{k_1} = R^{z_1} $. This implies that, if $ c $ is the last character of $ R $, then $ c $ is also the last character of $ \mathcal{T}_k $ and the last character of $ \mathcal{T}_{k_1} $, and $ \mathcal{T}_{\pred(k)}^\omega = c \mathcal{T}_k^\omega = c \mathcal{T}_{k_1}^\omega = \mathcal{T}_{\pred(k_1)}^\omega $, so $ \pred(k) $ and $ \pred(k_1) $ belong to the same $ D_{j'} $.

    $ (\Leftarrow) $ Since $ \pred(k) $ and $ \pred(k_1) $ belong to the same $ D_{j'} $, then $ T_{\pred(k)}^\omega = T_{\pred(k_1)}^\omega $, and if we define $ c = \mathcal{T}_{\pred(k)}^\omega[1] = \mathcal{T}_{\pred(k_1)}^\omega[1] $, we have $ c \mathcal{T}_k^\omega = T_{\pred(k)}^\omega = T_{\pred(k_1)}^\omega = c \mathcal{T}_{k_1}^\omega $, so $ \mathcal{T}_k^\omega = \mathcal{T}_{k_1}^\omega $, which means that $ k $ and $ k_1 $ belong to the same $ D_j $.
\end{proof}

\noindent{\textbf{Statement of Lemma \ref{lem:goodbehaviordj}}.}
Let $ \mathcal{T} $ be a dictionary, and let $ 1 \le j \le n' $. Then, there exists $ 1 \le j' \le n $ such that $ \pred(D_j) = D_{j'} $. Moreover, if $ c = \mathcal{S}_{j'} [1] $, we have $ \mathcal{S}_{j'} = c \mathcal{S}_j $.

\begin{proof}
    Fix any $ k'_1 \in \pred(D_j) $, and let $ 1 \le j' \le n $ be such that $ k'_1 \in D_{j'} $. We want to prove that $ \pred(D_j) = D_{j'} $. Let $ k_1 \in D_j $ be such that $ k'_1 = \pred (k_1) $. Let us prove that $ \pred(D_j) = D_{j'} $.

    $ (\subseteq) $ Pick $ k' \in \pred(D_j) $. Then, there exists $ k \in D_j $ such that $ k' = \pred (k) $. We have $ k, k_1 \in D_j $ and $\pred (k_1) = k'_1 \in D_{j'} $, hence by Lemma \ref{lem:consistencypredecessors} we have $ k' = \pred (k) \in D_{j'} $.

    $ (\supseteq) $ Pick $ k' \in D_{j'} $. We know that $ \pred $ is a permutation of $ \{1, 2, \dots, n \} $, so we can write $ k' = \pred(k) $ for some $ 1 \le k \le n $. Since $ \pred (k_1) = k'_1 \in D_{j'} $, $ \pred (k) = k' \in D_{j'} $ and $ k_1 \in D_j $, then by Lemma \ref{lem:consistencypredecessors} we obtain $ k \in D_j $, which implies $ k' \in \pred(D_j) $.

    Lastly, we have $ \mathcal{S}_{j'} = \mathcal{T}_{k'_1}^\omega = \mathcal{T}_{\pred (k_1)}^\omega = c \mathcal{T}_{k_1}^\omega = c \mathcal{S}_j $.
\end{proof}

\noindent{\textbf{Statement of Lemma \ref{lem:axioms1and2}}.}
Let $ \mathcal{T} $ be a dictionary. let $ 1 \le j_1, j'_1, j_2, j'_2 \le n' $ and $ c, d \in \Sigma $ such that $ \mathcal{S}_{j'_1} = c \mathcal{S}_{j_1} $ and $ \mathcal{S}_{j'_2} = d \mathcal{S}_{j_2} $.
    \begin{enumerate}
        \item (Property 1) If $ j'_1 < j'_2 $, then $ c \preceq d $.
        \item (Property 2) If $ c = d $ and $ j'_1 < j'_2 $, then $ j_1 < j_2 $.
    \end{enumerate}

\begin{proof}
    \begin{enumerate}
        \item If $ j'_1 < j'_2 $, then $ c \mathcal{S}_{j_1} = \mathcal{S}_{j'_1} \prec \mathcal{S}_{j'_2} = d \mathcal{S}_{j_2} $ and so $ c \preceq d $.
        \item If $ c = d $ and $ j'_1 < j'_2 $, then $ c \mathcal{S}_{j_1} = \mathcal{S}_{j'_1} \prec \mathcal{S}_{j'_2} = d \mathcal{S}_{j_2} = c \mathcal{S}_{j_2} $, so $ \mathcal{S}_{j_1} \prec \mathcal{S}_{j_2} $ and we conclude $ j_1 < j_2 $.
    \end{enumerate}
\end{proof}

\noindent{\textbf{Statement of Lemma \ref{lem:convex}}.}
    Let $ \mathcal{T} $ be a dictionary, let $ U \subseteq \{1, 2, \dots, n' \} $ and let $ c \in \Sigma $. If $ U $ is convex, then $ \back(U, c) $ is convex.

\begin{proof}
    Assume that $ 1 \le j'_1 < j' < j'_2 \le n' $ are such that $ j'_1, j'_2 \in   \back(U, c) $. We must prove that $ j' \in \back(U, c) $. Since $ j'_1, j'_2 \in \back(U, c) $, there exist $ j_1, j_2 \in U $ such that $ \mathcal{S}_{j'_1} = c \mathcal{S}_{j_1} $ and $ \mathcal{S}_{j'_2} = c \mathcal{S}_{j_2} $. Since $ \psi $ is a permutation, there exist $ 1 \le j \le n' $ and $ d \in \Sigma $ such that $ \mathcal{S}_{j'} = d \mathcal{S}_j $. From Property 1 of Lemma \ref{lem:axioms1and2} and $ j'_1 < j' < j'_2 $ we obtain $ c \preceq d \preceq c $, so $ d = c $. From Property 2 of Lemma \ref{lem:axioms1and2} and $ j'_1 < j' < j'_2 $ we obtain $ j_1 < j < j_2 $, so $ j \in U $ because $ j_1, j_2 \in U $ and $ U $ is convex. From $ j \in U $ and $ \mathcal{S}_{j'} = c \mathcal{S}_j $ we conclude $ j' \in \back(U, c) $.
\end{proof}

\noindent{\textbf{Statement of Theorem \ref{theor:FMindex}}.}
    Let $ \mathcal{T} $ be a dictionary.
    \begin{enumerate}
        \item $ \BWT $ is an encoding of $ G_\mathcal{T} $, that is, given $ \BWT $, we can retrieve $ G_\mathcal{T} $ up to isomorphism.
        \item There exists a data structure encoding $ G_\mathcal{T} $ of $ n' \log \sigma (1 + o(1)) + O(n') $ bits that supports the following queries in $ O(\log \log \sigma) $ time: (i) $ \access $, $\rank $, and $ \select $ queries on $ \BWT $, (ii) $ \bws(\ell, r, c) $: given $ 1 \le \ell \le r \le n' $ and $ c \in \Sigma $, decide if $ \back(\{\ell, \ell + 1, \dots, r \}, c) $ is empty and, if it is nonempty, return $ \ell' $ and $ r' $ such that $ \back(\{\ell, \ell + 1, \dots, r \}, c) = \{\ell', \ell' + 1, \dots, r' \} $, (iii) $ \prev(j) $: given $ 1 \le j \le n' $, return $ 1 \le j' \le n' $ such that $ \pred (D_j) = D_{j'} $, and (iv) $ \follow(j') $: given $ 1 \le j' \le n' $, return $ 1 \le j \le n' $ such that $ \pred (D_j) = D_{j'} $.
    \end{enumerate}

\begin{proof}
    We have already observed that Lemma \ref{lem:axioms1and2} can be interpreted as follows: if $ (D_{j'_1}, D_{j_1}, c) $ and $ (D_{j'_2}, D_{j_2}, c) $ are two edges of $ G_\mathcal{T} $, then (Property 1) if $ j'_1 < j'_2 $, then $ c \preceq d $, and (Property 2) if $ c = d $ and $ j'_1 < j'_2 $, then $ j_1 < j_2 $. Let us prove the lemma. Note that $ \bws(\ell, r, c) $ is well-defined by Lemma \ref{lem:convex}.
    \begin{enumerate}
        \item First, we have $ n' = |\BWT| $. Retrieving $ G_\mathcal{T} $ up to isomorphism is equivalent to determining for every $ 1 \le j, j' \le n' $ and for every $ c \in \Sigma $ whether $ (D_{j'}, D_j, c) $ is an edge of $ G_\mathcal{T} $. Indeed, with this information we can define the graph $ G^* = (V^*, E^*) $ such that $ V^* = \{1, 2, \dots, n' \} $ and for every $ 1 \le j, j' \le n' $ and for every $ c \in \Sigma $ we have $ (j', j, c) \in E $ if and only if $  (D_{j'}, D_j, c) $ is an edge of $ G_\mathcal{T} $. Then, $ G^* $ is isomorphic to $ G_\mathcal{T} $.
        
        For every $ c \in \Sigma $, let $ X_c = \{1 \le j \le n' \;|\; \text{ the edge leaving $ D_j $ is labeled with $ c $} \} $ and $ Y_c = \{1 \le j \le n' \;|\; \text{ the edge reaching $ D_j $ is labeled with $ c $} \} $. We can retrieve each $ Y_c $ from $ \BWT $ because $ \BWT[j] $ is the label of the edge reaching $ D_j $ for every $ 1 \le j \le n' $. Moreover, we know that $ \BWT^*[j] $ is the character leaving $ D_j $ for every $ 1 \le j \le n' $, and we can retrieve $ \BWT $ by sorting $ \BWT^*$. This means that we can also retrieve each set $ X_c $.
        
        Fix $ c \in \Sigma $. We must show how to retrieve the set of all edges labeled $ c $. The sets $ X_c $ and $ Y_c $ have the same size, so define $ w = |X_c| = |Y_c| $. For every $ 1 \le t \le w $, let $ x_t $ the $ t $-smallest element in $ X_c $ and let $ y_t $ be the $ t $-smallest element in $ Y_c $. Then, the set of all edges labeled $ c $ is $ \{(D_{x_t}, D_{y_t}, c) \;|\; 1 \le t \le w \} $, because otherwise there would exist two edges that contradict Property 2.

        \item Since the alphabet is effective, we can encode $ \BWT $ by using a data structure of $ n' \log \sigma (1 + o(1)) + O(n') $ bits which supports $ \access $, $ \rank $ and $ \select $ queries in $ O(\log \log \sigma) $ time \cite{navarro2016book}. Since $ \BWT $ is an encoding of $ G_\mathcal{T} $ by the previous point, then this new data structure is also an encoding of $ G_\mathcal{T} $. We are left with showing how to support $ \bws(\ell, r, c) $, $ \prev(j) $ and $ \follow(j') $ in $ O(\log \log \sigma) $ time.
        \begin{enumerate}
            \item Let us show how to support $ \bws(\ell, r, c) $ in $ O(\log \log \sigma) $ time. Notice that $ \back(\{\ell, \ell + 1, \dots, r \}, c) \not = \emptyset $ if and only if there exists $ \ell \le j \le r $ such that $ \BWT[j] = c $, because for every $ j $ we have that $ \BWT[j] $ labels the edge entering node $ D_j $. Formally, we have $ \back(\{\ell, \ell + 1, \dots, r \}, c) \not = \emptyset $ if and only if $ \mu (j) = c $ for some $ \ell \le j \le r $, if and only if $ \BWT[j] = c $ for some $ \ell \le j \le r $. As a consequence, we can check whether $ \back(\{\ell, \ell + 1, \dots, r \}, c) \not = \emptyset $ by using $ \rank $ queries, because there exists $ \ell \le j \le r $ such that $ \BWT[j] = c $ if and only if $ \rank_c(\BWT, \ell - 1) < \rank_c(\BWT, r) $.
        
            Let us assume in the following that $ \back(\{\ell, \ell + 1, \dots, r \}, c) \not = \emptyset $. We must show how to compute $ \ell' $ and $ r' $ such that  $ \back(\{\ell, \ell + 1, \dots, r \}, c) = \{\ell', \ell' + 1, \dots, r' \} $. From the proof of the previous point, we know that the set of all edges labeled $ c $ is $ \{(D_{x_t}, D_{y_t}, c) \;|\; 1 \le t \le w \} $. Calling $ 1 \le t^* \le w $ the smallest integer such that $ \ell \le y_{t^*} \le r $, we have $ x_{t^*} = \ell' $. Notice that $ t^* = \rank_c(\BWT, \ell - 1) + 1 $. We know that $ \{ \select_1 (B_2, c), \select_1 (B_2, c) + 1, \select_1 (B_2, c) + 2, \dots, \select_1 (B_2, c + 1) - 2, \select_1 (B_2, c + 1) - 1 \} $ consists of all $ 1 \le j' \le n' $ such that the edge leaving node $ D_{j'} $ is labeled with $ c $, so $ x_{t^*} = \select_1 (B_2, c) + t^* - 1 $. We conclude $ \ell' = \select_1 (B_2, c) + \rank_c(\BWT, \ell - 1) $. An analogous argument shows that $ r' = \select_1 (B_2, c) + \rank_c(\BWT, r) - 1 $.
            \item Let us show how to support $ \prev(j) $ in $ O(\log \log \sigma) $ time. Since $ (D_{\prev(j)}, D_j, \BWT [j]) $ is an edge of $ G_\mathcal{T} $, we only have to apply the backward search starting from $ \{j \} $. Formally, we know that $ \bws(j, j, \BWT [j]) = \{\prev(j) \}  $ because $ \mathcal{S}_{\prev(j)} = \BWT [j] \mathcal{S}_j $ by the definition of $ \BWT [j] $, so the conclusion follows from the previous point.
            \item Let us show how to support $ \follow(j') $ in $ O(\log \log \sigma) $ time. If $ j = \follow(j') $, then $ \BWT[j] = \BWT^*[j'] $ and $ \{ j'\} = \bws(j, j, \BWT[j]) = \bws(j, j, \BWT^*[j']) $. Let $ c = \BWT^*[j'] = \BWT[j] $, and consider again the set $ \{(D_{x_t}, D_{y_t}, c) \;|\; 1 \le t \le w \} $ of all edges labeled $ c $. Since the alphabet is effective, we know that $ c = \BWT^*[j'] = \rank_1 (B_2, j') $. Now we only need to reverse the argument used to compute $ \bws $. If $ j' $ is the $ t^* $-th smallest integer such that the edge leaving $ D_{j'} $ is $ \BWT^*[j'] $, then $ j' = x_{t^*} $ and $ j = y_{t^*} $. Notice that $ t^* = j' - \select_1 (B_2, c) + 1 $ and $ \follow(j') = j = y_{t^*} = \select_c (\BWT, t^*) $.
        \end{enumerate}
    \end{enumerate}
\end{proof}

\subsection{Proofs from Section \ref{sec:suffixarray}}

\noindent{\textbf{Statement of Theorem \ref{theor:sampledsuffixarray}}.}
Let $ \mathcal{T} $ be a dictionary. By storing $ o(n \log \sigma) + O(n) + O(d \log n) $ bits, we can compute each entry $ \SA[t] $ in $ O(\log n) $ time.

\begin{proof}
    Let $ s $ be a sampling factor (to be fixed later). We will sample all $ k $'s occurring in $ \SA $ (i.e., for which there exists $ t $ such that $ \SA[t] = k $) having the form $ k_h + s x $, where $ 1 \le h \le d$ and $ x \ge 0 $ is an integer such that $ k_h + s x \le k_h + |\rho_h| - 1 $, that is, $  0 \le x \le (|\rho_h| - 1) / s $. We sample at most $ ((|\rho_h| - 1) / s) + 1 $ values for each $ h $, and so we sample at most $ n/s + d $ values in total, which requires $ (n/s + d) \log n $ bits. We store the sampled values in an array $ \SA^* $ (respecting the order in $ \SA $), and we use a bitvector $ B_5 $ to mark the entries of $ \SA[k] $ that have been sampled (see Figure \ref{fig:suffixarray}, where we assume $ s = 2 $).
    
    Let us show how to compute an entry $ \SA[t] $. Let $ k = \SA[t] $ and let $ k \in D_j $. If $ B_5[t] = 1 $, we can immediately conclude $ \SA[t] = \SA^*[\rank_1 (B_5, t)] $. Assume that $ B_5[t] = 0 $. We know that $ j = \rank_1 (B_4, t) $, and in particular $ k $ is the $ q $-th smallest element in $ D'_j $, where $ q = t - \select_1 (B_4, j) + 1 $. Let $ 1 \le j' \le n $ be such that $ \pred(D_j) = D_{j'} $ (see Lemma \ref{lem:goodbehaviordj}). Then, $ j' = \prev(j) $, so by Theorem \ref{theor:FMindex} we can compute $ j' $ in $ O(\log \log \sigma) $ time. Since we have defined the array $ \SA $ by sorting the elements in $ D_{j'} $ and the elements in $ D'_{j'} $ according to the indexes of the strings to which they refer, we know that $ \pred(k) $ is the $ q $-th smallest element in $ D'_{j'} $, and so, if $ t' = \select_1 (B_4, j') + q - 1 $, then $ \SA[t'] = \pred(k) $. Notice that we in our sampling we sample the first character of each $ T_h $, and we are assuming $ B_5[t] = 0 $, so we must have $ \pred(k) = k - 1 $. We conclude that $ \SA[t'] = k - 1 $, and so $ \SA[t] = \SA[t'] + 1 $. We have thus reduced in $ O(\log \log \sigma) $ time the problem of computing $ \SA[t] $ to the problem of computing $ \SA[t'] $.

    We can now repeat the same argument starting from $ \SA[t'] $. After at most $ s $ steps, we must find a sampled value because of the definition of $ \SA^* $. We conclude that we can retrieve $ \SA[t] $ in $ O(s \log \log \sigma) $ time.

    The conclusion follows by choosing $ s = \log n / \log \log \sigma $.

\end{proof}

\subsection{Proofs from Section \ref{sec:lcp}}

\begin{lemma}\label{lem:goesdownbyone}
    Let $ \mathcal{T} $ be a dictionary. Let $ g \ge 1 $ be an integer. If there exists $ 2 \le j \le n' $ such that $ \LCP[j] = g $, then there exists $ 2 \le j' \le n' $ such that $ \LCP[j'] = g - 1 $.
\end{lemma}

\begin{proof}
    Since $ \psi $ is a permutation, there exist $ 1 \le j_1, j_2 \le n' $ and $ c_1, c_2 \in \Sigma $ such that $ \mathcal{S}_{j - 1} = c_1 \mathcal{S}_{j_1} $ and $ \mathcal{S}_j = c_2 \mathcal{S}_{j_2} $. From $ \LCP[j] \ge 1 $ we obtain $ c_1 = c_2 $, hence from $ \mathcal{S}_{j - 1} \prec \mathcal{S}_j $ we obtain $ \mathcal{S}_{j_1} \prec \mathcal{S}_{j_2} $ and so $ j_1 < j_2 $. We have $ g = \LCP[j] = \lcp(\mathcal{S}_{j - 1}, \mathcal{S}_j) = 1 + \lcp(\mathcal{S}_{j_1}, \mathcal{S}_{j_2}) = 1 + \min_{j_1 + 1 \le j' \le j_2} \lcp(\mathcal{S}_{j' - 1}, \mathcal{S}_{j'}) = 1 + \min_{j_1 + 1 \le j' \le j_2} \LCP[j']  $, so there exists $ j_1 + 1 \le j' \le j_2 $ such that $ \LCP[j'] = g - 1 $.
\end{proof}

\noindent{\textbf{Statement of Lemma \ref{lem:upperbound}}.}
    Let $ \mathcal{T} $ be a dictionary. Then, $ \LCP[j] \le n' - 2 $ for every $ 2 \le j \le n' $.

\begin{proof}
    Let $ g \ge 0 $ be an integer such that there exists $ 2 \le j \le n' $ for which $ \LCP[j] = g $. We must prove that $ g \le n' - 2 $. By Lemma \ref{lem:goesdownbyone} we know that for every $ 0 \le g' \le g $ there exists $ 2 \le j' \le n' $ such that $ \LCP[j'] = g' $. Since the array $ \LCP $ has length $ n' - 1 $, we must have $ g + 1 \le n' - 1 $, so $ g \le n' - 2 $.
\end{proof}

We now show how to prove Theorem \ref{theor:lcpsampling}. To define our sampled LCP array, we follow the idea of the proof of Lemma \ref{lem:goesdownbyone}. Suppose that we want to compute $ \LCP[j] = \lcp(\mathcal{S}_{j - 1}, \mathcal{S}_j) $. If $ B_2[j] = 1 $, we have $ \mathcal{S}_{j - 1}[1] \not = \mathcal{S}_j[1] $ and we conclude $ \LCP[j] = 0 $. Now assume that $ B_2[j] = 0 $. In this case $ \LCP[j] \ge 1 $. Let $ c = \mathcal{S}_{j - 1}[1] = \mathcal{S}_j[1] $, and let $ 1 \le j_1, j_2 \le n' $ be such that $ \mathcal{S}_{j - 1} = c \mathcal{S}_{j_1} $ and $ \mathcal{S}_j = c \mathcal{S}_{j_2} $. Then, as in the proof of Lemma \ref{lem:goesdownbyone}, we have $ j_1 < j_2 $ and $ \LCP[j] = 1 + \min_{j_1 + 1 \le j' \le j_2} \LCP[j'] $. Notice that by Theorem \ref{theor:FMindex} we have $ j_1 = \follow (j - 1) $ and $ j_2 = \follow (j) $, so $ j_1 $ and $ j_2 $ can be computed in $ O(\log \log \sigma) $ time. Then, we can compute the smallest $ j_1 + 1 \le j' \le j_2 $ such that $ \LCP[j] = 1 + \LCP[j'] $ in $ O(1) $ time because $ j' = \rmq_{\LCP} (j_1 + 1, j_2) $. We conclude that in $ O(\log \log \sigma) $ time we have reduced the problem of computing $ \LCP[j] $ to the problem of computing $ \LCP[j'] $.

Consider the unlabeled directed graph $ Q $ defined as follows (see Figure \ref{subfig:samplelcp}). The set of nodes is $ \{D_2, D_3, \dots, D_{n'} \} $. Every node has at most one outgoing edge. More precisely, if $ \LCP[j] = 0 $, then $ D_j $ has no outgoing edges, and if $ \LCP[j] \ge 1 $, then in $ Q $ there is the edge $ (D_j, D_{j'}) $, where $ j' $ is the value for which $ \LCP[j] = 1 + \LCP[j'] $ that we computed above. Notice that $ Q $ is acyclic because every path in $ Q $ can contain at most $ n' - 2 $ edges by Lemma \ref{lem:upperbound}.

We will now sample some nodes of $ Q $ in such a way that there is some guarantee about the distance of each node from a sampled node. We will use the following result \cite{cotumaccio2023spire}.

\begin{lemma}\label{lem:samplingnodes}
    Consider an unlabeled directed graph with $ n $ nodes in which every node has at most one outgoing edge. Let $ s $ be a sampling parameter. Then, there exists a subset of nodes of size at most $ n / s $ such that for every node $ u $ there exists $ 0 \le i \le 2s - 2 $ for which, if we follow the (unique) path starting from $ u $, then the path contains at least $ i $ edges and, after following $ i $ edges, we reach a node $ v $ such that one of the following is true:
    \begin{enumerate}
        \item $ v $ does not have an outgoing edge.
        \item $ v $ is a sampled node.
        \item we have already encountered before $ v $ along the path.
    \end{enumerate}
\end{lemma}

If we apply Lemma \ref{lem:samplingnodes} to $ Q $, then case 3 cannot occur because $ Q $ is acyclic. Let us consider an array $ \LCP^*[j] $ obtained by storing only the values $ \LCP[j] $ such that $ D_j $ is a sampled node in $ Q $ (respecting the order in $ \LCP $, see Figure \ref{subfig:tablelcp}). Then, $ \LCP^* $ is an array of length $ n / s $ (where $ s $ is the sampling parameter), and we can store each entry of $ \LCP^* $ using at most $ \log n $ bits, for a total of at most $ (n / s) \log n $ bits. We also store a bitvector $ B_6 $ to mark the sampled entries of $ \LCP $ (see Figure \ref{subfig:tablelcp}). Then, we sampled a value $ \LCP[j] $ if and only if $ B_6[j] = 1 $ and, in this case, the sampled value is $ \LCP^*[\rank_1 (B_6, j)] $. After performing at most $ 2s - 2 $ steps of our algorithm for computing $ \LCP[j] $, either we encounter a sampled node $ D_{j'} $, and so we can retrieve the value $ \LCP[j'] $, or we encounter a node $ D_{j'} $ with no outgoing edges, and so $ \LCP[j'] = 0 $. Since every step takes $ O(\log \log \sigma) $ time, the total time required to compute any value $ \LCP[j] $ is $ O(s \log \log \sigma) $. By choosing $ s = \log n / \log \log \sigma $, we obtain Theorem \ref{theor:lcpsampling}.

\subsection{Proofs from Section \ref{sec:topologysuffixtree}}

The suffix tree of a string is the compressed trie of its suffixes. In our setting, it would be natural to define the suffix tree of $ \mathcal{T} $ as the compressed trie of the $ \mathcal{S}_j $'s, but we need some care because the $ \mathcal{S}_j $'s are infinite strings. We proceed as follows. The \emph{trie} of the $ \mathcal{S}_j $'s is the \emph{infinite} trie for the set $ \mathcal{P} = \{\mathcal{S}_j[1, t] \;|\; 1 \le j \le n', t \ge 0 \} $, which is an infinite set of finite strings (note that $ \epsilon \in \mathcal{P} $). Then, the \emph{suffix tree} $ \Suff(\mathcal{T}) $ of $ \mathcal{T} $ is obtained by compressing all (finite or infinite) unary paths (see Figure \ref{subfig:suffixtreeexample} for an example).

The set $ \mathcal{P} $ is closed under prefixes and suffixes, as shown in the following lemma.

\begin{lemma}\label{lem:Pprefixessuffixes}
    Let $ \mathcal{T} $ be a dictionary. If $ P \in \mathcal{P} $, then every suffix and every prefix of $ P $ are in $ \mathcal{P} $.
\end{lemma}

\begin{proof}
    Since $ P \in \mathcal{P} $, there exists $ 1 \le j \le n' $ such that $ P = \mathcal{S}_j[1, |P|] $. We can assume $ |P| \ge 1 $ otherwise the conclusion is immediate.
    
    Let $ P' $ be any prefix of $ P $. We must prove that $ P' \in \mathcal{P} $. We have $ P' = P[1, |P'|] = (\mathcal{S}_j[1, |P|])[1, |P'|] = \mathcal{S}_j[1, |P'|] $, so $ P' \in \mathcal{P} $.

    To prove that any suffix of $ P $ is in $ \mathcal{P} $, we only have to prove that $ P[2, |P|] \in \mathcal{P} $, because then we can argue by induction. Since $ \psi $ is a bijection, there exists $ 1 \le j_1 \le n $ such that $ \mathcal{S}_j = c \mathcal{S}_{j_1} $, where $ c = \mathcal{S}_j[1] $. Then, $ P[2, |P|] = (\mathcal{S}_j[1, |P|])[2, |P|] = \mathcal{S}_j[2, |P|] = (c \mathcal{S}_{j_1})[2, |P|] = \mathcal{S}_{j_1}[1, |P| - 1] $, so $ P[2, |P|] \in \mathcal{P} $.
\end{proof}

Notice that, even if the trie of the $ \mathcal{S}_j $'s is infinite, the suffix tree in Figure \ref{subfig:suffixtreeexample} is finite. We aim to show that $ \Suff (\mathcal{T}) $ is \emph{always} a finite tree. To this end, it will be expedient to characterize the nodes of $ \Suff (\mathcal{T}) $. Once again, to guide our intuition, we can rely on the graph $ G_\mathcal{T} $. Let $ Y_P $ be the set of all $ 1 \le j \le n' $ such that in $ G_\mathcal{T} $ there is an occurrence of $ P $ starting from node $ D_j $. Formally, in our graph-free setting:

\begin{definition}
    Let $ \mathcal{T} $ be a dictionary and let $ P \in \Sigma^* $. Define:
    \begin{equation*}
        Y_P = \{1 \le j \le n' \;|\; \mathcal{S}_j[1, |P|] = P \}.
    \end{equation*}
\end{definition}

Note that for every $ P \in \Sigma^* $ we have $ Y_P \not = \emptyset $ if and only if $ P \in \mathcal{P} $ (recall that $ \mathcal{P} = \{\mathcal{S}_j[1, t] \;|\; 1 \le j \le n', t \ge 0 \} $). We also have the following lemma.

\begin{lemma}\label{lem:alphabetatleasttwo}
    Let $ \mathcal{T} $ be a dictionary, with $ \sigma = |\Sigma| $.
    \begin{itemize}
        \item If $ \sigma \ge 2 $, then for every $ P \in \mathcal{P} $ we have $ Y_P = \{1, 2, \dots, n' \} $ if and only if $ P = \epsilon $.
        \item  If $ \sigma = 1 $, then $ \mathcal{P} = \Sigma^* $, $ n' = 1 $ and $ Y_P = \{1 \} $ for every $ P \in \mathcal{P} $.
    \end{itemize}
\end{lemma}

\begin{proof}
    \begin{itemize}
        \item Note that $ \mathcal{S}_j[1, 0] = \epsilon $ for every $ 1 \le j \le n' $, so $ Y_\epsilon = \{1, 2, \dots, n' \} $. Now, let $ P \in \mathcal{P} $ such that $ |P| \ge 1 $, and let $ c = P[1] $. Since $ \sigma \ge 2 $, there exists $ c' \in \Sigma \setminus \{c \} $, and there exists $ 1 \le j \le n' $ such that $ \mathcal{S}_j[1] = c' $ because the alphabet is effective. Then, $ j \not \in Y_P $, so $ Y_P \not = \{1, 2, \dots, n' \} $.
        \item Let $ \Sigma = \{c \} $. Since the alphabet is effective, we have $ \mathcal{S}_j = c^\omega $ for every $ 1 \le j \le n' $. The $ \mathcal{S}_j $'s are pairwise distinct, so $ n' = 1 $. Lastly, if $ P \in \Sigma^* $, then $ P = c^{|P|} $, so $ \mathcal{S}_1[1, |P|] = c^{|P|} = P $ and $ P \in \mathcal{P} $, with $ Y_P = \{1 \} $.
    \end{itemize}
\end{proof}

Let $ P \in \Sigma $ and let $ c \in \Sigma $. Since $ Y_P $ is the set of all $ 1 \le j \le n' $ such that in $ G_\mathcal{T} $ there is an occurrence of $ P $ starting from node $ D_j $, and $ Y_{cP} $ is the set of all $ 1 \le j' \le n' $ such that in $ G_\mathcal{T} $ there is an occurrence of $ cP $ starting from node $ D_{j'} $, then $ Y_{cP} = \back (Y_P, c) $. Formally:

\begin{lemma}\label{lem:YcP}
    Let $ \mathcal{T} $ be a dictionary, let $ P \in \Sigma^* $ and let $ c \in \Sigma $. Then, $ Y_{cP} = \back (Y_P, c) $.
\end{lemma}

\begin{proof}
Fix any $ 1 \le j' \le n $, and let $ 1 \le j \le n $ be such that $ \mathcal{S}_{j'} = \mu (j) \mathcal{S}_j $. Then, we have $ j' \in Y_{cP} $ if and only if $ \mathcal{S}_{j'}[1, |cP|] = cP $, if and only if $ \mu(j) = c $ and $ \mathcal{S}_j[1, |P|] = P $, if and only if $ j \in Y_P $ and $ \mathcal{S}_{j'} = c \mathcal{S}_j $, if and only if $ j' \in \back (Y_P, c) $.
\end{proof}

We immediately obtain the following corollaries.

\begin{corollary}\label{cor:Pcontainment}
    Let $ \mathcal{T} $ be a dictionary, let $ P_1, P_2 \in \Sigma^* $ and let $ c \in \Sigma $. If $ Y_{P_1} \subseteq Y_{P_2} $, then $ Y_{cP_1} \subseteq Y_{cP_2} $. 
\end{corollary}

\begin{proof}
    Since $ Y_{P_1} \subseteq Y_{P_2} $, from Lemma \ref{lem:YcP} we obtain $ Y_{cP_1} = \back (Y_{P_1}, c) \subseteq \back (Y_{P_2}, c) = Y_{cP_2} $.
\end{proof}

\begin{corollary}\label{cor:ypconvex}
    Let $ \mathcal{T} $ be a dictionary and let $ P \in \Sigma^* $. Then, $ Y_P $ is convex.
\end{corollary}

\begin{proof}
    We proceed by induction on $ |P| $. If $ |P| = 0 $, then $ P $ is the empty string $ \epsilon $ and we have $ Y_P = Y_\epsilon = \{1, 2, \dots, n' \} $, which is convex. Now assume that $ |P| \ge 1 $. We can write $ P = cP' $, where $ c \in \Sigma $ and $ P' \in \Sigma^* $. By the inductive hypothesis, we know that $ Y_{P'} $ is convex, and from Lemma \ref{lem:convex} and Lemma \ref{lem:YcP} we conclude that $ Y_P = \back (Y_{P'}, c) $ is convex.
\end{proof}

For every $ 1 \le \ell \le r \le n' $, let $ [\ell, r] = \{\ell, \ell + 1, \dots, r - 1, r \} $. We can then give the following definition.

\begin{definition}
    Let $ \mathcal{T} $ be a dictionary and let $ P \in \mathcal{P} $. Let $ 1 \le \ell_P \le r_P \le n' $ be such that $ Y_P = [\ell_P, r_P] $-
\end{definition}

Note that $ \ell_P $ and $ r_P $ are well defined because (1) $ Y_P $ is convex by Corollary \ref{cor:ypconvex} and (2) $ Y_P \not = \emptyset $ if (and only if) $ P \in \mathcal{P} $.  From Lemma \ref{lem:YcP} and Theorem \ref{theor:FMindex} we immediately obtain the following corollary.

\begin{corollary}\label{cor:bacward-cP}
    Let $ \mathcal{T} $ be a dictionary, let $ P \in \mathcal{P} $ and let $ c \in \Sigma $. Given $ \ell_P $ and $ r_P $, in $ O(\log \log \sigma) $ time we can decide whether $ cP  \in \mathcal{P} $ and, if $ cP  \in \mathcal{P} $, we can also return $ \ell_{cP} $ and $ r_{cP} $.
\end{corollary}

\begin{proof}
    Since $ P \in \mathcal{P} $, then $ \ell_P $ and $ r_P $ are well defined, and $ Y_P = \{\ell_P, \ell_P + 1, \dots, r_P - 1, r_P \} $. Analogously, if $ cP  \in \mathcal{P} $, then $ \ell_{cP} $ and $ r_{cP} $ are well defined, and $ Y_{cP} = \{\ell_{cP}, \ell_{cP} + 1, \dots, r_{cP} - 1, r_{cP} \} $. By Lemma \ref{lem:YcP}, we have $ Y_{cP} = \back (Y_P, c) $, so the conclusion follows from Theorem \ref{theor:FMindex} by computing $ \bws(\ell_P, r_P, c) $.
\end{proof}

Recall that $ \Suff (\mathcal{T}) $ is defined starting from the trie of the $ \mathcal{S}_j $'s. Let $ P \in \mathcal{P} $, and let $ u $ be the node reached by reading $ P $ starting from the root. Then, the number of children of $ u $ in the trie is $ |\{c \in \Sigma \;|\; Pc \in \mathcal{P} \}| $. Note that $ u $ has at least one child: since $ P \in \mathcal{P} $, there exists $ 1 \le j \le n' $ such that $ \mathcal{S}_j[1, |P|] = P $, so if $ c = \mathcal{S}_j[|P| + 1] $, we have $ \mathcal{S}_j[1, |Pc|] = Pc $ and $ Pc \in \mathcal{P} $. For every $ j \in Y_P $ there exists exactly one $ c $ such that $ j \in Y_{Pc} $, so $ Y_P = \bigcup_{c \in \Sigma, Pc \in \mathcal{P}} Y_{Pc} $ and the union is disjoint.

For every node $ u $ in the trie, let $ P_u $ the string in $ \mathcal{P} $ that can be read from the root to $ u $. Then:
\begin{itemize}
    \item If node $ u $ is the parent of node $ v $, then there exists $ c \in \Sigma $ such that $ P_v = P_u c $, so from the previous discussion we obtain that $ Y_{P_v} \subseteq Y_{P_u} $ and $ v $ is the unique child of $ u $ if and only if  $ Y_{P_v} = Y_{P_u} $.
    \item If node $ u $ is an ancestor of node $ v $, then $ Y_{P_v} \subseteq Y_{P_u} $, and there is a unary path from $ u $ to $ v $ if and only if $ Y_{P_v} = Y_{P_u} $. This follows inductively from the previous point.
    \item If two nodes $ u $ and $ v $ of the trie are not one an ancestor of the other, then $ Y_{P_u} \cap Y_{P_v} = \emptyset $. Indeed, let $ w $ be the lowest common ancestor of $ u $ and $ v $, let $ u' $ be the child of $ w $ on the path from $ w $ to $ u $ and let $ v' $ be the child of $ w $ on the path from $ w $ to $ v $. Then $ u' \not = v' $ and there exist distinct $ c, d \in \Sigma $ such that $ P_{u'} = P_{w} c $ and $ P_{v'} = P_{w} d $, so from the previous discussion we have $ Y_{P_{w} c} \cap Y_{P_{w} d} = \emptyset $, or equivalently, $ Y_{P_{u'}} \cap Y_{P_{v'}} $. Since $ u' $ is an ancestor of $ u $, we have $ Y_{P_u} \subseteq Y_{P_{u'}} $, and analogously $ Y_{P_v} \subseteq Y_{P_{v'}} $, so we conclude $ Y_{P_u} \cap Y_{P_v} = \emptyset $.
\end{itemize}
Consequently, when we build $ \Suff (\mathcal{T}) $ from the trie, we collapse exactly all nodes $ u, v $ in the trie such that $ Y_{P_u} = Y_{P_v} $. This means that we can unambiguously associate to each node $ u $ of $ \Suff(\mathcal{T}) $ a set $ [\ell, r] $ such that, for every $ P \in \mathcal{P} $, we have $ [\ell, r] = [\ell_P, r_P] $ if and only if $ P $ can be read on $ \Suff (\mathcal{T}) $ starting from the root and ending at $ u $, where we assume that we can stop after reading any nonempty prefix of the last edge label. For example, in Figure \ref{subfig:suffixtreeexample} we have $ [\ell_P, r_P] = [3, 4] $ if and only if $ P \in \{bca, bcab, bcabc \} $.

In the following, we will identify a node $ u $ and the corresponding $ [\ell, r] $, as we do in Figure \ref{subfig:suffixtreeexample}. We can then give the following definition.

\begin{definition}
    Let $ \mathcal{T} $ be a dictionary.
    \begin{itemize}
        \item The set $ \mathcal{N} $ of all nodes of $ \Suff (\mathcal{T}) $ is $ \mathcal{N} = \{[\ell_P, r_P] \;|\; P \in \mathcal{P} \} $.
        \item For every $ [\ell, r] \in \mathcal{N} $, let $ \Str([\ell, r]) = \{|P| \;|\; P \in \mathcal{P}, [\ell_P, r_P] = [\ell, r] \} $.
    \end{itemize}
\end{definition}

For example, in Figure \ref{subfig:suffixtreeexample} we have $ \Str([3, 4]) = \{3, 4, 5 \} $. It is now clear that $ \Suff(\mathcal{T}) $ is finite because the set $ \mathcal{N} $ must be finite. We can prove a stronger result: $ \Suff(\mathcal{T}) $ contains exactly $ n' $ leaves and at most $ 2n' - 1 $ nodes in total. This follows from the following lemma, where we also describe some additional properties of $ \Suff (\mathcal{T}) $ that can be easily deduced by the properties of the trie of the $ \mathcal{S}_j $'s.

\begin{lemma}\label{lem:propertiesnodessuffixtree}
    Let $ \mathcal{T} $ be a dictionary.
    \begin{enumerate}
        \item The root of $ \Suff(\mathcal{T}) $ is $ [1, n'] $.\label{item:propertiesnodessuffixtree1}
        \item Every internal node of $ \Suff(\mathcal{T}) $ contains at least two children.\label{item:propertiesnodessuffixtree2}
        \item Let $ [\ell, r] \in \mathcal{N} $ be an internal node of $ \Suff(\mathcal{T}) $ with $ t $ children, and let $ [\ell_i, r_i] $ be the $ i $-th child of $ [\ell, r] $, for $ 1 \le i \le t $. Then (i) $ \ell_1 = \ell $, (2) $ \ell_i = r_{i - 1} + 1 $ for $ 2 \le i \le t $ and (iii) $ r_t = r $.\label{item:propertiesnodessuffixtree3}
        \item Let $ [\ell, r], [\ell', r'] \in \mathcal{N} $. If $ [\ell, r] $ is a proper ancestor of $ [\ell', r'] $ in $ \Suff(\mathcal{T}) $, then $ [\ell', r'] \subsetneqq [\ell, r] $.\label{item:propertiesnodessuffixtree4}
        \item Let $ [\ell, r], [\ell', r'] \in \mathcal{N} $. If $ [\ell, r] $ is not an ancestor of $ [\ell', r'] $ in $ \Suff(\mathcal{T}) $ and $ [\ell', r'] $ is not an ancestor of $ [\ell, r] $ in $ \Suff (\mathcal{T}) $, then $ [\ell, r] $ and $ [\ell', r'] $ are disjoint.\label{item:propertiesnodessuffixtree5}
        \item $ \Suff(\mathcal{T}) $ contains exactly $ n' $ leaves, namely, $ [j, j] $ for every $ 1 \le j \le n' $.\label{item:propertiesnodessuffixtree6}
        \item $ |\mathcal{N}| \le 2n' - 1 $.   \label{item:propertiesnodessuffixtree7}
        \item For every $ [\ell, r] \in \mathcal{N} $, the leftmost leaf of $ [\ell, r] $ in $ \Suff(\mathcal{T}) $ is $ [\ell, \ell] $ and the righmost leaf of $ [\ell, r] $ in $ \Suff (\mathcal{T}) $ is $ [r, r] $, and in particular $ [\ell, r] $ is the lowest common ancestor of $ [\ell, \ell] $ and $ [r, r] $ in $ \Suff (\mathcal{T}) $. \label{item:propertiesnodessuffixtree8}
    \end{enumerate}
\end{lemma}

\begin{proof}
    \begin{enumerate}
        \item The root is $ [\ell_\epsilon, r_\epsilon] = [1, n'] $.
        \item It is sufficient to recall that $ \Suff (\mathcal{T}) $ is obtained by compressing the unary paths in the trie of the $ \mathcal{S}_j $'s.
        \item Since $ [\ell, r] $ is an \emph{internal} node, then node $ [\ell, r] $ is obtained by compressing a \emph{finite} unary path in the trie of the $ \mathcal{S}_j $'s. Let $ u $ be the deepest node in this finite unary path, and let $ P = P_u $. Let $ c_1 \prec c_2 \prec \dots \prec c_t $ all characters in $ \Sigma $ such that $ Pc_i \in \mathcal{P} $ for every $ 1 \le i \le t $. Then we know that $ [\ell, r] $ has exactly $ t $ children, and for $ 1 \le i \le t $ the $ i $-th child of $ [\ell, r] $ is $ [\ell_{Pc_i}, r_{Pc_i}] $. Moreover, from the previous discussion, we know that $ [\ell, r] $ is the disjoint union of the $ [\ell_{Pc_i}, r_{Pc_i}] $'s. Lastly, for $ 1 \le i \le t - 1 $ we have $ \mathcal{S}_{\ell_{Pc_i}}[1, Pc_i] = Pc_i \prec Pc_{i + 1} = \mathcal{S}_{\ell_{Pc_{i + 1}}}[1, Pc_{i + 1}] $, so $ \ell_{Pc_i} < \ell_{Pc_{i + 1}} $ for every $ 1 \le i \le t - 1 $. The conclusion follows.
        \item From the previous point we know that if $ [\ell, r] $ is the parent of $ [\ell', r'] $, then $ [\ell', r'] \subseteq [\ell, r] $, and the containment must be strict because $ [\ell, r] $ has at least two children by the second point. The conclusion follows by arguing recursively.

        \item Let $ [\ell^*, r^*] $ be the lowest common ancestor of $ [\ell, r] $ and $ [\ell', r] $ in $ \Suff (\mathcal{T}) $. Let $ [\ell_1, r_1] $ be the child of $ [\ell^*, r^*] $ on the path from $ [\ell^*, r^*] $ to $ [\ell, r] $, and let $ [\ell'_1, r'_1] $ be the child of $ [\ell^*, r^*] $ on the path from $ [\ell^*, r^*] $ to $ [\ell', r'] $. Then, $ [\ell_1, r_1] $ and $ [\ell'_1, r'_1] $ are disjoint because they are distinct children of $ [\ell^*, r^*] $. This implies that $ [\ell, r] $ and $ [\ell', r'] $ are disjoint because by the previous point we have $ [\ell, r] \subseteq [\ell_1, r_1] $ and $ [\ell', r'] \subseteq [\ell'_1, r'_1] $.        
        \item We only have to prove that $ [j, j] \in \mathcal{N} $ for every $ 1 \le j \le n' $, because then from points 2 and 3 we obtain that each $ [j, j] $ is not an internal node (so it must be a leaf) and from point 5 we obtain that every node is an ancestor of some $ [j, j] $ (because a node $ [\ell, r] $ is not disjoint from the leaf $ [\ell, \ell] $).
    
        Fix $ 1 \le j \le n' $. Let us prove that $ [j, j] \in \mathcal{N} $. Consider the string $ P = \mathcal{S}_j[1, n' - 1] $. We have $ P \in \mathcal{P} $ and $ j \in [\ell_P, r_P] $. It will suffice to show that $ [\ell_P, r_P] = [j, j] $. To this end, we only have to show that (i) if $ j \ge 2 $, then $ j - 1 \not \in [\ell_P, r_P]  $, and (ii) if $ j \le n' - 1 $, then $ j + 1 \not \in [\ell_P, r_P] $. We only prove statement (i) because the proof of statement (ii) is analogous. We have to show that $ P \not = \mathcal{S}_{j - 1}[1, n' - 1] $, or equivalently $ \mathcal{S}_{j - 1}[1, n' - 1] \not = \mathcal{S}_j[1, n' - 1] $. This follows from Lemma \ref{lem:upperbound} because $ \lcp(\mathcal{S}_{j - 1}, \mathcal{S}_j) = \LCP[j] \le n' - 2 $.

        \item  By point 6, we know that $ \Suff(\mathcal{T}) $ contains exactly $ n' $ leaves and by point 2 every internal node contains at least two children, so the conclusion follows. 

        \item First, notice that $ [\ell, \ell] $ and $ [r, r] $ are leaves by point 6. Then, we obtain that $ [\ell, \ell] $ is the leftmost leaf and $ [r, r] $ is the rightmost leaf by applying point 3 inductively.
    \end{enumerate}
\end{proof}

For every $ 1 \le \ell \le r \le n' $, let $ \lambda_{\ell, r} = \lcp (\mathcal{S}_\ell, \mathcal{S}_r) $. The following lemma highlights some properties of $ \Str([\ell, r]) $. In particular, we show that, if $ [\ell, r] $ is an internal node, then the longest string reaching node $ [\ell, r] $ has length $ \lambda_{\ell, r} $.

\begin{lemma}\label{lem:nodesSuff}
    Let $ \mathcal{T} $ be a dictionary.
    \begin{enumerate}
        \item Let $ P_1, P_2 \in \mathcal{P} $. If $ P_1 $ is a prefix of $ P_2 $, then $ [\ell_{P_1}, r_{P_1}] $ is an ancestor of $ [\ell_{P_2}, r_{P_2}] $ in $ \Suff (\mathcal{T}) $. In particular, if $ |P_2| = |P_1| + 1 $, then either $ [\ell_{P_1}, r_{P_1}] = [\ell_{P_2}, r_{P_2}] $ or $ [\ell_{P_1}, r_{P_1}] $ is the parent of $ [\ell_{P_2}, r_{P_2}] $ in $ \Suff (\mathcal{T}) $. \label{item:nodesSuff1}
        \item Let $ P_1, P_2 \in \mathcal{P} $. If $ [\ell_{P_1}, r_{P_1}] $ is a strict ancestor of $ [\ell_{P_2}, r_{P_2}] $ in $ \Suff (\mathcal{T}) $, then $ P_1 $ is a strict prefix of $ P_2 $.\label{item:nodesSuff2}
        \item $ \Str([\ell, r]) $ is convex for every $ [\ell, r] \in \mathcal{N} $, and for every $ h \in \Str([\ell, r]) $ there exists exactly one $ P \in \mathcal{P} $ such that $ |P| = h $ and $ [\ell_P, r_P] = [\ell, r] $.\label{item:nodesSuff3}
        \item For every $ [\ell, r] \in \mathcal{N} $, If $ \ell \ge 2 $, then $ \LCP[\ell] < \lambda_{\ell, r} $, and if $ r \le n' - 1 $, then $ \LCP[r + 1] < \lambda_{\ell, r} $.\label{item:nodesSuff4}
        \item For every $ [\ell, r] \in \mathcal{N} $, $ \Str([\ell, r]) $ is finite if and only if $ \ell < r $, and in this case the largest element of $ \Str([\ell, r]) $ is $ \lambda_{\ell, r} $.\label{item:nodesSuff5}
    \end{enumerate}
\end{lemma}

\begin{proof}
    The first three points follow immediately because $ \Suff(\mathcal{T}) $ is defined by compressing all unary paths in the trie of the $ \mathcal{S}_j $'s.

    Let us prove point 4. We only prove that if $ \ell \ge 2 $, then $ \LCP[\ell] < \lambda_{\ell, r} $ (the proof of the other statement is analogous). Fix any $ P \in \Sigma^* $ such that $ [\ell_P, r_P] = [\ell, r] $. We know that $ P = \mathcal{S}_\ell [1, |P|] = \mathcal{S}_r [1, |P|] $, so $ \lambda_{\ell, r} = \lcp (\mathcal{S}_\ell, \mathcal{S}_r) \ge |P| $. Since $ \ell - 1 \not \in [\ell, r] $, then $ P \not = \mathcal{S}_{\ell - 1}[1, |P|] $, or equivalently, $ \mathcal{S}_\ell [1, |P|] \not = \mathcal{S}_{\ell - 1}[1, |P|] $, which implies $ |P| > \lcp(\mathcal{S}_{\ell - 1}, \mathcal{S}_\ell) = \LCP[\ell] $. We conclude $ \LCP[\ell] < |P| \le \lambda_{\ell, r} $.
    
    Let us prove point 5. Notice that in the trie of the $ \mathcal{S}_j $'s every node is an ancestor of infinitely many nodes. As a consequence, $ \Str([\ell, r]) $ is finite is and only if $ [\ell, r] $ is not a leaf of $ \Suff(\mathcal{T}) $. By Lemma \hyperref[item:propertiesnodessuffixtree6]{\ref*{lem:propertiesnodessuffixtree} (\ref*{item:propertiesnodessuffixtree6})} we conclude that $ \Str([\ell, r]) $ is finite if and only if $ \ell < r $. Now assume that $ \ell < r $. We want to prove that the largest element of $ \Str([\ell, r]) $ is $ \lambda_{\ell, r} $. To this end, it will suffice to show that $ [\ell_{\mathcal{S}_\ell [1, \lambda_{\ell, r}]}, r_{\mathcal{S}_\ell [1, \lambda_{\ell, r}]}] = [\ell, r] $ but $ [\ell_{\mathcal{S}_\ell [1, \lambda_{\ell, r} + 1]}, r_{\mathcal{S}_\ell [1, \lambda_{\ell, r} + 1]}] \not = [\ell, r] $.
    
    Let us prove that $ [\ell_{\mathcal{S}_\ell [1, \lambda_{\ell, r}]}, r_{\mathcal{S}_\ell [1, \lambda_{\ell, r}]}] = [\ell, r] $. Fix any $ 1 \le j \le n' $. We know that $ [\ell_{\mathcal{S}_\ell [1, \lambda_{\ell, r}]}, r_{\mathcal{S}_\ell [1, \lambda_{\ell, r}]}] $ is convex, so it will suffice to show that (i) $ \mathcal{S}_\ell [1, \lambda_{\ell, r}] = \mathcal{S}_r [1, \lambda_{\ell, r}] $, (ii) if $ \ell \ge 2 $, then $ \mathcal{S}_{\ell - 1} [1, \lambda_{\ell, r}] \not = \mathcal{S}_\ell [1, \lambda_{\ell, r}] $ and (iii) if $ r \le n' - 1 $, then $ \mathcal{S}_r [1, \lambda_{\ell, r}] \not = \mathcal{S}_{r + 1} [1, \lambda_{\ell, r}] $. Notice that statement (i) is true because $ \lambda_{\ell, r} = \lcp (\mathcal{S}_\ell, \mathcal{S}_r) $, statement (ii) is true because $ \LCP[\ell] < \lambda_{\ell, r} $ by point 4, and statement (iii) is true because $ \LCP[r + 1] < \lambda_{\ell, r} $ again by point 4.

    Let us prove that $ [\ell_{\mathcal{S}_\ell [1, \lambda_{\ell, r} + 1]}, r_{\mathcal{S}_\ell [1, \lambda_{\ell, r} + 1]}] \not = [\ell, r] $. It will suffice to show that $ \mathcal{S}_\ell [1, \lambda_{\ell, r} + 1] \not = \mathcal{S}_r [1, \lambda_{\ell, r} + 1] $. This follows from $ \lambda_{\ell, r} = \lcp (\mathcal{S}_\ell, \mathcal{S}_r) $.
\end{proof}

In view of Lemma \hyperref[item:nodesSuff5]{\ref*{lem:nodesSuff} (\ref*{item:nodesSuff5})}, let us show how to compute $ \lambda_{\ell, r} $ quickly.

\begin{corollary}\label{cor:computingelllr}
    Let $ \mathcal{T} $ be a dictionary. Given $ 1 \le \ell \le r \le n' $, we can compute $ \lambda_{\ell, r} $ in $ O(\log n) $ time.
\end{corollary}

\begin{proof}
    If $ \ell = r $, we have $ \lambda_{\ell, r} = \lcp (\mathcal{S}_\ell, \mathcal{S}_r) = + \infty $. Now assume $ \ell < r $. We have $ \lambda_{\ell, r} = \lcp (\mathcal{S}_\ell, \mathcal{S}_r) = \min_{\ell + 1 \le j \le r} \LCP[j] $. In $ O(1) $ time we can compute $ j^* = \rmq_{\LCP}(\ell + 1, r) $ (see Section \ref{sec:lcp}), and we have $ \lambda_{\ell, r} = \LCP[j^*] $. The conclusion follows from Theorem \ref{theor:lcpsampling}.
\end{proof}

Storing $ \Suff(\mathcal{T}) $ explicitly would require more than $ O(n) $ bits so, as in Sadakane's compressed suffix tree, we will only store its topology by using standard techniques (see \cite{navarro2016book} for an introduction). Given an ordinal tree $ T $ with $ t $ nodes, the \emph{balanced parenthesis} representation of $ T $ is the bit array $ Z_T[1, 2t] $ that we can build incrementally as follows. Visit all nodes of the tree in depth-first search order (respecting the order of the children of each node), starting from the root. Every time we encounter a new node $ u $ we append an $ 1 $, and as soon as we leave the subtree rooted at $ u $ we add a $ 0 $. Hence $ Z_T[1, 2t] $ is a bit array containing exactly $ t $ values equal to $ 1 $ and $ t $ values equal to $ 0 $ (see Figure \ref{subfig:tabletrees} for $ Z_{\Suff (\mathcal{T})} $). It is easy to see that $ Z_T $ is an encoding of the ordinal tree $ T $. This encoding is called the balanced representation of $ T $ because, if we interpret each $ 1 $ as an opening parenthesis `(' and each $ 0 $ as a closing parenthesis`)', then it is possible to capture the subtree-containment hierarchy of the nodes, which supports quick navigation of the tree (as we will see in Theorem \ref{theor:balancedparenthesisnavarrosadakane}).

If $ u $ is a node of $ T $, we say that the \emph{BP-index} of $ u $ is the unique integer $ 1 \le i \le 2t $ corresponding to the first time that we encounter $ u $ in the depth-first search order (in particular, $ Z_T[i] = 1 $). Equivalently, the BP-index of $ u $ is $ \select_1(Z_T[1, 2t], q) $, where $ q $ is such that $ u $ is the $ q $-th node in the depth-first traversal of $ T $. For example, in Figure \ref{subfig:suffixtreeexample} the BP-index of the root is $ 1 $, the BP-index of node $ [1, 2] $ is $ 2 $, the BP-index of node $ [1, 1] $ is $ 3 $, the BP-index of node $ [2, 2] $ is $ 5 $, and so on. If $ u $ is a node of $ T $, we will also refer to the \emph{BP$^\#$-index} of $ u $, namely, the unique integer $ 1 \le i \le 2t $ corresponding to when we leave the subtree rooted at $ u $ (in particular, $ Z_T[i] = 0 $). For example, in Figure \ref{subfig:suffixtreeexample} the BP$^\#$-index of the node $ [1, 2] $ is $ 7 $. Then, we have the following result \cite{navarrosadakane2014trees}.

\begin{theorem}\label{theor:balancedparenthesisnavarrosadakane}
    Let $ T $ be an ordinal tree with $ t $ nodes. There exists a data structure of $ 2t + o(t) $ bits that supports the following operations in $ O(1) $ time:
    \begin{itemize}
        \item $ \access (Z_T, i) $, for $ 1 \le i \le 2t $.
        \item $ \rank_c (Z_T, i) $, for $ c \in \{0, 1 \} $, $ 1 \le i \le 2t $.
        \item $ \select_c (Z_T, i) $, for $ c \in \{0, 1 \} $ and $ 1 \le i \le \rank_c (Z_T, 2t) $.
        \item $ \open(T, i) $: given the BP$^\#$-index $ i $ of a node $ u $, return the BP-index of $ u $.
        \item $ \parent(T, i) $: given the BP-index $ i $ of a node $ u $, return the BP-index of the parent of $ u $ (if $ u $ is the root, then the query returns $ \perp $).
        \item $ \lca(T, i, i') $: given the BP-indexes $ i $ and $ i' $ of two nodes $ u $ and $ v $, return the BP-index of the lowest common ancestor of $ u $ and $ v $.
        \item $ \leftmost(T, i) $: given the BP-index $ i $ of a node $ u $, return the BP-index of the leftmost leaf of $ u $.
        \item $ \rightmost(T, i) $: given the BP-index $ i $ of a node $ u $, return the BP-index of the rightmost leaf of $ u $.
    \end{itemize}
\end{theorem}

We store the data structure of Theorem \ref{theor:balancedparenthesisnavarrosadakane} for the ordinal tree $ \Suff(\mathcal{T}) $. By Lemma \hyperref[item:propertiesnodessuffixtree7]{\ref*{lem:propertiesnodessuffixtree} (\ref*{item:propertiesnodessuffixtree7})} we know that $ \Suff(\mathcal{T}) $ has at most $ 2n' - 1 $ nodes, so our data structure requires $ O(n) $ bits.

Each node $ [\ell, r] $ of $ \Suff(\mathcal{T}) $ is uniquely determined by $ \ell $ and $ r $, but the data structure of Theorem \ref{theor:balancedparenthesisnavarrosadakane} requires the BP-index $ i $ of $ [\ell, r] $ to support the navigation in the suffix tree. Consequently, we need to quickly switch from one representation of the nodes to the other one.  We store a bitvector $ B_7 $ of length $ 2|\mathcal{N}| $ such that for every $ 1 \le t \le 2|\mathcal{N}| $ we have $ B_7 [t] = 1 $ if and only if $ t $ is the BP-index of a leaf in $ \Suff(\mathcal{T}) $, see Figure \ref{subfig:tabletrees}. Note that for every $ 1 \le t \le 2|\mathcal{N}| $, if $ B_7 [t] = 1 $, then $ Z_{\Suff (\mathcal{T})}[t] = 1 $. We can now show how to switch from one representation to the other.

\begin{lemma}\label{lem:fromintervaltobpindex}
    Let $ \mathcal{T} $ be a dictionary. In $ O(1) $ time we can solve the following queries:
    \begin{itemize}
        \item $ \frominter(\ell, r) $: given $ 1 \le \ell \le r \le n' $ such that $ [\ell, r] \in \mathcal{N} $, return the BP-index of $ [\ell, r] $ in $ \Suff(\mathcal{T}) $.
        \item $ \tointer(i) $: given the BP-index $ i $ of a node $ [\ell, r] \in \mathcal{N} $, return $ \ell $ and $ r $.
    \end{itemize}
\end{lemma}

\begin{proof}
    First, notice that from Lemma \hyperref[item:propertiesnodessuffixtree6]{\ref*{lem:propertiesnodessuffixtree} (\ref*{item:propertiesnodessuffixtree6})} we know that $ \{[j, j] \;|\; 1 \le j \le n' \} $ is the set of all leaves, and by applying Lemma \hyperref[item:propertiesnodessuffixtree3]{\ref*{lem:propertiesnodessuffixtree} (\ref*{item:propertiesnodessuffixtree3})} recursively we obtain that, if $ 1 \le j < j' \le n' $, then the BP-index of $ [j, j] $ is smaller than the BP-index of $ [j', j'] $. As a consequence, for every $ 1 \le j \le n' $, if $ t $ is the BP-index of $ [j, j] $, then $ t = \select_1 (B_7, j) $ and $ j = \rank_1 (B_7, t) $. We can now prove the lemma.
    \begin{itemize}
        \item The BP-index $ t $ of $ [\ell, \ell] $ is $ \select_1 (B_7, \ell) $ and the BP-index $ t' $ of $ [r, r] $ is $ \select_1 (B_7, r) $. By Lemma \hyperref[item:propertiesnodessuffixtree8]{\ref*{lem:propertiesnodessuffixtree} (\ref*{item:propertiesnodessuffixtree8})}, we know that $ [\ell, r] $ is the lowest common ancestor of $ [\ell, \ell] $ and $ [r, r] $ in $ \Suff (\mathcal{T}) $, so by Theorem \ref{theor:balancedparenthesisnavarrosadakane} the BP-index of $ [\ell, r] $ is $ \lca(\Suff (\mathcal{T}), t, t') $.
        \item We only show how to compute $ \ell $ because $ r $ can be computed analogously. By Lemma \hyperref[item:propertiesnodessuffixtree8]{\ref*{lem:propertiesnodessuffixtree} (\ref*{item:propertiesnodessuffixtree8})} we know that $ [\ell, \ell] $ is the leftmost leaf of $ [\ell, r] $, so by Theorem \ref{theor:balancedparenthesisnavarrosadakane} the BP-index $ t $ of $ [\ell, \ell] $ is $ \leftmost(\Suff (\mathcal{T}), t) $ and we conclude $ \ell = \rank_1 (B_7, t) $.
    \end{itemize}
\end{proof}

\section{Proofs from Section \ref{sec:solvingcirculad}}

\noindent{\textbf{Statement of Lemma \ref{lem:climbingsuffixtree}}.}
    Let $ \mathcal{T} $ be a dictionary, let $ P \in \Sigma^* $, let $ 1 \le i \le m $ and let $ 1 \le j \le n' $.
    \begin{enumerate}
        \item Assume that $ j \in  [\ell_{P[i, t_i]}, r_{P[i, t_i]}] $. Then, for every $ k \in D_j $ we have that $ \mathcal{T}_k $ occurs in $ P $ at position $ i $ if and only if $ |\mathcal{T}_k| \le t_i - i + 1 $.
        \item Assume that $ j \not \in  [\ell_{P[i, t_i]}, r_{P[i, t_i]}] $. Let $ [\ell, r] $ be the nearest ancestor of $ [\ell_{P[i, t_i]}, r_{P[i, t_i]}] $ in $ \Suff (\mathcal{T}) $ for which $ j \not \in [\ell, r] $. Then, $ [\ell, r] $ is not the root of $ \Suff (\mathcal{T}) $. Moreover, let $ [\ell', r'] $ be the parent of $ [\ell, r] $ in $ \Suff (\mathcal{T}) $. Then, $ j \in [\ell', \ell - 1] \cup [r + 1, r'] $, and for every $ k \in D_j $ we have that $ \mathcal{T}_k $ occurs in $ P $ at position $ i $ if and only if $ |\mathcal{T}_k| \le \lambda_{\ell', r'} $.
    \end{enumerate}

\begin{proof}
        Let $ k \in D_j $. We have that $ \mathcal{T}_k $ occurs in $ P $ at position $ i $ if and only if $ |\mathcal{T}_k| \le m - i + 1 $ and $ \mathcal{T}_k = P[i, i + |\mathcal{T}_k| - 1] $. Moreover, we know that $ \mathcal{T}_k = \mathcal{S}_j[1, |\mathcal{T}_k|] $. Let us prove the theorem.
    \begin{enumerate}
        \item  Since $ j \in  [\ell_{P[i, t_i]}, r_{P[i, t_i]}] $, we have $ \mathcal{S}_j[1, t_i - i + 1] = P[i, t_i] $. 
        \begin{itemize}
            \item Assume that $ |\mathcal{T}_k| \le t_i - i + 1 $. Then, $ |\mathcal{T}_k| \le m - i - 1 $ because $ t_i \le m $, and $ \mathcal{T}_k = \mathcal{S}_j[1, |\mathcal{T}_k|] = P[i, i +  |\mathcal{T}_k| - 1] $, which proves that $ \mathcal{T}_k $ occurs in $ P $ at position $ i $.
            \item Assume that $ |\mathcal{T}_k| > t_i - i + 1 $. If $ |\mathcal{T}_k| > m - i + 1 $ we immediately conclude that $ \mathcal{T}_k $ does not occur in $ P $ at position $ i $, so we can assume $ t_i - i + 1 < |\mathcal{T}_k| \le m - i + 1 $, which implies $ t_i < m $. By the maximality of $ t_i $ we have $ P[i, t_i + 1] \not \in \mathcal{P} $, so $ \mathcal{S}_j[1, t_i - i + 2] \not = P[i, t_i + 1] $ (because $ \mathcal{S}_j[1, t_i - i + 2] \in \mathcal{P} $) and $ \mathcal{T}_k = \mathcal{S}_j[1, |\mathcal{T}_k|] \not = P[i, i +  |\mathcal{T}_k| - 1] $, which proves that $ \mathcal{T}_k $ does not occur in $ P $ at position $ i $.
        \end{itemize}
        \item  First, $ [\ell, r] $ is well defined because $ j \not \in  [\ell_{P[i, t_i]}, r_{P[i, t_i]}] $, and $ [\ell, r] $ is not the root of $ \Suff (\mathcal{T}) $ because $ j \not \in [\ell, r] $ and the root is $ [1, n'] $ (Lemma \hyperref[item:propertiesnodessuffixtree1]{\ref*{lem:propertiesnodessuffixtree} (\ref*{item:propertiesnodessuffixtree1})}). This implies that $ [\ell', r'] $ is well defined. Next, we have $ j \in [\ell', \ell - 1] \cup [r + 1, r'] $ because $ [\ell, r] $ is the nearest ancestor of $ [\ell_{P[i, t_i]}, r_{P[i, t_i]}] $ in $ \Suff (\mathcal{T}) $ for which $ j \not \in [\ell, r] $ and because $ [\ell, r] \subsetneqq [\ell', r'] $ by Lemma \hyperref[item:propertiesnodessuffixtree4]{\ref*{lem:propertiesnodessuffixtree} (\ref*{item:propertiesnodessuffixtree4})}, which also implies $ \ell' < r' $.

        Let $ P^* $ be the longest string in $ \mathcal{P} $ such that $ [\ell_{P^*}, r_{P^*}] = [\ell', r'] $ (which is well defined by Lemma \hyperref[item:nodesSuff3]{\ref*{lem:nodesSuff} (\ref*{item:nodesSuff3})} and Lemma \hyperref[item:nodesSuff5]{\ref*{lem:nodesSuff} (\ref*{item:nodesSuff5})}). Since $ [\ell', r'] $ is a strict ancestor of $ [\ell_{P[i, t_i]}, r_{P[i, t_i]}] $, then $ P^* $ is a strict prefix of $ P[i, t_i] $ by Lemma \hyperref[item:nodesSuff2]{\ref*{lem:nodesSuff} (\ref*{item:nodesSuff2})}. By Lemma \hyperref[item:nodesSuff5]{\ref*{lem:nodesSuff} (\ref*{item:nodesSuff5})}, we have $ |P^*| = \lambda_{\ell', r'} $, so $ P^* = P[i, i + \lambda_{\ell', r'} - 1] $ and $ \lambda_{\ell', r'} = |P^*| < |P[i, t_i]| = t_i - i + 1 $. Since $ j \in [\ell', r'] = [\ell_{P^*}, r_{P^*}] $, we have $ \mathcal{S}_j[1, |P^*|] = P^* $, or equivalently, $ \mathcal{S}_j[1, \lambda_{\ell', r'}] = P[i, i + \lambda_{\ell', r'} - 1] $.
        \begin{itemize}
            \item Assume that $ |\mathcal{T}_k| \le \lambda_{\ell', r'} $. In particular, $ |\mathcal{T}_k| \le \lambda_{\ell', r'} < t_i - i + 1 \le m - i + 1 $. Moreover, $ \mathcal{T}_k = \mathcal{S}_j[1, |\mathcal{T}_k|] = P[i, i +  |\mathcal{T}_k| - 1] $, which proves that $ \mathcal{T}_k $ occurs in $ P $ at position $ i $.
            \item Assume that $ |\mathcal{T}_k| > \lambda_{\ell', r'} $. If $ |\mathcal{T}_k| > m - i + 1 $ we immediately conclude that $ \mathcal{T}_k $ does not occur in $ P $ at position $ i $, so we can assume $ \lambda_{\ell', r'} < |\mathcal{T}_k| \le m - i + 1 $. Since $ P^* $ is a strict prefix of $ P[i, t_i] $, there exists $ c \in \Sigma $ such that $ P^*c $ is a prefix of $ P[i, t_i] $. Then, from Lemma \hyperref[item:nodesSuff1]{\ref*{lem:nodesSuff} (\ref*{item:nodesSuff1})} and the maximality of $ P^* $, we obtain $  [\ell_{P^*c}, r_{P^*c}] = [\ell, r] $. Since $ j \not \in [\ell, r] = [\ell_{P^*c}, r_{P^*c}] $, we have $ \mathcal{S}_j[1, |P^*c|] \not =  P^*c $, or equivalently, $ \mathcal{S}_j[1, \lambda_{\ell', r'} + 1] \not = P[i, i + \lambda_{\ell', r'}] $. We conclude $ \mathcal{T}_k = \mathcal{S}_j[1, |\mathcal{T}_k|] \not = P[i, i +  |\mathcal{T}_k| - 1] $, which proves that $ \mathcal{T}_k $ does not occur in $ P $ at position $ i $.
        \end{itemize}
    \end{enumerate}
\end{proof}

Let us present a simple property of the $ t_i $'s.

\begin{lemma}\label{lem:t_imonotone}
    Let $ \mathcal{T} $ be a dictionary and let $ P \in \Sigma^* $. We have $ t_i \le t_{i + 1} $ for every $ 1 \le i \le m - 1 $.
\end{lemma}

\begin{proof}
    By the definitions of $ t_i $ and $ t_{i + 1} $, we have $ t_i \ge i - 1 $ and $ t_{i + 1} \ge i $. If $ t_i \le i $ the conclusion is immediate, so we can assume $ t_i \ge i + 1 $. By the definition of $ t_i $ we have $ P[i, t_i] \in \mathcal{P} $, so by Lemma \ref{lem:Pprefixessuffixes} we have $ P[i + 1, t_i] \in \mathcal{P} $ and by the maximality of $ t^*_i $ we conclude $ t_i \le t^*_i $.
    \end{proof}

We can now show how to compute the $ t_i $'s, the $ \ell_{P[i, t_i]} $'s and the $ r_{P[i, t_i]} $'s efficiently. 

\begin{lemma}\label{lem:computing3mvalues}
    Let $ \mathcal{T} $ be a dictionary and let $ P \in \Sigma^* $. In $ O(m \log n) $ we can compute all the following $ 3m $ values: (i) $ t_i $ for $ 1 \le i \le m $, (ii) $ \ell_{P[i, t_i]} $ for $ 1 \le i \le m $, (iii) $ r_{P[i, t_i]} $ for $ 1 \le i \le m $.
\end{lemma}

\begin{proof}
    Let us give an informal description of our algorithm. We compute some values $ t^*_m $, $ t^*_{m - 1} $, $ t^*_{m - 2} $, $ \dots $, $ t^*_1 $ (in this order) such that $ t^*_i = t_i $ for $ 1 \le i \le m $. We start from $ P[m + 1, m] = \epsilon $, and we determine all $ 1 \le i \le m $ for which $ P[i, m] \in \mathcal{P} $ by using Corollary \ref{cor:bacward-cP}. This is exactly the set of all $ i $'s for which $ t_i = m $. Now consider the largest $ i $ for which $ P[i - 1, m] \not \in \mathcal{P} $ (and so $ P[i, m] \in \mathcal{P} $). Then, $ t_{i - 1} < m $, and $ t_{i - 1} $ is the largest integer for which $ P[i - 1, t_{i - 1}] \in \mathcal{P} $. For every prefix $ P[i, j] $ of $ P[i, m] $ we have $ [\ell_{P[i, m]}, r_{P[i, m]}] \subseteq [\ell_{P[i, j]}, r_{P[i, j]}] $, and there exists $ m' < m $ such that $ [\ell_{P[i, m]}, r_{P[i, m]}] \subsetneqq [\ell_{P[i, m']}, r_{P[i, m']}] $ but $ [\ell_{P[i, m]}, r_{P[i, m]}] = [\ell_{P[i, j]}, r_{P[i, j]}] $ for every $ m' + 1 \le j \le m $. In particular, $ P[i, m'] $ must be the longest string reaching the parent of node $ [\ell_{P[i, m]}, r_{P[i, m]}] $, so $ m' $ can be computed by using Lemma \hyperref[item:nodesSuff5]{\ref*{lem:nodesSuff} (\ref*{item:nodesSuff5})}. For every $ m' + 1 \le j \le m $ we have $ P[i - 1, j] \not \in \mathcal{P} $ because $ P[i - 1, m] \not \in \mathcal{P} $ and $ [\ell_{P[i, m]}, r_{P[i, m]}] = [\ell_{P[i, j]}, r_{P[i, j]}] $, so we must have $ t_{i - 1} \le m' $. Consequently, we can now consider $ P[i, m'] $, and we can continue in the same way: if $ P[i - 1, m'] \in \mathcal{P} $, we conclude $ t_{i - 1} = m' $ and we compute all $ 1 \le i' \le i - 1 $ for which $ t_{i'} = m' $ by using Corollary \ref{cor:bacward-cP}; otherwise, we determine a smaller $ m'' < m' $. The algorithm ends when we consider a string of the form $ P[1, j] $ and we can thus determine $ t_1 $.
    
    We can now give a formal description of our algorithm. At any time, the algorithm maintains the quadruple $ (i, j, \ell, r) $. We will maintain the invariants $ j \ge i - 1 $, $ P[i, j] \in \mathcal{P} $, $ \ell = \ell_{P[i, j]} $ and $ r = r_{P[i, j]} $. A step of the algorithm consists of updating $ (i, j, \ell, r) $ conveniently. We start from $ (m + 1, m, 1, n') $ (for which the invariants are true because $ P[m + 1, m] = \epsilon $), and the algorithm ends when we reach a quadruple $ (i, j, \ell, r) $ for which $ i = 1 $. 
    
    Assuming that at the beginning of a step we have $ (i, j, \ell, r) $, let us show how to compute the new quadruple. We compute $ \bws (\ell, r, P[i - 1]) $ in $ O(\log \log \sigma) $ time by using Theorem \ref{theor:FMindex}. Note by the invariants and Lemma \ref{lem:YcP} we have $ \back([\ell, r], P[i - 1]) = \back([\ell_{P[i, j]}, r_{P[i, j]}], P[i - 1]) = \back(Y_{P[i, j]}, P[i - 1]) = Y_{P[i - 1, j]} $. We distinguish two cases.
    \begin{itemize}
        \item Assume that $ \back([\ell, r], P[i - 1]) \not = \emptyset $, or equivalently, $ P[i - 1, j] \in \mathcal{P} $. Then, $ \bws (\ell, r, P[i - 1]) $ also returns $ 1 \le \ell' \le' r' \le n' $ such that $ \back([\ell, r], P[i - 1]) = [\ell', r'] $. We define $ t^*_{i - 1} = j $, and the new quadruple is $ (i - 1, j, \ell', r') $. Note that we maintain the invariants because $ j \ge i - 1 $ implies $ j \ge (i - 1) - 1 $, and $ \ell = \ell_{P[i, j]} $, $ r = r_{P[i, j]} $ and $ \back([\ell, r], P[i - 1]) \not = \emptyset $ imply $ P[i - 1, j] \in \mathcal{P} $, $ \ell' = \ell_{P[i - 1, j]} $ and $ r' = r_{P[i - 1, j]} $.
        \item Assume that $ \back([\ell, r], P[i - 1]) = \emptyset $, or equivalently, $ P[i - 1, j] \not \in \mathcal{P} $. By the invariant, we have $ j \ge i - 1 $. We distinguish two subcases.
        \begin{itemize}
            \item Assume that $ j = i - 1 $. Then, we define $ t^*_{i - 1} = i - 2 $ and the new quadruple is $ (i - 1, i - 2, 1, n') $. We maintain the invariant because $ i - 2 \ge (i - 1) - 1 $ and $ P[i - 1, i -2] = \epsilon $.
            \item Assume that $ j > i - 1 $. We know that $ P[i, j] \in \mathcal{P} $ by the invariants, so every prefix of $ P[i, j] $ is also in $ \mathcal{P} $ by Lemma \ref{lem:Pprefixessuffixes}. Let $ j' $ be the largest integer such that $ i - 1 \le j' < j $ and $ [\ell_{P[i, j]}, r_{P[i, j]}] \subsetneqq [\ell_{P[i, j']}, r_{P[i, j']}] $. Note that $ j' $ is well defined: since $ j > i - 1 $, we have $ P[i, j] \not = \epsilon $ and so $ [\ell_{P[i, j]}, r_{P[i, j]}] \not = [1, n'] $ by Lemma \ref{lem:alphabetatleasttwo} (we have $ \sigma \ge 2 $ because the alphabet is effective and $ P[i - 1, j] \not \in \mathcal{P} $), which implies that $ j' $ is well defined because $ P[i, i - 1] = \epsilon $ and so $ [\ell_{P[i, i - 1]}, r_{P[i, i - 1]}] = [1, n'] $. In particular, $ P[i, j'] \in \mathcal{P} $. Let us show that we can determine $ \ell_{P[i, j']} $ and $ r_{P[i, j']} $ in $ O(1) $ time and we can determine $ j' $ in $ O(\log n) $ time. By the invariant, we have $ \ell_{P[i, j]} = \ell $ and $ r_{P[i, j]} = r $. By Lemma \hyperref[item:propertiesnodessuffixtree4]{\ref*{lem:propertiesnodessuffixtree} (\ref*{item:propertiesnodessuffixtree4})} and Lemma \hyperref[item:nodesSuff1]{\ref*{lem:nodesSuff} (\ref*{item:nodesSuff1})}, we know that $ [\ell_{P[i, j']}, r_{P[i, j']}] $ must be the parent of $ [\ell_{P[i, j]}, r_{P[i, j]}] $ in $ \Suff(\mathcal{T}) $, so $ \ell_{P[i, j']} $ and $ r_{P[i, j']} $ can be computed in $ O(1) $ time as follows: we first compute the BP-index $ t $ of $ [\ell_{P[i, j]}, r_{P[i, j]}] $ from $ \ell_{P[i, j]} $ and $ r_{P[i, j]} $ by using Lemma \ref{lem:fromintervaltobpindex}, then we compute the BP-index $ t' = \parent(\Suff(\mathcal{T}), t) $ of $ [\ell_{P[i, j']}, r_{P[i, j']}] $ by using Theorem \ref{theor:balancedparenthesisnavarrosadakane}, and we finally compute $ \ell_{P[i, j']} $ and $ r_{P[i, j']} $ from $ t' $ by using again Lemma \ref{lem:fromintervaltobpindex}. Moreover, from Lemma \hyperref[item:nodesSuff2]{\ref*{lem:nodesSuff} (\ref*{item:nodesSuff2})}, the maximality of $ j' $ and Lemma \hyperref[item:nodesSuff5]{\ref*{lem:nodesSuff} (\ref*{item:nodesSuff5})} we obtain $ |P[i, j']| = \lambda_{\ell_{P[i, j']}, r_{P[i, j']}} $, so $ j' = \lambda_{\ell_{P[i, j']}, r_{P[i, j']}} + i - 1 $ and we can compute $ j' $ in $ O(\log n) $ time by Corollary \ref{cor:computingelllr}. The new quadruple is $ (i, j', \ell_{P[i, j']}, r_{P[i, j']}) $. We maintain the invariants because $ j' \ge i - 1 $.
        \end{itemize}
    \end{itemize}
    Let us show that the algorithm terminates after $ O(m) $ steps. Note that, when the quadruple $ (i, j, \ell, r) $ is updated to $ (i', j',\ell', r) $, then (in all cases) we have $ i' \le i $, $ j' \le j $ and $ (i < i') \lor (j' < j) $, which implies $ i' + j' < i + j $. From the invariant $ j \ge i - 1 $ we obtain that at some point we must necessarily have $ i' = 1 $ and the algorithm terminates. Moreover, at the beginning of the algorithm we have $ i + j = 2m + 1 $ and at the end of the algorithm we have $ i + j \ge 1 $, so the algorithm terminates after $ O(m) $ steps. Each step takes $ O(\log n) $ time in the worst case (we assume that the alphabet is effective, so $ O(\log n) $ dominates $ O(\log \log \sigma) $), and we conclude that the running time of the algorithm is $ O(m \log n) $.

    Let us prove that at any step of the algorithm, if we are processing the quadruple $ (i, j , \ell, r) $, then $ t_{i - 1} \le j $ (we only process quadruples for which $ 2 \le i \le m + 1 $ because when we reach $ i = 1 $ the algorithm terminates, so $ t_{i - 1} $ is always well defined). At the beginning we process $ (m + 1, m, 1, n') $ and we have $ t_m \le m $. Now, assume that at some point we process $ (i, j, \ell, r) $ for which $ t_{i - 1} \le j $. We must prove that $ t_{i' - 1} \le j' $, where $ (i', j', \ell', r') $ is the new quadruple (assuming $ i' \ge 2 $). We distinguish two cases.
    \begin{itemize}
        \item Assume that $ P[i - 1, j] \in \mathcal{P} $. In this case, $ i' = i - 1 $ and $ j' = j $, so we must prove that $ t_{i - 2} \le j $. From Lemma \ref{lem:t_imonotone} we conclude $ t_{i - 2} \le t_{i - 1} \le j $.
        \item Assume that $ P[i - 1, j] \not \in \mathcal{P} $. We distinguish two subcases.
        \begin{itemize}
            \item Assume that $ j = i - 1 $. In this case, $ i' = i - 1 $ and $ j' = i - 2 $, so we must prove that $ t_{i - 2} \le i - 2 $. By Lemma \ref{lem:Pprefixessuffixes}, it will suffice to prove that $ P[i - 2, i - 1] \not \in \mathcal{P} $. This follows from $ P[i - 1, i - 1] \not \in \mathcal{P} $ and Lemma \ref{lem:Pprefixessuffixes}.
            \item Assume that $ j > i - 1 $. In this case, $ i' = i $ and $ j' $ is the largest integer such that $ i - 1 \le j' < j $ and $ [\ell_{P[i, j]}, r_{P[i, j]}] \subsetneqq [\ell_{P[i, j']}, r_{P[i, j']}] $, so we must prove that $ t_{i - 1} \le j' $. By Lemma \ref{lem:Pprefixessuffixes}, it will suffice to prove that $ P[i - 1, j' + 1] \not \in \mathcal{P} $. From Lemma \hyperref[item:nodesSuff1]{\ref*{lem:nodesSuff} (\ref*{item:nodesSuff1})}, Lemma \hyperref[item:propertiesnodessuffixtree4]{\ref*{lem:propertiesnodessuffixtree} (\ref*{item:propertiesnodessuffixtree4})} and the maximality of $ j' $ we conclude $ [\ell_{P[i, j]}, r_{P[i, j]}] = [\ell_{P[i, j' + 1]}, r_{P[i, j' + 1]}] $, or equivalently, $ Y_{P[i, j]} = Y_{P[i, j' + 1]} $. From Corollary \ref{cor:Pcontainment} we obtain $ Y_{P[i - 1, j]} = Y_{P[i - 1, j' + 1]} $, so from $ P[i - 1, j] \not \in \mathcal{P} $ we conclude $ P[i - 1, j' + 1] \not \in \mathcal{P} $.
        \end{itemize}
    \end{itemize}
    
    Let us show that the algorithm correctly computes all the $ m $ values $ t_1 $, $ t_2 $, $ \dots $, $ t_m $. We have to show that $ t^*_i $ is defined and equal to $ t_i $, for every $ 1 \le i \le m $. Note that, when the quadruple $ (i, j, \ell, r) $ is updated to $ (i', j',\ell', r) $, we have $ i - 1 \le i' \le i $. At the beginning $ i = m + 1 $ and at the end $ i = 1 $. Notice that we define $ t^*_{i - 1} $ exactly when we go from $ i $ to $ i - 1 $, so $ t^*_i $ is defined for every $ 1 \le i \le m $.
    
    Let us prove that $ t^*_i = t_i $ for every $ 1 \le i \le m $. We distinguish two cases.
    \begin{itemize}
        \item We define $ t^*_{i - 1} = j $ when the algorithm is processing a quadruple $ (i, j, \ell, r) $ for which $ P[i - 1, j] \in \mathcal{P} $. In this case, we must prove that $ t_{i - 1} = j $. Note that by the invariants we have $ j \ge i - i $. Since the current quadruple is $ (i, j, \ell, r) $, we have $ t_{i - 1} \le j $, so from $ P[i - 1, j] \in \mathcal{P} $ and the maximality of $ t_{i - 1} $ we conclude $ t_{i - 1} = j $.
        \item We define $ t^*_{i - 1} = i - 2 $ when the algorithm is processing a quadruple $ (i, i - 1, \ell, r) $ for which $ P[i - 1, i - 1] \not \in \mathcal{P} $. In this case, we must prove that $ t_{i - 1} = i - 2 $. Since the current quadruple is $ (i, i - 1, \ell, r) $, we have $ t_{i - 1} \le i - 1 $, so from $ P[i - 1, i - 1] \not \in \mathcal{P} $ we conclude $ t_{i - 1} = i - 2 $.
    \end{itemize}

    Let us show that the algorithm correctly computes $ \ell_{P[i, t_i]} $ and $ r_{P[i, t_i]} $ for every $ 1 \le i \le m $. We again distinguish two cases.
    \begin{itemize}
        \item We define $ t^*_{i - 1} = j $ when the algorithm is processing the quadruple $ (i, j, \ell, r) $. In this case, $ t_{i - 1} = j $ and the new quadruple is $ (i - 1, j, \ell', r') $ for some $ \ell' $ and $ r' $. By the invariants we must have $ \ell' = \ell_{P[i - 1, j]} = \ell_{P[i - 1, t_{i - 1}]} $ and $ r' = r_{P[i - 1, j]} = r_{P[i - 1, t_{i - 1}]} $.
        \item We define $ t^*_{i - 1} = i - 2 $ when the algorithm is processing the quadruple $ (i, i - 1, \ell, r) $. Then, we have $ t_{i - 1} = i - 2 $, so $ \ell_{P[i - 1, t_{i - 1}]} = \ell_\epsilon = 1 $ and $ r_{P[i - 1, t_{i - 1}]} = r_\epsilon = n' $. Notice that, similarly to the previous case, the new quadruple is $ (i - 1, i - 2, 1, n') $, so one may also argue by using the invariants.
    \end{itemize}
\end{proof}   

Let us prove that in $ O(1) $ time (i) we can move from $ \Suff (\mathcal{T}) $ to $ \Suff^* (\mathcal{T}) $ and from $ \Suff^* (\mathcal{T}) $ to $ \Suff (\mathcal{T}) $ and (ii) we can determine the nearest marked ancestor of a node.

\begin{lemma}\label{lem:bpindexrespected}
    Let $ \mathcal{T} $ be a dictionary, and let $ u $ be a marked node in $ \Suff (\mathcal{T}) $.
    \begin{itemize}
        \item If the BP-index (resp. BP$^\#$-index) of $ u $ in $ \Suff (\mathcal{T}) $ is $ i $, then the BP-index (resp. BP$^\#$-index) of $ u $ in $ \Suff^* (\mathcal{T}) $ is $ \rank_1(B_8, i) $.
        \item If the BP-index (resp. BP$^\#$-index) of $ u $ in $ \Suff^* (\mathcal{T}) $ is $ i $, then the BP-index (resp. BP$^\#$-index) of $ u $ in $ \Suff (\mathcal{T}) $ is $ \select_1(B_8, i) $.
    \end{itemize}
\end{lemma}

\begin{proof}
    Recall that $ \Suff^* (\mathcal{T}) $ is defined from $ \Suff (\mathcal{T}) $ following the nodes in depth-first search order, respecting the order of the children of each node. Since $ Z_{\Suff (\mathcal{T})} $, $ B_8 $ and $ Z_{\Suff^* (\mathcal{T})} $ are defined following the same order, the conclusion follows.
\end{proof}

\begin{lemma}\label{lem:fromtreetoauxiliary}
    Let $ \mathcal{T} $ be a dictionary. In $ O(1) $ time we can solve the following queries:
    \begin{itemize}
        \item $ \toaux(i) $: given the BP-index $ i $ of a marked node $ u $ in $ \Suff (\mathcal{T}) $, return the BP-index of $ u $ in $ \Suff^* (\mathcal{T}) $.
        \item $ \fromaux(i) $: given the BP-index $ i $ of a node $ u $ in $ \Suff^* (\mathcal{T}) $, return the BP-index of $ u $ in $ \Suff (\mathcal{T}) $.
        \item $ \nma(i) $: given the BP-index $ i $ of a node $ u $ in $ \Suff (\mathcal{T}) $, return the BP-index in $ \Suff^* (\mathcal{T}) $ of the nearest marked ancestor $ v $ of $ u $ in $ \Suff (\mathcal{T}) $ (in particular, if $ u $ is marked, then $ v = u $).
    \end{itemize}
\end{lemma}

\begin{figure}
\captionsetup[subfigure]{justification=centering}
	\centering
    \begin{subfigure}[b]{0.49 \textwidth}
        \centering
        \scalebox{0.9}{
\begin{tikzpicture}[->,>=stealth', semithick, auto, scale=1]

\node[] (0)    at (-1,1)	{};
\node[circle, draw, fill=orange] (1)    at (0,0)	{};
\node[circle, draw] (2)    at (-1,-1)	{};
\node[circle, draw, fill=orange, label=above:{$ v $}] (3)    at (1,-1)	{};
\node[circle, draw] (4)    at (2,-2)	{};
\node[circle, draw, label=above:{$ u $}] (5)    at (3,-3)	{};
\node[circle, draw] (6)    at (2,-3)	{};
\node[circle, draw] (7)    at (0,-2)	{};
\node[circle, draw] (8)    at (1,-2)	{};
\node[circle, draw] (9)    at (0,-3)	{};
\node[circle, draw] (10)    at (1,-3)	{};
\node[circle, draw, fill=orange] (11)    at (-1,-2)	{};
\node[circle, draw] (12)    at (-1,-3)	{};

\draw (0) edge [dashed] node [] {} (1);
\draw (1) edge [] node [] {} (2);
\draw (1) edge [] node [] {} (3);
\draw (3) edge [] node [] {} (4);
\draw (4) edge [] node [] {} (5);
\draw (4) edge [] node [] {} (6);
\draw (3) edge [] node [] {} (7);
\draw (3) edge [] node [] {} (8);
\draw (7) edge [] node [] {} (9);
\draw (7) edge [] node [] {} (10);
\draw (2) edge [] node [] {} (11);
\draw (11) edge [] node [] {} (12);
\end{tikzpicture}
}
\caption{}
\end{subfigure}
    \begin{subfigure}[b]{0.49 \textwidth}
        \centering
        \scalebox{0.9}{
\begin{tikzpicture}[->,>=stealth', semithick, auto, scale=1]

\node[] (-1)    at (-2,2)	{};
\node[circle, draw, fill=orange, label=above:{$ v $}] (0)    at (-1,1)	{};
\node[circle, draw] (1)    at (0,0)	{};
\node[circle, draw] (2)    at (-1,-1)	{};
\node[circle, draw] (3)    at (1,-1)	{};
\node[circle, draw] (4)    at (2,-2)	{};
\node[circle, draw, label=above:{$ u $}] (5)    at (3,-3)	{};
\node[circle, draw, fill = orange, label=above:{$ \bar{v} $}] (6)    at (2,-3)	{};
\node[circle, draw, fill = orange] (7)    at (0,-2)	{};
\node[circle, draw] (8)    at (1,-2)	{};
\node[circle, draw] (9)    at (0,-3)	{};
\node[circle, draw, fill = orange] (10)    at (1,-3)	{};
\node[circle, draw, fill=orange] (11)    at (-1,-2)	{};
\node[circle, draw] (12)    at (-1,-3)	{};
\node[circle, draw] (13)    at (-2, 0)	{};
\node[circle, draw] (14)    at (-2, -1)	{};
\node[circle, draw, fill=orange] (15)    at (-2, -2)	{};
\node[circle, draw] (16)    at (1, -4)	{};
\node[circle, draw] (17)    at (2, -4)	{};

\draw (-1) edge [dashed] node [] {} (0);
\draw (0) edge [] node [] {} (1);
\draw (1) edge [] node [] {} (2);
\draw (1) edge [] node [] {} (3);
\draw (3) edge [] node [] {} (4);
\draw (4) edge [] node [] {} (5);
\draw (8) edge [] node [] {} (6);
\draw (3) edge [] node [] {} (7);
\draw (3) edge [] node [] {} (8);
\draw (7) edge [] node [] {} (9);
\draw (7) edge [] node [] {} (10);
\draw (2) edge [] node [] {} (11);
\draw (11) edge [] node [] {} (12);
\draw (0) edge [] node [] {} (13);
\draw (13) edge [] node [] {} (14);
\draw (14) edge [] node [] {} (15);
\draw (6) edge [] node [] {} (16);
\draw (6) edge [] node [] {} (17);
\end{tikzpicture}
}
\caption{}
\end{subfigure}
\caption{Proof of Lemma \ref{lem:fromtreetoauxiliary}. Both pictures represent $ \Suff (\mathcal{T}) $, and all marked nodes are orange. (a) Case $ Z_{\Suff^*(\mathcal{T})}[j^*] = 1 $. (b) Case $ Z_{\Suff^*(\mathcal{T})}[j^*] = 0 $.}
\label{fig:computingnearestmarkedancestor}
\end{figure}
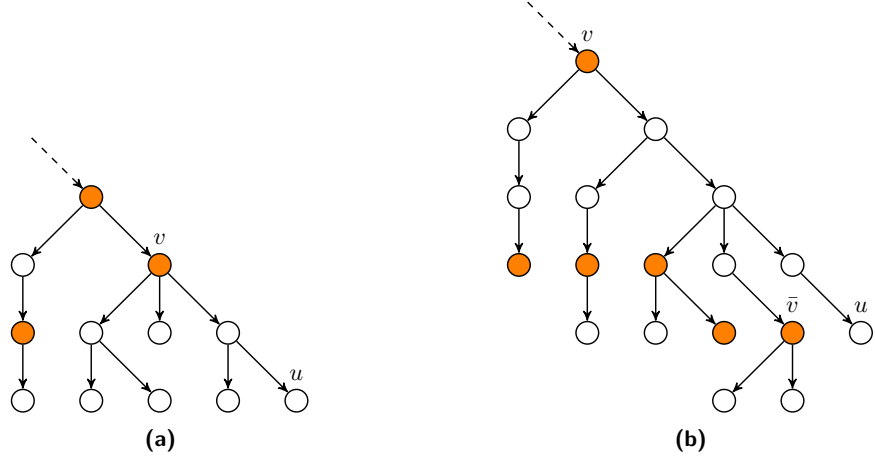

\begin{proof}
    By Lemma \ref{lem:bpindexrespected}, we have $ \toaux(i) = \rank_1(B_8, i) $ and $ \fromaux(i) = \select_1 (B_8, i) $.

    Let us how to compute $ \nma (i) $. First, notice that $ \nma (i) $ is always well defined: the nearest marked ancestor is defined for every node of $ \Suff (\mathcal{T}) $ because the root is marked. Compute $ j^* = \rank_1 (B_8, i) $. Then, $ 1 \le j^* \le 2 |\mathcal{N}^*| $ and $ B_8[k] = 0 $ for $ \select_1 (B_8, j^*) + 1 \le k \le i $. We distinguish two cases (see Figure \ref{fig:computingnearestmarkedancestor}).

        \begin{itemize}
            \item Assume that $ Z_{\Suff^*(\mathcal{T})}[j^*] = 1 $. We will prove that $ \nma(i) = j^* $. Let $ v $ be the node of $ \Suff^*(\mathcal{T}) $ having BP-index $ j^* $ in $ \Suff^* (\mathcal{T}) $, let $ j $ be the BP-index of $ v $ in $ \Suff (\mathcal{T}) $ and let $ t $ be the BP$^{\#}$-index of $ v $ in $ \Suff(\mathcal{T}) $. Then, $ j < t $, $ j \le i $ (because by Lemma \ref{lem:bpindexrespected} we have $ j = \select_1 (B_8, j^*) $) and $ B_8[k] = 0 $ for every $ j + 1 \le k \le i $ (because $ j = \select_1 (B_8, j^*) $). To prove the claim, we must prove that $ v $ is the nearest marked ancestor of $ u $ in $ \Suff (\mathcal{T}) $.
            \begin{itemize}
                \item Node $ v $ is marked because $ v $ is defined by using $ \Suff^*(\mathcal{T}) $.
                \item We have $ B_8[t] = 1 $ because $ v $ is marked.
                \item  We have $ i < t $ because $ j < t $, $ B_8[t] = 1 $ and $ B_8[k] = 0 $ for every $ j + 1 \le k \le i $.
                \item Let us prove that $ v $ is a (marked) ancestor of $ u $ in $ \Suff (\mathcal{T}) $. Since $ \Suff (\mathcal{T}) $ is defined through a depth-first search, the conclusion follows from $ j \le i < t $, because this implies that when we visit $ u $ we are in the subtree rooted at $ v $.
                \item Let us prove that $ v $ is the \emph{nearest} marked ancestor of $ u $ in $ \Suff (\mathcal{T}) $. Let $ v' $ the nearest marked ancestor of $ u $ in $ \Suff (\mathcal{T}) $ and let $ j' $ be the BP-index of $ v' $ in $ \Suff(\mathcal{T}) $. To prove that $ v = v' $, it will suffice to show that $ j = j' $. Since $ v' $ is an ancestor of $ u $, we have $ j' \le i $. Moreover, since $ v' $ is the nearest marked ancestor of $ u $ and $ v $ is a marked ancestor of $ u $, we have $ j \le j' $. Since $ v' $ is marked, we have $ B[j'] = 1 $, so $ j' \le j $ because $ j' \le i $ and $ B_8[k] = 0 $ for every $ j + 1 \le k \le i $. From $ j \le j' $ and $ j' \le j $ we conclude $ j = j' $.
            \end{itemize}
            \item Assume that $ Z_{\Suff^*(\mathcal{T})}[j^*] = 0 $. We will prove that:
            \begin{equation*}
                \nma(i) = \parent (\Suff^*(\mathcal{T}), \open(\Suff^*(\mathcal{T}), j^*)).
            \end{equation*}
             Let $ \bar{v} $ be the node of $ \Suff^*(\mathcal{T}) $ with BP$^\#$-index $ j^* $, let $ h^* = \open(\Suff^*(\mathcal{T}), j^*) $ be the BP-index of $ \bar{v} $ in $ \Suff^*(\mathcal{T}) $, let $ h $ be the BP-index of $ \bar{v} $ in $ \Suff (\mathcal{T}) $, and let $ j $ be the BP$^\#$-index of $ \bar{v} $ in $ \Suff (\mathcal{T}) $. Then, $ h^* < j^* $, $ h < j $, $ j < i $ (we have $ j \le i $ because $ j = \select_1 (B_8, j^*) $ by Lemma \ref{lem:bpindexrespected}, and we have $ j \not = i $ because $ j $ is a BP$^\#$-index and $ i $ is a BP-index) and $ B_8[k] = 0 $ for every $ j + 1 \le k \le i $ (because $ j = \select_1 (B_8, j^*) $). Let $ v $ be the parent of $ \bar{v} $ in $ \Suff^*(\mathcal{T}) $. Notice that $ v $ is well-defined because $ \bar{v} $ is not the root of $ \Suff^*(\mathcal{T}) $: indeed, since $ i $ is a BP-index in $ \Suff (\mathcal{T}) $, we have $ i < 2|\mathcal{N}| $, so $ j < i < 2|\mathcal{N}| $, which means that $ \bar{v} $ is not the root of $ \Suff (\mathcal{T}) $ and so $ \bar{v} $ is not the root of $ \Suff^* (\mathcal{T}) $. Moreover, let $ s $ be the BP-index of $ v $ in $ \Suff(\mathcal{T}) $ and let $ t $ be the BP$^\#$-index of $ v $ in $ \Suff(\mathcal{T}) $; we have $ s < t $. To prove the claim, we must prove that $ v $ is the nearest marked ancestor of $ u $ in $ \Suff (\mathcal{T}) $.
            \begin{itemize}
                \item Nodes $ \bar{v} $ and $ v $ are marked because they are defined by using $ \Suff^*(\mathcal{T}) $.
                \item Since $ v $ is the parent of $ \bar{v} $ in $ \Suff^*(\mathcal{T}) $, then $ v $ is a strict ancestor of $ \bar{v} $ in $ \Suff(\mathcal{T}) $, so $ s < h < j < t $.
                \item From $ s < h $, $ h < j $ and $ j < i $ we conclude $ s < i $.
                \item We have $ B_8[t] = 1 $ because $ v $ is marked.
                \item We have $ i < t $ because $ j < t $, $ B_8[t] = 1 $ and $ B_8[k] = 0 $ for every $ j + 1 \le k \le i $.
                 \item Let us prove that $ v $ is a (marked) proper ancestor of $ u $ in $ \Suff (\mathcal{T}) $. Since $ \Suff (\mathcal{T}) $ is defined through a depth-first search, the conclusion follows from $ s < i < t $, because this implies that $ v \not = u $ and when we visit $ u $ we are in the subtree rooted at $ v $.
                \item Let us prove that $ v $ is the \emph{nearest} marked ancestor of $ u $ in $ \Suff (\mathcal{T}) $. Let $ v' $ the nearest marked ancestor of $ u $ in $ \Suff (\mathcal{T}) $, let $ s' $ be the BP-index of $ v' $ in $ \Suff(\mathcal{T}) $ and let $ t' $ be the BP$^\#$-index of $ v' $ in $ \Suff(\mathcal{T}) $ (and so $ s' < t' $). To prove that $ v = v' $, it will suffice to show that $ s = s' $. Since $ v' $ is an ancestor of $ u $, we have $ s' \le i < t' $. Moreover, since $ v' $ is the nearest marked ancestor of $ u $ and $ v $ is a marked ancestor of $ u $, we have $ s \le s' $. Note that (i) $ j < t' $ because $ j < i $ and $ i < t' $ and (ii) $ h < t' $ because $ h < j $, $ j < i $ and $ i < t' $. Since $ v' $ is marked, we have $ B[s'] = 1 $, so $ s' \le j $ because $ s' \le i $ and $ B_8[k] = 0 $ for every $ j + 1 \le k \le i $. Moreover, we must have $ s' < h $, otherwise we would have $ h \le s' \le j $, which would mean that $ v' $ is located in the subtree rooted at $ \bar{v} $, and so we would conclude $ t' \le j $, which contradicts $ j < t' $. We conclude $ s' < h < t' $, which means that $ v' $ is a marked ancestor of $ \bar{v} $ in $ \Suff (\mathcal{T}) $. Since $ v $ is the nearest marked ancestor of $ \bar{v} $ in $ \Suff (\mathcal{T}) $ (because $ v $ is the parent of $ \bar{v} $ in $ \Suff^* (\mathcal{T}) $), we conclude $ s' \le s $. From $ s \le s' $ and $ s' \le s $ we conclude $ s = s' $. 
            \end{itemize}
        \end{itemize}
\end{proof}

Let us present some auxiliary results. We will use them to show that in Theorem \ref{theor:main} (our main result) we can achieve running time proportional to the number of occurrences.

\begin{lemma}\label{lem:existencet^*}
    Let $ \mathcal{T} $ be a dictionary. Given $ 1 \le t_1, t_2 \le n^* $ (where $ n^* $ is the length of $ \SA $) and an integer $ x $, in $ O(\log n) $ time we can determine if there exists $ t_1 \le t^* \le t_2 $ such that $ |\mathcal{T}_{\SA[t^*]}| \le x $ and, if the answer is affirmative, return both one such $ t^* $ and the value $ \SA[t^*] $.
\end{lemma}

\begin{proof}
    If $ t_1 > t_2 $, we immediately conclude that there exists no $ t^* $ with the desired property, so in the following we assume $ t_1 \le t_2 $. In $ O(1) $ time, we compute $ t' = \rmq_\Len (t_1, t_2) $. We have $ |\mathcal{T}_{\SA[t']}| = \Len[t'] \le \Len[t] = |\mathcal{T}_{\SA[t]}| $ for every $ t_1 \le t' \le t_2 $.  We compute $ \SA[t'] $ in $ O(\log n) $ time by using Theorem \ref{theor:sampledsuffixarray}. Next, we can compute $ |\mathcal{T}_{\SA[t']}| $ in $ O(1) $ time, because (i) we know that $ \SA[t'] $ refers to string $ T_{h'} $, where $ h' = \rank_1 (B_1, \SA[t']) $ (see Section \ref{sec:suffixarray}), (ii) we can compute $ |T_{h'}| $ in $ O(1) $ time (see Section \ref{sec:BWTFM}) and (iii) $ |\mathcal{T}_{\SA[t']}| = |T_{h'}| $. If $ x < |\mathcal{T}_{\SA[t']}| $, we conclude $ x < |\mathcal{T}_{\SA[t']}| \le |\mathcal{T}_{\SA[t]}| $ for every $ t_1 \le t \le t_2 $ and so there exists no $ t^* $ with the desired properties. If $ |\mathcal{T}_{\SA[t']}| \le x $, we can return $ t^* = t' $ and $ \SA[t^*] = \SA[t'] $.
\end{proof}

\begin{lemma}\label{lem:computingY}
    Let $ \mathcal{T} $ be a dictionary. Given $ 1 \le \ell, r \le n' $ and an integer $ y $, in $ O((1 + |Y|) \log n) $ time we can compute the set $ Y $ of all $ 1 \le k \le n $ such that (i) $ k \in \bigcup_{j = \ell}^r D_j $ and (ii) $ |\mathcal{T}_k| \le y $.
\end{lemma}

\begin{proof}
    By the definition of $ \SA $, we have:
    \begin{equation*}
        \bigcup_{j = \ell}^r D'_j = \{\SA[t] \;|\; \select_1 (B_4, \ell) \le t \le \select_1 (B_4, r + 1) - 1 \}
    \end{equation*}
    which implies:
    \begin{equation}\label{eq:secondstep}
        \bigcup_{j = \ell}^r D_j = \bigcup_{j = \ell}^r \bigcup_{k \in D'_j} [k]_{\sim} = \bigcup_{\substack{\select_1 (B_4, \ell) \le t \\ \le \select_1 (B_4, r + 1) - 1}} [\SA[t]]_\sim. 
    \end{equation}
We know that for every $ 1 \le k \le n $ all elements in $ [k]_\sim $ refers to the same string $ T_h $, hence $ |\mathcal{T}_{k'}| = |T_h| = |\mathcal{T}_k| $ for every $ k' \in [k]_\sim $. Consequently, from Equation \ref{eq:secondstep} we obtain:
\begin{equation}\label{eq:thirdstep}
    Y = \{k \in \bigcup_{j = \ell}^r D_j \;|\; |\mathcal{T}_k| \le y \} = \bigcup_{\substack{\select_1 (B_4, \ell) \le t \\ \le \select_1 (B_4, r + 1) - 1; \\ |\mathcal{T}_{\SA[t]}| \le y}} [\SA[t]]_\sim.
\end{equation}

Note that in Equation \ref{eq:thirdstep} we know that $ Y $ is the \emph{disjoint} union of the $ [\SA[t]]_\sim $'s, because in general if $ t \not = t' $, then $ \SA[t] \not \sim \SA[t'] $ (either $ \SA[t] $ and $ \SA[t] $ belong to distinct $ D_j $'s, or they belong to the same $ D_j $ but they refer to distinct strings $ T_h $'s). This means that, to compute $ Y $, we can compute each $ [\SA[t]]_\sim $ separately, and we will never report the same occurrence twice.

Consider now the instance of Lemma \ref{lem:existencet^*} where $ t_1 = \select_1 (B_4, \ell) $, $ t_2 = \select_1 (B_4, r + 1) - 1 $ and $ x = y $. In $ O(\log n) $ time, we can decide whether there exists $ t_1 \le t^* \le t_2 $ such that $ |\mathcal{T}_{\SA[t^*]}| \le x $ and, if the answer is affirmative, we can compute both one such $ t^* $ and the value $ \SA[t^*] $. If such a $ t^* $ does not exist, by Equation \ref{eq:thirdstep} we conclude $ Y = \emptyset $ (so $ |Y| = 0 $) and, if fact, we have computed $ Y $ in $ O(\log n) = O((1 + |Y|) \log n) $ time. Now assume that such a $ t^* $ exists. In this case, we know both $ t^* $ and $ \SA[t^*] $. Then, by Equation \ref{eq:thirdstep} we obtain:
\begin{equation}\label{eq:fourthstep}
    Y = [\SA[t^*]]_\sim \cup \Big( \bigcup_{\substack{\select_1 (B_4, \ell) \le t \le t^* - 1; \\ |\mathcal{T}_{\SA[t]}| \le y}}  [\SA[t]]_\sim \Big) \cup \Big( \bigcup_{\substack{t^* + 1 \le \select_1 (B_4, r + 1) - 1; \\ |\mathcal{T}_{\SA[t]}| \le y}}  [\SA[t]]_\sim \Big).
\end{equation}
We now recursively process the two new intervals $ \select_1 (B_4, \ell) \le t \le t^* - 1 $ and $ t^* + 1 \le \select_1 (B_4, r + 1) - 1 $ identified by Equation \ref{eq:fourthstep}. More precisely, we consider the instance of Lemma \ref{lem:existencet^*} where $ t_1 = \select_1 (B_4, \ell) $, $ t_2 = t^* - 1 $ and $ x = y $, and we consider the instance of Lemma \ref{lem:existencet^*} where $ t_1 = t^* + 1 $, $ t_2 = \select_1 (B_4, r + 1) - 1 $ and $ x = y $. For each of these instances, in $ O(\log n) $ time we can decide whether there exists $ t_1 \le t^{**} \le t_2 $ such that $ |\mathcal{T}_{\SA[t^{**}]}| \le x $ and, if the answer is affirmative, we can compute both one such $ t^{**} $ and the value $ \SA[t^{**}] $. If such a $ t^ {**} $ does not exist, we stop processing the corresponding interval, otherwise, we create two new subintervals, one with right endpoint $ t^{**} - 1 $, and one with left endpoint $ t^{**} + 1 $. The algorithm terminates when there is no interval left.

The time required to process all intervals is $ O(|\mathcal{I}| \log n) $, where $ \mathcal{I} $ is the set of all intervals generated in the process. Let $ \mathcal{I}_1 $ be the set of all generated intervals such that the corresponding instance of Lemma \ref{lem:existencet^*} finds an element $ t^* $ of the interval for which $ |\mathcal{T}_{\SA[t^*]}| \le x $, and let $ \mathcal{I}_2 $ be the set of all generated intervals such that the corresponding instance of Lemma \ref{lem:existencet^*} does not find an element $ t^* $ of the interval for which $ |\mathcal{T}_{\SA[t^*]}| \le x $. Then, $ |\mathcal{I}| = |\mathcal{I}_1| + |\mathcal{I}_2| $. Notice that for every interval in $ \mathcal{I}_1 $, the corresponding $ t^* $ satisfies $ \SA[t^*] \in Y $, and distinct intervals in $ \mathcal{I}_1 $ yield distinct $ t^* $'s, so $ |\mathcal{I}_1| \le |Y| $. Moreover, every interval in $ \mathcal{I}_2 $ is one of the two subintervals generated by some elements in $ \mathcal{I}_1 $, so $ |\mathcal{I}_2| \le 2 |\mathcal{I}_1| \le 2 |Y| $. We conclude that $ |\mathcal{I}| \le 3 |Y| $, so the time required to process all interval is $ O(|\mathcal{I}| \log n) \subseteq O(|Y| \log n) $.

After processing all intervals, we know the set $ T $ of all $ \select_1 (B_4, \ell) \le t  \le \select_1 (B_4, r + 1) - 1 $ such that $ Y = \bigcup_{t \in T} [\SA[t]]_\sim $, and we also know $ \SA[t] $ for every $ t \in T $. We saw in Section \ref{sec:suffixarray} that for every $ t \in T $ we have $ [\SA[t]]_\sim = \{\SA[t] + w |\rho_h| \;|\; 0 \le w \le (|T_h| / |\rho_h|) - 1 \} $, where $ h = \rank_1(B_1, \SA[t]) $ and $ |\rho_h| $ and $ |T_h| $ can be computed in $ O(1) $ time (see Section \ref{sec:suffixarray}). Consequently, we can compute $ [\SA[t]]_\sim $ in $ O(|[\SA[t]]_\sim|) $ time for every $ t \in T $, and so we can compute $ Y $ in $ \sum_{t \in T} O(|[\SA[t]]_\sim|) = O(Y) $ time.
\end{proof}

We can now prove the main result.

\bigskip

\noindent{\textbf{Statement of Theorem \ref{theor:main}}.}
    Let $ \mathcal{T} = (T_1, T_2, \dots, T_d) $ be a dictionary of total length $ n $. Then, $ \mathcal{T} $ can be encoded using a data structure of $ n \log \sigma (1 + o(1)) + O(n) + O(d \log n) $ bits such that, given a string $ P $ of length $ m $, we can compute $ \Cdm(\mathcal{T}, P) $  in $ O((m + occ) \log n) $ time.

\begin{proof}
    As anticipated in Section \ref{sec:circularpatternmatching}, we store the FM-index of $ \mathcal{T} $ using $ n \log \sigma (1 + o(1)) + O(n) $ bits (Theorem \ref{theor:FMindex}), the sampled suffix array using $ o(n \log \sigma) + O(n) + O(d \log n) $ bits (Theorem \ref{theor:sampledsuffixarray}), the sampled LCP array using $ o(n \log \sigma) + O(n) $ bits (Theorem \ref{theor:lcpsampling}) and some auxiliary data structures using $ O(n) $ bits: the $ 8 $ bitvectors $ B_1 $-$ B_8 $, a range minimum query data structure for both $ \LCP $ and $ \Len $ (Section \ref{sec:topologysuffixtree} and Section \ref{sec:solvingcirculad}), and the data structure of Theorem \ref{theor:balancedparenthesisnavarrosadakane} for both $ \Suff (\mathcal{T}) $ and $ \Suff^* (\mathcal{T}) $. The total space is $ n \log \sigma (1 + o(1)) + O(n) + O(d \log n) $ bits.

    Let us show that the data structure is an encoding of $ \mathcal{T} $. We retrieve the number $ d $ of strings in the dictionary from $ B_1 $, and we retrieve $ n' $ from $ B_6 $. We retrieve $ \BWT $ by using Theorem \ref{theor:FMindex}, and we retrieve $ \BWT^* $ by sorting the elements in $ \BWT $. We know that $ \BWT^*[j] = \mathcal{S}_j[1] $ for every $ 1 \le j \le n' $. For every $ 1 \le j \le n' $, we retrieve $ D_j $ by using $ \SA $ (see Section \ref{sec:suffixarray}). We conclude that for every $ 1 \le j \le n' $ and for every $ k \in D_j $ we have $ (T_1 T_2 \dots T_d)[k] = \mathcal{S}_j[1] = \BWT^*[j] $ (where $ T_1 T_2 \dots T_d $ is the concatenation of $ T_1 $, $ T_2 $, $ \dots $, $ T_d $), so we can retrieve $ T_1 T_2 \dots T_d $. By using $ B_1 $, we can finally retrieve $ T_1 $, $ T_2 $, $ \dots $, $ T_d $ and so $ \mathcal{T}$.

    Now, let us consider a string $ P $ of length $ m $. We must show that we can compute $ \Cdm(\mathcal{T}, P) $  in $ O((m + occ) \log n) $ time.
    
     For every $ [\ell, r] \in \mathcal{N} $ distinct from the root of $ \Suff (\mathcal{T}) $, let $ \Cdm_{\ell, r} (\mathcal{T}) $ be the set of all $ 1 \le k \le n $ such that (i) $ k \in D_j $, with $ j \in [\ell', \ell - 1] \cup [r + 1, r'] $ and (ii) $ |\mathcal{T}_k| \le \lambda_{\ell', r'} $, where $ [\ell', r'] $ is the parent of $ [\ell, r] $ in $ \Suff (\mathcal{T}) $. By definition, we have $ |\Cdm_{\ell, r} (\mathcal{T})| \ge 1 $ if and only if $ [\ell, r] $ is a marked node.
     
    By Lemma \ref{lem:computing3mvalues}, in $ O(m \log n) $ time we can compute all the values $ t_i $'s, $ \ell_{P[i, t_i]} $'s and $ r_{P[i, t_i]} $'s. As explained in Section \ref{sec:solvingcirculad}, to prove that we can compute $ \Cdm (\mathcal{T}, P) $ in $ O((m + occ) \log n) $ time, it will suffice to show how to compute $ \Cdm_i (\mathcal{T}, P) $ in $ O((1 + occ_i) \log n) $ time for every $ 1 \le i \le m $.

    Fix $ 1 \le i \le m $. Let $ \Cdm^0_i (\mathcal{T}, P) $ the set of all $ 1 \le k \le n $ such that (i) $ \mathcal{T}_k $ occurs in $ P $ at position $ i $ and (ii) $ k \in D_j $, where $ j \in [\ell_{P[i, t_i]}, r_{P[i, t_i]} ] $. Moreover, let $ [\ell_1, r_1] $, $ [\ell_2, r_2] $, $ \dots $, $ [\ell_d, r_d] $ all marked ancestor of $ [\ell_{P[i, t_i]}, r_{P[i, t_i]} ] $ in $ \Suff (\mathcal{T}) $ distinct from the root, where $ d \ge 0 $ and $ [\ell_s, r_s] \subsetneqq [\ell_{s + 1}, r_{s + 1}] $ for $ 1 \le s \le d - 1 $ (see Lemma \hyperref[item:propertiesnodessuffixtree4]{\ref*{lem:propertiesnodessuffixtree} (\ref*{item:propertiesnodessuffixtree4})}). Note that, if $ d \ge 1 $, then $ [\ell_{P[i, t_i]}, r_{P[i, t_i]}] \subseteq [\ell_1, r_1] $, and $ [\ell_{P[i, t_i]}, r_{P[i, t_i]}] = [\ell_1, r_1] $ if and only if $ [\ell_{P[i, t_i]}, r_{P[i, t_i]}] $ is marked. Define $ \Cdm^s_i (\mathcal{T}, P) = \Cdm_{\ell_s, r_s} (\mathcal{T}) $ for $ 1 \le s \le d $. Notice that from the definition of $ \Cdm^s_i (\mathcal{T}, P) $ and Lemma \ref{lem:climbingsuffixtree} we conclude that, for every $ 1 \le s \le d $, $ \Cdm^s_i (\mathcal{T}, P) $ is the set of all $ 1 \le k \le n $ such that (i) $ \mathcal{T}_k $ occurs in $ P $ at position $ i $ and (ii) $ k \in D_j $ and $ j \in [\ell'_s, \ell_s - 1] \cup [r_s + 1, r'_s] $, where $ [\ell'_s, r'_s] $ is the parent of $ [\ell_s, r_s] $ in $ \Suff (\mathcal{T}) $. From Lemma \hyperref[item:propertiesnodessuffixtree4]{\ref*{lem:propertiesnodessuffixtree} (\ref*{item:propertiesnodessuffixtree4})} we also obtain that $ [\ell'_s, r'_s] \subseteq [\ell_{s + 1}, r_{s + 1}] $ for every $ 1 \le s \le d - 1 $.
    
    Let us prove that $ \Cdm_i (\mathcal{T}, P) = \bigcup_{s = 0}^d \Cdm_i^s (\mathcal{T}, P) $ and the union is disjoint.
    \begin{itemize}
        \item Let us prove that $ \Cdm_i^0 (\mathcal{T}, P) $, $ \Cdm_i^1 (\mathcal{T}, P) $, $ \dots $, $ \Cdm_i^d (\mathcal{T}, P) $ are pairwise disjoint. Fix $ 0 \le s < s' \le d $ and $ 1 \le k \le n $ such that $ k \in \Cdm_i^s (\mathcal{T}, P) $. We must show that $ k \not \in \Cdm_i^{s'} (\mathcal{T}, P) $. Let $ k \in D_j $. We distinguish two cases.
        \begin{itemize}
            \item Assume that $ s = 0 $. Then, $ s' \ge 1 $ and $ j \in [\ell_{P[i, t_i]}, r_{P[i, t_i]}] \subseteq [\ell_1, r_1] \subseteq [\ell_2, r_2] \subseteq \dots \subseteq [\ell_{s'}, r_{s'}] $, so $ j \not \in [\ell'_{s'}, \ell_{s'} - 1] \cup [r_{s'} + 1, r'_{s'}] $ and $ k \not \in \Cdm_i^{s'} (\mathcal{T}, P) $.
            \item Assume that $ s \ge 1 $. Then, $ j \in [\ell'_s, \ell_s - 1] \cup [r_s + 1, r'_s] \subseteq [\ell'_s, r'_s] \subseteq [\ell_{s + 1}, r_{s + 1}] \subseteq [\ell_{s + 2}, r_{s + 2}] \subseteq \dots \subseteq [\ell_{s'}, r_{s'}] $, so $ j \not \in [\ell'_{s'}, \ell_{s'} - 1] \cup [r_{s'} + 1, r'_{s'}] $ and $ k \not \in \Cdm_i^{s'} (\mathcal{T}, P) $.
        \end{itemize}
        \item Let us prove that $ \Cdm_i (\mathcal{T}, P) = \bigcup_{s = 0}^d \Cdm_i^s (\mathcal{T}, P) $.

        $ (\supseteq) $ Fix $ 0 \le s \le d $. We must prove that $ \Cdm_i^s (\mathcal{T}, P) \subseteq \Cdm_i (\mathcal{T}, P) $. Fix $ k \in \Cdm_i^s (\mathcal{T}, P) $. We know that $ \mathcal{T}_k $ occurs in $ P $ at position $ i $, so $ k \in \Cdm_i (\mathcal{T}, P) $.

        $ (\subseteq) $. Fix $ k \in \Cdm_i (\mathcal{T}, P) $. We know that $ \mathcal{T}_k $ occurs in $ P $ at position $ i $. Let $ k \in D_j $. If $ j \in [\ell_{P[i, t_i]}, r_{P[i, t_i]}] $, then $ k \in \Cdm_i^0 (\mathcal{T}, P)$. Now assume that $ j \not \in [\ell_{P[i, t_i]}, r_{P[i, t_i]}] $. Let $ [\ell, r] $ be the nearest ancestor of $ [\ell_{P[i, t_i]}, r_{P[i, t_i]}] $ in $ \Suff (\mathcal{T}) $ for which $ j \not \in [\ell, r] $. Then, by Lemma \ref{lem:climbingsuffixtree} we have that $ [\ell, r] $ is not the root of $ \Suff (\mathcal{T}) $ and $ j \in [\ell', \ell - 1] \cup [r + 1, r'] $, where $ [\ell', r'] $ is the parent of $ [\ell, r] $ in $ \Suff (\mathcal{T}) $. The existence of $ k $ proves that $ [\ell, r] $ is a marked node of $ \Suff (\mathcal{T}) $, so since $ [\ell, r] $ is distinct from the root, there exists $ 1 \le s \le d $ such that $ [\ell, r] = [\ell_s, r_s] $. Hence, $ j \in [\ell'_s, \ell_s - 1] \cup [r_s + 1, r'_s] $ and we conclude $ k \in \Cdm_i^s (\mathcal{T}, P) $.
    \end{itemize}
    
    For every $ 0 \le s \le d $, define $ occ_i^s = |\Cdm^s_i (\mathcal{T}, P)| $. Since $ \Cdm_i (\mathcal{T}, P) = \bigcup_{s = 0}^d \Cdm_i^s (\mathcal{T}, P) $ and the union is disjoint, we have $ occ_i = \sum_{s = 0}^d occ_i^s $. Moreover, for every $ 1 \le s \le d $, we know that $ [\ell_s, r_s] $ is a marked node in $ \Suff (\mathcal{T}) $, so $ occ_i^s \ge 1 $ (while possibly $ occ_i^0 = 0 $). This implies that $ d \le \sum_{s = 1}^d occ_i^s \le occ_i $.

    Let us show that in $ O(1 + d) \subseteq O(1 + occ_i) $ time we can determine $ d $ and we can compute all values $ \ell_s $'s and $ r_s $'s for $ 1 \le s \le d $. First notice the that for every $ 1 \le s < d $ we know that $ [\ell_{s + 1}, r_{s + 1}] $ is the parent of $ [\ell_s, r_s] $ in $ \Suff^* (\mathcal{T}) $. Moreover, $ d = 0 $ if and only if the nearest marked ancestor of $ [\ell_{P[i, t_i]}, r_{P[i, t_i]}] $ in $ \Suff (\mathcal{T}) $ is the root, and if $ d \ge 1 $ the parent of $ [\ell_d, r_d] $ in $ \Suff^* (\mathcal{T}) $ is the root. As a consequence, we can proceed as follows. We first compute the BP-index $ i = \frominter (\ell_{P[i, t_i]}, r_{P[i, t_i]}) $ of $ [\ell_{P[i, t_i]}, r_{P[i, t_i]}] $ in $ \Suff (\mathcal{T}) $. Next, we compute the BP-index $ i^* = \nma (i) $ in $ \Suff^* (\mathcal{T}) $ of the nearest marked ancestor of $ [\ell_{P[i, t_i]}, r_{P[i, t_i]}] $ in $ \Suff (\mathcal{T}) $. If $ i^* = 1 $, then the nearest marked ancestor of $ [\ell_{P[i, t_i]}, r_{P[i, t_i]}] $ in $ \Suff (\mathcal{T}) $ is the root and we conclude $ d = 0 $. If $ i^* > 1 $, then $ d \ge 1 $, the nearest marked ancestor of $ [\ell_{P[i, t_i]}, r_{P[i, t_i]}] $ in $ \Suff (\mathcal{T}) $ is $ [\ell_1, r_1] $, and we know the BP-index $ x^*_1 $ of $ [\ell_1, r_1] $ in $ \Suff^* (\mathcal{T}) $. We now repeatedly compute $ x^*_2 = \parent(\Suff^* (\mathcal{T}), x^*_1) $, $ x^*_3 = \parent(\Suff^* (\mathcal{T}), x^*_2) $, and so on, until we reach the root of $ \Suff^* (\mathcal{T}) $. This requires $ 1 + d $ steps, and in the end we know $ d $ and the BP-index $ x^*_s $ of $ [\ell_s, r_s] $ in $ \Suff^* (\mathcal{T}) $ for every $ 1 \le s \le d $. We finally compute the BP-index $ x_s = \fromaux (x^*_s) $ of $ [\ell_s, r_s] $ in $ \Suff (\mathcal{T}) $ for every $ 1 \le s \le d $, and we retrieve $ \ell_s $ and $ r_s $ by computing $ \tointer(x_s) $.

    For every $ 1 \le i \le d $, we can compute $ \ell'_s $ and $ r'_s $ in $ O(1) $ time. Indeed, we know the BP-index $ x_s $ of $ [\ell_s, r_s] $ in $ \Suff (\mathcal{T}) $, so $ x'_s = \parent(\Suff (\mathcal{T}), x_s) $ is the BP-index of $ [\ell_s, r_s] $ in $ \Suff (\mathcal{T}) $, and we can retrieve $ \ell'_s $ and $ r'_s $ by computing $ \tointer(x'_s) $. We conclude that we can compute all the $ \ell'_s $'s and all the $ r'_s $'s in $ O(d) \subseteq O(occ_i) $ time.

    For every $ 1 \le i \le d $, we can compute $ \lambda_{\ell'_s, r'_s} $ in $ O(\log n) $ time by Corollary \ref{cor:computingelllr}. We conclude that we can compute all values $ \lambda_{\ell'_s, r'_s} $'s in $ O(d \log n) \subseteq O(occ_i \log n) $ time.

    We know that $ \Cdm_i (\mathcal{T}, P) = \bigcup_{s = 0}^d \Cdm_i^s (\mathcal{T}, P) $ and the union is disjoint, so to compute $ \Cdm_i (\mathcal{T}, P) $ it will suffice to compute $ \Cdm_i^s (\mathcal{T}, P) $ for every $ 0 \le s \le d $. Let us show that we can compute $ \Cdm_i^0 (\mathcal{T}, P) $ in $ O((1 + occ_i^0) \log n) $ time and, for $ 1 \le s \le d $, we can compute $ \Cdm_i^s (\mathcal{T}, P) $ in $ O(occ_i^s \log n) $ time (we can write $ O(occ_i^s \log n) $ instead of $ O((1 + occ_i^s) \log n) $ because we know that $ occ_i^s \ge 1 $ for $ 1 \le s \le d $). We will then conclude that we can compute $ \Cdm_i (\mathcal{T}, P) $ in $ O((1 + occ_i^0 + \sum_{s = 1}^d occ_i^s) \log n) = O((1 + occ_i) \log n) $ time.
    \begin{itemize}
        \item By Lemma \ref{lem:climbingsuffixtree}, we know that $ \Cdm_i^0 (\mathcal{T}, P) $ is the set of all $ 1 \le k \le n $ such that (i) $ k \in D_j $, with $ j \in [\ell_{P[i, t_i]}, r_{P[i, t_i]} ] $ and (ii) $ |\mathcal{T}_k| \le t_i - i + 1 $, so by choosing the instance of Lemma \ref{lem:computingY} where $ \ell = \ell_{P[i, t_i]} $, $ r = _{P[i, t_i]} $ and $ y = t_i - i + 1 $ we obtain $ Y = \Cdm_i^0 (\mathcal{T}, P) $, and we conclude that we can compute $ \Cdm_i^0 (\mathcal{T}, P) $ in $ O((1 + |Y|) \log n) = O((1 + occ_i^0) \log n) $ time.
        \item Fix $ 1 \le s \le d $. We know that $ \Cdm_s^0 (\mathcal{T}, P) $ is the set of all $ 1 \le k \le n $ such that (i) $ k \in D_j $, with $ j \in [\ell'_s, \ell_s - 1] \cup [r_s + 1, r'_s] $ and (ii) $ |\mathcal{T}_k| \le \lambda_{\ell'_s, r'_s} $. Now, we do not consider a single set $ Y $, but we consider two sets $ Y_1 $ and $ Y_2 $. The set $ Y_1 $ is defined by choosing the instance of Lemma \ref{lem:computingY} where $ \ell = \ell'_s $, $ r = \ell_s - 1 $ and $ y = \lambda_{\ell'_s, r'_s} $, and the set $ Y_2 $ is defined by choosing the instance of Lemma \ref{lem:computingY} where $ \ell = r_s + 1 $, $ r = r'_s $ and $ y = \lambda_{\ell'_s, r'_s} $. Then, $ \Cdm_s^0 (\mathcal{T}, P) $ is the disjoint union of $ Y_1 $ and $ Y_2 $, and we conclude that we can compute $ \Cdm_i^s (\mathcal{T}, P) $ in $ O((1 + |Y_1|) \log n) + O((1 + |Y_2|) \log n) \subseteq O((1 + occ_i^s) \log n) $  time.
    \end{itemize}
\end{proof}

\begin{remark}\label{rem:occlarge}
    Let us show that the time bound in Theorem \ref{theor:main} can be improved to $ O(m \log n + \min \{occ \log n, n \log n + occ \}) $, without needing to know $ occ $ in advance.

    First, let us prove that we can compute $ \Cdm(\mathcal{T}, P) $ in $ O((m + n) \log n + occ) $ time. We explicitly compute the arrays $ \SA $ and $ \LCP $ as follows. The array $ |\SA| $ has length $ n^* = \sum_{h = 1}^d |\rho_h| \le n $, and the array $ \LCP $ has length $ n' - 1 < n $, so we can retrieve $ \SA $ and $ \LCP $ in $ O(n \log n) $ time because retrieving each entry takes $ O(\log n) $ time (Theorem \ref{theor:sampledsuffixarray} and Theorem \ref{theor:lcpsampling}). Next, proceeding as in the proof of Theorem \ref{theor:main}, in $ O(m \log n) $ time we compute all the values $ t_i $'s, $ \ell_{P[i, t_i]} $'s and $ r_{P[i, t_i]} $'s.

    Let us show how to compute $ \Cdm(\mathcal{T}, P) $ in $ O(m + occ) $ time. To this end, it will suffice to show how to compute each $ \Cdm_i(\mathcal{T}, P) $ in $ O(1 + occ_i) $ time. By proceeding as in the proof of Theorem \ref{theor:main}, in $ O(1 + occ_i) $ time we can compute $ d $, all the values $ \ell_s $'s and $ r_s $'s, and all the values $ \ell'_s $'s and $ r'_s $'s.
    
    For every $ 1 \le i \le d $, we can compute $ \lambda_{\ell'_s, r'_s} $ in $ O(1) $ time (and not $ O(\log n) $ time) because the $ O(\log n) $ term in Corollary \ref{cor:computingelllr} only comes from accessing $ \LCP $, and we have explicitly computed the array $ \LCP $. We conclude that we can compute all values $ \lambda_{\ell'_s, r'_s} $'s in $ O(d) \subseteq O(occ_i) $ time.
    
    Let us show that we can compute $ \Cdm_i^0 (\mathcal{T}, P) $ in $ 1 + occ_i^0 $ time and, for $ 1 \le d \le s $, we can compute $ \Cdm_i^s (\mathcal{T}, P) $ in $ occ_i^s $ time. Then, we will conclude that we can compute $ \Cdm_i (\mathcal{T}, P) $ in $ O(1 + occ_i^0 + \sum_{s = 1}^d occ_i^s) = O(1 + occ_i) $ time. Notice that the $ O(\log n) $ time bound in the statement of Lemma \ref{lem:existencet^*} only comes from accessing $ \SA $, which can now be achieved in $ O(1) $ time because we have explicitly computed $ \SA $. The bound $ O((1 + |Y|) \log n) $ in Lemma \ref{lem:computingY} comes from applying Lemma \ref{lem:existencet^*} for $ O(1 + |Y|) $ steps, so it can be now replaced with the bound $ O(1 + |Y|) $. The conclusion follows by arguing as in the proof of Theorem \ref{theor:main}.

    We are now ready to prove that in $ O(m \log n + \min \{occ \log n, n \log n + occ \}) $ we can compute $ \Cdm(\mathcal{T}, P) $, without needing to know $ occ $ in advance. We can determine whether $ occ \ge n $ at runtime as follows. We start with the $ O((m + occ) \log n) $ algorithm, and if at some point we find out that $ occ \ge n $, we switch to the $ O((m + n) \log n + occ) $ algorithm. More precisely, we compute $ \Cdm_1(\mathcal{T}, P) $, $ \Cdm_2(\mathcal{T}, P) $, $ \dots $, $ \Cdm_m(\mathcal{T}, P) $ in this order. Assume that for some $ 1 \le i \le m $ we have $ \sum_{j = 1}^{i - 1} occ_j < n $. We now start computing $ \Cdm_i(\mathcal{T}, P) $. We determine $ d $. If $ d + \sum_{j = 1}^{i - 1} occ_i \ge n $, then $ occ \ge \sum_{j = 1}^i occ_j \ge d + \sum_{j = 1}^{i - 1} occ_i \ge n $ and we immediately switch to the $ O((m + n) \log n + occ) $ algorithm. If $ d + \sum_{j = 1}^{i - 1} occ_i \le n $, we start computing all the $ occ_i $ elements of $ \Cdm_i(\mathcal{T}, P) $. If at some point the number of elements in $ \Cdm_i(\mathcal{T}, P) $ that we have already computed is $ occ^*_i $, with $ occ^*_i + \sum_{j = 1}^{i - 1} occ_i \ge n $, we switch to the $ O((m + n) \log n + occ) $ algorithm. Notice that when we switch to the $ O((m + n) \log n + occ) $ algorithm we have only performed $ O((m + n) \log n) $ steps because we have only computed $ n $ occurrences of $ \Cdm(\mathcal{T}, P) $. This proves that we can compute $ \Cdm(\mathcal{T}, P) $ in $ O(m \log n + \min \{occ \log n, n \log n + occ \}) $ time, without needing to know $ occ $ in advance.
\end{remark}

\end{document}